%% file: main.tex
\documentclass[twocolumn]{aastex63}

\usepackage{afterpage}

\usepackage{booktabs} 
\let\tablenum\relax 
\usepackage{siunitx} 
\usepackage{amsmath,amstext,color}
\usepackage{gensymb}
\usepackage{natbib}
\usepackage{hyperref}
\usepackage{longtable}
\usepackage{threeparttablex}
\usepackage{booktabs}

\usepackage{color}
\usepackage{graphicx}
\usepackage{amsmath}
\usepackage{amssymb}

\usepackage[]{appendix}

\begin{document}
\title{A Hubble Space Telescope Search for $r$-Process Nucleosynthesis in Gamma-ray Burst Supernovae}
\correspondingauthor{Jillian Rastinejad}
\email{jillianrastinejad2024@u.northwestern.edu}

\shorttitle{Searching for r-Process in GRB-SNe}
\shortauthors{Rastinejad et al.}
\input{affiliation.tex}
\input{authors.tex}

\begin{abstract}
The existence of a secondary (in addition to compact object mergers) source of heavy element ($r$-process) nucleosynthesis, the core-collapse of rapidly-rotating and highly-magnetized massive stars, has been suggested by both simulations and indirect observational evidence. Here, we probe a predicted signature of $r$-process enrichment, a late-time ($\gtrsim 40$~days post-burst) distinct red color, in observations of GRB-supernovae (GRB-SNe) which are linked to these massive star progenitors. We present optical to near-IR color measurements of four GRB-SNe at $z \lesssim 0.4$, extending out to $> 500$~days post-burst, obtained with the {\em Hubble Space Telescope} and large-aperture ground-based telescopes. Comparison of our observations to models indicates that GRBs\,030329, 100316D and 130427A are consistent with both no enrichment and producing $0.01 - 0.15 M_{\odot}$ of $r$-process material if there is a low amount of mixing between the inner $r$-process ejecta and outer SN layers. GRB\,190829A is not consistent with any models with $r$-process enrichment $\geq 0.01 M_{\odot}$. Taken together the sample of GRB-SNe indicates color diversity at late times.
Our derived yields from GRB-SNe may be underestimated due to $r$-process material hidden in the SN ejecta (potentially due to low mixing fractions) or the limits of current models in measuring $r$-process mass. We conclude with recommendations for future search strategies to observe and probe the full distribution of $r$-process produced by GRB-SNe.
\end{abstract}


\section{Introduction}

For decades, the formation sites for many of the Universe's heaviest elements remained unknown. These elements, including many of those that enable life on Earth such as Thorium and Iodine, are widely believed to form via rapid neutron capture (``$r$-process'') nucleosynthesis \citep{Burbidge+57,Cameron57}. This process requires origin sites capable of producing high neutron abundance fractions that are common enough to create the observed yields within a Hubble time.
The traditional core-collapse of massive stars, in which material close to the remnant is neutronized (e.g., \citealt{Cowan+91,Woosley+94,HoffmanWoosley97,Thielemann+11,Cowan+21}), was initially proposed as an origin site but recently has fallen out of favor (as discussed further below). Binary neutron star (BNS) or neutron star-black hole mergers, in which a high neutron fraction is naturally achieved through tidal stripping of the outer layers of the neutron star in the final orbits before merging (e.g., \citealt{LattimerSchramm74,Rosswog+99}) or through outflows from the post-merger accretion disk (e.g., \citealt{Metzger+09}), became the favored channel to produce $r$-process elements. In 2017, BNS mergers were confirmed to produce at least a fraction of the Universe's $r$-process elements with the first BNS merger detected through gravitational waves (GW170817; \citealt{gw170817mma}), coincident short gamma-ray burst (GRB\,170817A; \citealt{gw170817_grb170817a,Goldstein+2017,Savchenko+17}) and thermal kilonova AT\,2017gfo \citep{Arcavi+17,Coulter+17,Lipunov+17,Tanvir+17,Soares-Santos+17,Valenti+17}. 

Despite the spectacular confirmation of $r$-process in BNS mergers, it remains an open question as to whether other heavy element formation channels are needed to explain the mass and abundance pattern of $r$-process elements observed in our Solar System. Indirect evidence in the form of early heavy element enrichment in our Galaxy provides a convincing argument for a site tied to the collapse of massive stars (e.g., \citealt{Cote+17,Hotokezaka+18,Naidu+22}). Observations of $r$-process-enhanced metal-poor stars in the Milky Way, some ultra-faint dwarf galaxies (e.g., \citealt{Ji+16,Hansen+17,Frebel+18}), and globular clusters (e.g., \citealt{Zevin+19,Kirby+20,Kirby+23}) suggest the existence of a heavy element formation channel with a short delay from star formation (\citealt{Kirby+23} find a delay $\lesssim 0.8$~Myr). Current estimates of the minimum delay time of BNS mergers, inferred from short GRB host galaxy offsets, stellar population ages, and star formation histories (e.g., \citealt{Fong+22,Nugent+22,O'Connor+22,Zevin+22,Nugent+23}) and predicted by simulations (e.g., \citealt{Belczynski+02,Dominik+12,Tauris+17,Mandhai+22}) remain uncertain, but it is unlikely that these events can provide a ubiquitous $r$-process enrichment source as quickly as massive star channels. 

Core-collapse SNe (CCSNe), associated with the deaths of massive stars and the formation of compact objects, provide a natural source with a short delay from star formation (on a stellar evolutionary timescale). In the 1990s, it was hypothesized that ordinary, jet-less CCSNe could produce neutron-rich material due to the high-entropy neutrino wind formed around the proto-neutron star remnant (e.g., \citealt{Woosley+94}). However, high event rates and observations of ordinary CCSNe have led to a general consensus that these events do not significantly contribute to the Universe's $r$-process budget (\citealt{MaciasRamirezRuiz18,Wallner+15,Wallner+21}; although see \citealt{Tsujimoto+01} who find that SN\,1987A's Ba/Sr ratio is more consistent with $r$-process than slow neutron capture, $s$-process). In addition, simulations have struggled to create or eject any neutron-rich material before accretion onto the remnant object (e.g., \citealt{ARcones+07,Fischer+10,MartinezPinedo+12,MartinezPinedo+14}). 

One suggested CCSNe source that could potentially decrease the event rates and overcome the ejection problem is magneto-rotational SNe (MR SNe; also termed jet-driven SNe or magneto-rotational hypernovae). In these MR SNe, rapid rotation of the iron core amplifies the magnetic field (e.g., \citealt{Cameron+03,Mosta+15}) launching jets and subsequent magneto-centrifugal winds (e.g., \citealt{Thompson+04,Metzger+07}) that provide a mechanism to eject neutron-rich material (see however \citealt{Mosta+14}). More recent simulations of MR SNe have found that strong ($B \approx 10^{13}$~G) and efficient magnetic fields are critical for this pathway to the $r$-process, though it is unknown how common magnetic fields of this strength are in nature (e.g., \citealt{HaleviMosta18,Mosta+18,Thompson+18}). At present, no models for observational signatures of $r$-process produced through the MR SN production channel have been published.

Recently, \cite{SiegelBarnesMetzger2019} demonstrated that accretion disks following rapidly-rotating massive stars undergoing core-collapse (``collapsars'') may also create and successfully eject $r$-process elements. Using magnetohydrodynamic (MHD) simulations they determined that the post-collapse disk favors weak interactions that produce neutron-rich material capable of synthesizing the heaviest elements. Their simulations also demonstrate that heating due to the disk's magnetic turbulence is sufficient to unbind the neutron-rich material, ejecting material in winds that will undergo the $r$-process, mix with the associated SN's outer layers and produce a red color signature analogous to the reddening of a kilonova (though on longer timescales; \citealt{SiegelBarnesMetzger2019,Zenati+20}). Additional simulations have further explored the dependency of $r$-process yields on neutrino treatment and accretion rate, and have disfavored collapsars as sources of heavy (e.g., lanthanide-rich) $r$-process material \citep{Miller+20,Fujibayashi+22,Just+22}. However, given the uncertainties on these parameters (e.g., assumed range of accretion rates, treatment of disk viscosity, assumed stellar structure and rotation profiles of progenitor models), $r$-process enrichment from collapsars remains plausible. While the MR SN and collapsar mechanisms represent distinct stages in the process of stellar core collapse, they are not mutually exclusive for a given event and in fact are associated with similar progenitor stars.

\citet{BarnesMetzger22} develop a semi-analytic light curve model for collapsar SNe enriched with $r$-process, which predicts their photometric color evolution for sufficiently large $r$-process enrichment levels. Their models produce a distinguishable red excess that emerges several weeks to months following the initial explosion. The model suite spans a range of $r$-process enrichment masses and degrees of mixing between the inner disk ejecta and SN outer layers, which imprint themselves on the light curves in the form of distinct optical to near-IR (NIR) colors. 

Despite extensive work on the theoretical end, few observational searches have been performed for $r$-process enhancement in SNe associated with collapsars. Searches for radio flares following long-duration GRBs (typically the product of collapsars) have been performed, a possible signature of interaction of the collapsar's wind ejecta with the surrounding medium \citep{Lee+22}. While no late-time radio flares were uncovered, there are several potential alternate sources that could explain radio emission in the event of such a discovery. Additionally, the high and sustained photospheric velocities inferred from observations of SN\,2020bvc have been interpreted as power by a heavy element mixing source such as the $r$-process \citep{Li+23}. However, other interpretations such as interaction with circumstellar material (CSM) or shock cooling remain plausible explanations for this event (e.g., \citealt{Izzo+20,Ho+20,Jin+21}) and there are no signs of reddening in the SN light curve.

\citet{Anand+24} perform a comprehensive search for $r$-process signatures in collapsar SNe using contemporaneous optical-NIR color measurements for a sample of 25 nearby broad-lined, stripped-envelope SNe (SNe Ic-BL) mostly discovered by the Zwicky Transient Facility (including one associated with a GRB, GRB\,190829A). SNe Ic-BL are often associated with collapsars \citep{MacFadyenWoosley99}, but may be explained with other mechanisms (e.g., \citealt{Kashiyama+16}). Their sample of SNe Ic-BL light curves was best fit by $r$-process-free models, favoring no or low $r$-process yields from nearby SNe Ic-BL detected mostly without GRBs.

The SNe Ic-BL associated with long GRBs (GRB-SNe) are generally considered the best targets for observable $r$-process enrichment in collapsars, in part due to the high angular momentum required to produce large accretion disks capable of launching the GRB jet (e.g., \citealt{MacFadyenWoosley99,SiegelBarnesMetzger2019,BarnesDuffell23}). In addition, \citet{BarnesDuffell23} find that GRB jets are likely to increase mixing between the inner $r$-process ejecta and the outer layers, thus producing a more prominent red color. As this effect is likely enhanced closer to the jet axis, GRB-SNe, which have relatively pole-on orientations, are strong candidates for observing $r$-process signatures. \citet{Blanchard+23} do not find signs of $r$-process enrichment from a late-time NIR spectrum of the SN counterpart to GRB\,221009A, though this observation was complicated by a bright afterglow. 

The absence of a previous search for photometric signatures of $r$-process enrichment in a sample of GRB-SNe is in part due to the low rates of events within the requisite volume to detect the faint signatures (e.g., $z\lesssim0.4$) . For many past low-redshift events, no published late-time NIR data exists. The paucity of these measurements reflects the NIR sensitivity often required to study even low-redshift SNe on the timescales of these signatures ($30 \lesssim \delta t \lesssim 300$~days, where $\delta t$ is the time since the GRB trigger).

Here, we provide late-time {\em Hubble Space Telescope} ({\it HST}) and large-aperture ground-based color measurements for a sample of four nearby GRB-SNe extending to $\gtrsim 500$~days following the GRB trigger. In Section~\ref{sec:obs} we describe our sample selection and detail the observations. In Section~\ref{sec:sn_corr} we describe our process of ascertaining the intrinsic SNe colors. In Section~\ref{sec:datacomp} we compare our observations to the $r$-process enriched SNe models of \citet{BarnesMetzger22}. In Section~\ref{sec:discussion} we review the implications of our work and discuss future observing strategies. Throughout, we assume a cosmology of $H_{0}$ = 69.6~km~s$^{-1}$~Mpc$^{-1}$, $\Omega_{M}$ = 0.286, $\Omega_{vac}$ = 0.714 \citep{Bennett+14} and report magnitudes in the AB system.

\input{grbs.tex}

\section{Observations}
\label{sec:obs}

\subsection{Sample Selection and Data Description}

As a starting point, we utilize the comprehensive GRB-SNe compilation of \citet{Dainotti+22}, which includes 58 long-duration GRBs with claimed SNe observed from 1990--2021. Following our motivation to identify NIR photometric excesses in the SN light curve due to the presence of $r$-process material, we narrow this sample to GRB-SNe events (i) for which late-time ($\delta t \gtrsim 30$~days), nearly simultaneous optical-NIR observations are available, (ii) identified after the discovery of SN\,1998bw, as we do not expect previous GRB-SNe to have well-sampled light curves, (iii) at $z<0.4$, the approximate distance out to which {\it HST} is capable of observing the predicted color evolution of $r$-process enriched GRB-SNe, and (iv) not associated with a putative kilonova (e.g., GRBs\,211211A and 230307A; \citealt{Rastinejad+22,Troja+22,Yang+22,Gillanders+23,Levan+23_230307a,Yang+23}). 
Applying these criteria, our final sample comprises four GRB-SNe: GRB\,030329 (SN\,2003dh), GRB\,100316D (SN\,2010bh), GRB\,130427A (SN\,2013cq), and GRB\,190829A (SN\,2019oyw). We also considered but ultimately did not include GRB\,221009A in our sample, due to a combination of high Galactic extinction (e.g., \citealt{Williams+23}), sparse late-time sampling of observations, and afterglow contamination in early {\it HST} epochs \citep{Levan+23_221009a}. We list the basic properties of these bursts in Table~\ref{tab:grbprops} and describe each GRB and our data reduction further in Sections~\ref{sec:030329}-\ref{sec:190829a}. Throughout this work, we refer to each event, including the SN, with its GRB name.

\begin{figure*}
\centering
\includegraphics[angle=0,width=.7\textwidth]{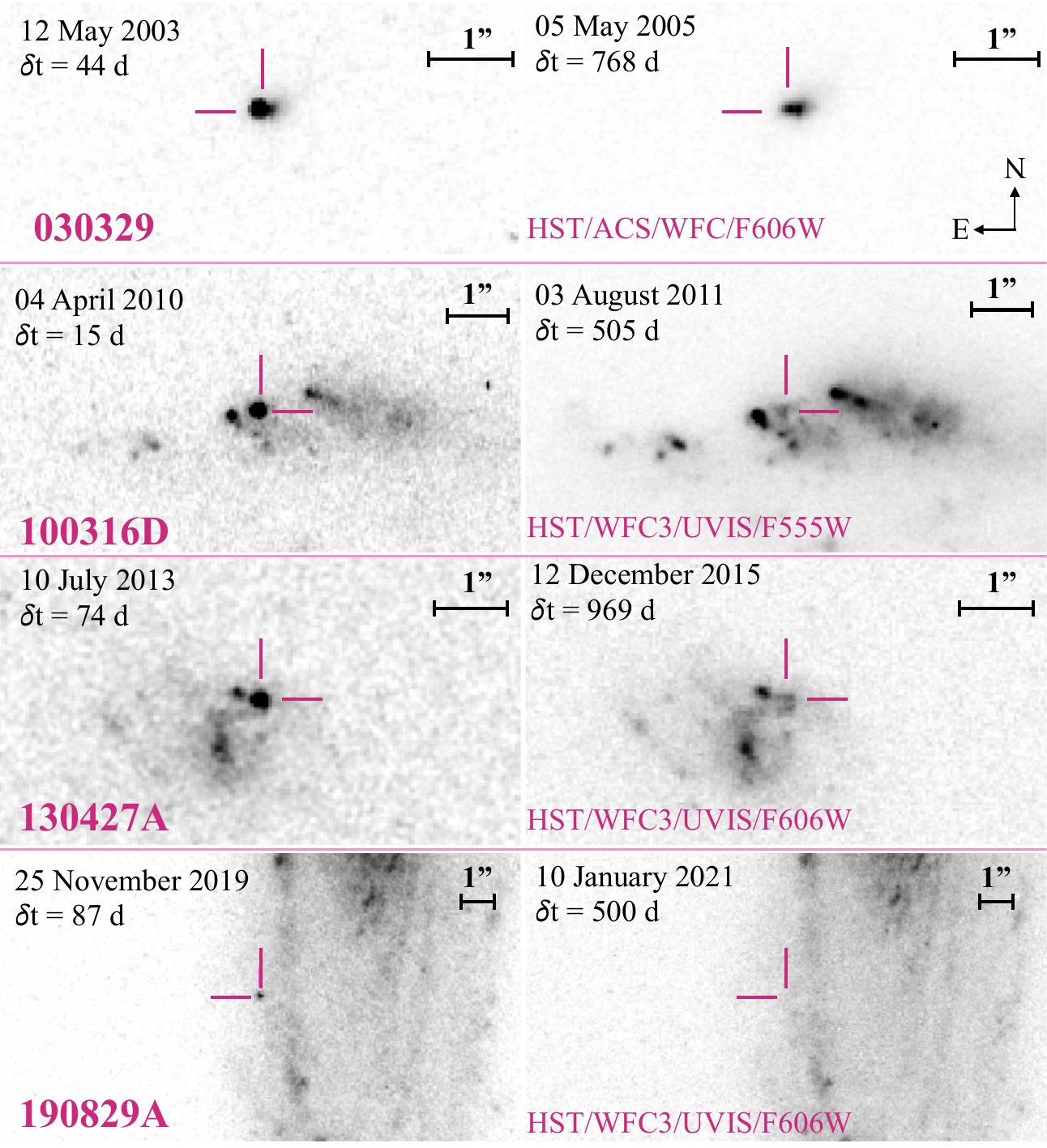}
\caption{Example {\it HST} images of the fields of the four GRBs in our sample in which the SN is detected (left) and the late-time templates (right). In each panel we show the position of the SN with pink crosshairs and note the time since GRB detection. We do not expect the SN or afterglow to be significantly contributing to any of our templates, with the potential exception of GRB\,130427A in F160W (see Section~\ref{sec:templates}).}
\label{fig:IMAGING}
\end{figure*}

In total, we collect 79 {\it HST}  observations obtained with the Advanced Camera for Surveys (ACS) Wide Field Channel (WFC), Wide Field Camera 3 (WFC3) Ultraviolet-Visible (UVIS) and Infrared (IR) channels, and the Near Infrared Camera and Multi-Object Spectrometer (NICMOS) Camera 2 (NIC2) instruments from the MAST archive\footnote{https://mast.stsci.edu/search/ui/\#/hst} and the Hubble Legacy Archive (HLA)\footnote{https://hla.stsci.edu}. The observations span $8 < \delta t < 969$~days post-burst (including template observations) and seven {\it HST} filters (Table~\ref{tab:grbprops}). We also collect and reduce VLT/X-shooter (acquisition camera), VLT/HAWK-I and MMT/Binospec observations of GRB\,190829A (Section~\ref{sec:190829a}). With the exception of the NICMOS/NIC2 imaging, all reported photometry is performed on image subtractions with a late-time ($\delta t \gtrsim 420$~days) template (see Section~\ref{sec:templates} for further discussion on potential template contamination). In Figure~\ref{fig:IMAGING} we show representative {\it HST} images where the SN is detected (left column) and the template image used for image subtraction (right column). We report all photometry in Table~\ref{tab:hstobs} and plot the observations in Figure~\ref{fig:lc_hst_sn}. In Section~\ref{sec:localdust} we discuss our corrections for Galactic and local dust extinction. 

To complement the photometry analyzed in this work, we gather additional relevant data from the literature. As our goal is to compare the optical-NIR color evolution of the GRB-SNe in our sample to relevant models, we collect only host-subtracted photometry in the $rRiIJHK$-bands of GRBs\,030329 \citep{Matheson+03}, 100316D \citep{Olivares+12}, 130427A \citep{Perley+14} and 190829A \citep{Hu+21}. This results in an additional 519 observations from the literature. The vast majority of these measurements occur at $0.1 \lesssim \delta t \lesssim 40$~days, extending our dataset at early times.

\subsection{GRB\,030329}
\label{sec:030329}

GRB\,030329 was discovered by the High Energy Transient Explorer II (HETE-II) at 11:14:14.67 UT on 29 March 2003 with a duration of $21$~s \citep{030329_gcn}. A bright optical afterglow counterpart was quickly localized. The burst's redshift of $z=0.1685$ was identified through afterglow spectroscopy \citep{Greiner+03}. Subsequent spectroscopic and photometric observations of the counterpart revealed evidence of an SN Ic-BL \citep{Hjorth+03,Kawabata+03,Matheson+03,Mazzali+03,Stanek+03,Lipkin+04}.

GRB\,030329 was observed with ACS/WFC in the F606W and F814W filters at several nearly contemporaneous epochs (within $\approx 24$~hours) over $17 \lesssim \delta t \lesssim 228$~days, and at $\delta t = 428$~days in F606W only (Program 9405; PI: Fruchter). We download the flat-fielded, dark-subtracted and CTE-corrected images and combine them using \texttt{astrodrizzle} \citep{DrizzlePac+12} with a pixel scale of $0.05\arcsec$. We note that while the F814W image observed on 12 November 2003 was partially contaminated by internally scattered light from the WFPC2 internal lamp, the position of the SN is not affected by the uneven background. 

We perform subpixel alignments between coadded images using \texttt{tweakreg} \citep{DrizzlePac+12} and align in image coordinates using standard IRAF tasks. We employ \texttt{HOTPANTS} \citep{becker15}, which uses point-spread function (PSF) convolution, and IRAF/\texttt{imarith} for image subtraction. In general, the subtraction methods produce consistent photometry. We use \texttt{HOTPANTS} as our default image subtraction software throughout this work as it often produces higher signal-to-noise residuals than \texttt{imarith}, likely because of the treatment of the slight variations in PSF due to focus and orientation changes. In select cases we find that \texttt{HOTPANTS} returned a pattern that does not resemble a point-source residuals, motivating us to employ \texttt{imarith}.
For each residual image, we detect a $\gtrsim 3\sigma$ residual consistent with the position of the SN \citep{Matheson+03} in the subtractions. We utilize the tabulated {\it HST} zeropoints to calibrate our images and perform aperture photometry with a 3--4 pixel aperture on the subtracted images (corresponding to $\sim 1 \times$ FWHM), accounting for the appropriate encircled energy corrections \citep{Bohlin16}.

GRB\,030329 was also observed in the F110W and F160W filters with NICMOS/NIC2 at $\delta t \approx 17, 23, 44$ and 228~days. We download the drizzled images from the HLA. We discard the F110W and F160W images observed at $\delta t = 17$~days due to saturation of the SN. As noted in previous works, these images suffer from known artifacts \citep{Ostlin+08} and have a narrow field-of-view, preventing robust alignment and thus, reliable image subtraction. Thus, we perform relative photometry to obtain our measurements by subtracting the flux of the host galaxy in the epoch at $\delta t = 228$~days from those in each of the initial three epochs. We correct our photometry for the NICMOS/NIC2 encircled energy corrections listed in the NICMOS Handbook\footnote{https://www.stsci.edu/hst/instrumentation/legacy/nicmos}.

\subsection{GRB\,100316D}
\label{sec:100316D}

The sub-energetic, low-luminosity GRB\,100316D was discovered with the Neil Gehrels {\it Swift} Observatory \citep[\textit{Swift};][]{Gehrels+04} at 12:44:50 UT on 16 March 2010. The burst duration was $277$~s and its spectrum was noticeably soft \citep{Sakamoto+!0}. The Swift X-Ray Telescope (XRT) promptly identified and localized an X-ray counterpart. Shortly thereafter, multiple telescopes observed an optical counterpart embedded in a galaxy at $z=0.0591$
(e.g., \citealt{Chornock+10,Starling+11}). Spectroscopic features of a SN Ic-BL were observed in the optical counterpart a few days following the initial trigger \citep{Chornock+10,Starling+11,Bufano+12,Olivares+12}.

GRB\,100316D was observed with WFC3/UVIS in the F555W and F814W filters, and with WFC3/IR in the F110W and F160W filters (Programs 11709, 12323; PI: Bersier) over $8 \lesssim \delta t \lesssim 505$~days. We reduce the {\it HST} data in the same manner as described in Section~\ref{sec:030329}. We perform image subtraction for each epoch using \texttt{HOTPANTS} and photometry on the \texttt{HOTPANTS} residual images using a 3--4 pixel aperture ($\sim 1-2 \times$ FWHM) and account for the appropriate encircled energy corrections\footnote{www.stsci.edu/hst/instrumentation/wfc3/data-analysis/photometric-calibration/uvis-encircled-energy \\ www.stsci.edu/hst/instrumentation/wfc3/data-analysis/photometric-calibration/ir-encircled-energy}.  We note that the SN in the F125W and F160W images on 4 April 2010 ($\delta t \approx 16$~days) is saturated and we do not include these images in our analysis.
We find our early photometry is in reasonable agreement with the analysis of this data reported in \citet{Cano+11}. At $\delta t \gtrsim 24$~days, we find our values are fainter by $\sim 0.3 - 1.2$~mag compared to their values. This difference can be ascribed to our process of image subtraction in comparison to their direct photometry on the host-embedded SN \citep{Cano+11}.

\subsection{GRB\,130427A}
\label{sec:130427a}

GRB\,130427A was detected by {\it Swift} and the {\it Fermi Space Telescope} (\textit{Fermi}; \citealt{Meegan+09}) on 27 April 2013 at 07:47:06.42 UT. Its gamma-ray properties were unprecedented: at the time, the GRB had the highest observed fluence and most energetic photon recorded to date (\citealt{GBM_130427a,Ackerman+14}; holding the fluence record until the recent discovery of GRB\,221009A; e.g., \citealt{Burns+23}). The burst duration as seen by {\it Swift} was 244~s. Prompt optical afterglow spectroscopy identified the GRB at $z=0.3399$ \citep{Levan+13_130427a} and multi-wavelength follow-up revealed an extraordinarily luminous afterglow (e.g., \citealt{Laskar+13,Maselli+14,Perley+14}). Spectroscopy of the counterpart confirmed the presence of an associated SN Ic-BL \citep{Xu+13,Melandri+14,Levan+14}.

GRB\,130427A was observed with ACS/WFC and WFC3/UVIS in the F606W filter and WFC3/IR in the F160W filter for 7 contemporaneous epochs over $74 \lesssim \delta t \lesssim 969$~days (Programs 13110, 13117, 13230, 13951; PIs: Fruchter, Levan). We perform image subtractions between the initial six observations and the template image observed at $\delta t = 969$~days, then perform photometry on the residual with a 3 pixel aperture. Due to striping on the F160W image observed on 2014 December 5, we do not see a $\gtrsim 3 \sigma$ significant residual and discard the image. For all other epochs, we perform photometry at the position of the residual with a 3 pixel aperture and correct for the encircled energy. 

\subsection{GRB\,190829A}
\label{sec:190829a}

GRB\,190829A was identified by the {\it Swift}, {\it Fermi} and Konus-Wind satellites on 29 August 2019 at 19:55:53 UT (time as discovered by Fermi) and promptly localized to a galaxy with a known redshift of $z=0.0785$ \citep{dichiara_190829a,fermi_190829a,konus_190829a}. The burst had a gamma-ray duration of $53$~s. The early afterglow was detected across the electromagnetic spectrum (e.g., \citealt{Rhodes+20,HESS+21,Dichiara+22}), and later spectroscopic observations revealed features consistent with an SN Ic-BL \citep{Hu+21,Anand+24}.

GRB\,190829A was observed with {\it HST} with WFC3/UVIS in F606W and WFC3/IR in F140W over six contemporaneous epochs spanning $87 \lesssim \delta t \lesssim 500$~days (Programs 15510, 16042, 16320; PIs: Levan, Tanvir). We also incorporate early ($29 < \delta t < 58$~days) images observed in the F110W and F160W filters (Program 15089; PI: Troja). We download, combine and align the F606W and F140W images using the procedure described above, and perform image subtractions with \texttt{HOTPANTS}. A residual at the SN position is detected at $\gtrsim 3 \sigma$ significance in all subtracted images. We perform photometry on all residual images (and directly on the F110W and F160W images as host contamination is insignificant at these epochs) with a 3--4 pixel aperture and correct for the encircled energy using the tabulated values.

In addition, we present VLT and MMT imaging of GRB\,190829A observed over $25 < \delta t < 141$~days. VLT observations were obtained in the $r$-band with the X-shooter acquisition camera and in the $JHK_s$-band with the HAWK-I instrument (Programs 0103.D-0819(A), 2103.D-5067(A), 105.20N7.001, 0104.D-0600(E), 103.202P.002; PIs: Levan, Tanvir). We retrieve the X-shooter images from the ESO archive facility and reduce $r$-band images using standard IRAF/\texttt{ccdproc} tasks. We use the fully reduced HAWK-I images from the ESO archive. We reduce the $r$-band MMT/BINOSPEC (Program 2019C-UAO-G199, UAO-G205-23B; PIs: Fong, Rastinejad) images using a custom Python pipeline\footnote{https://github.com/CIERA-Transients/POTPyRI}. For all bands, we utilize a template image obtained with the same instrument at $\delta t \gtrsim 500$~days for {\tt HOTPANTS} image subtractions. We calibrate our images using SDSS ($r$; \citealt{Alam+15}) or 2MASS ($JHK$; \citealt{2MASS}) stars in the field. Finally, where a residual at the position of the SN is detected at $> 3 \sigma$ level, we perform aperture photometry with IRAF. We report all our observations in Table~\ref{tab:hstobs}.

\subsection{Assessing Transient Contamination in Template Images}
\label{sec:templates}

Across all events, we consider the potential for afterglow or SN contamination in our {\it HST} template images, which may affect image subtraction residuals and bias our reported colors. Given that our sources are embedded in their host galaxies and likely to occur in star-forming regions (e.g., \citealt{Blanchard+16,Lyman+17}), thus making simple visual inspection difficult, we apply an analytic model to determine the expected magnitude of the SN or afterglow at the time of the template image. 

\begin{figure*}
\centering
\includegraphics[angle=0,width=0.9\textwidth]{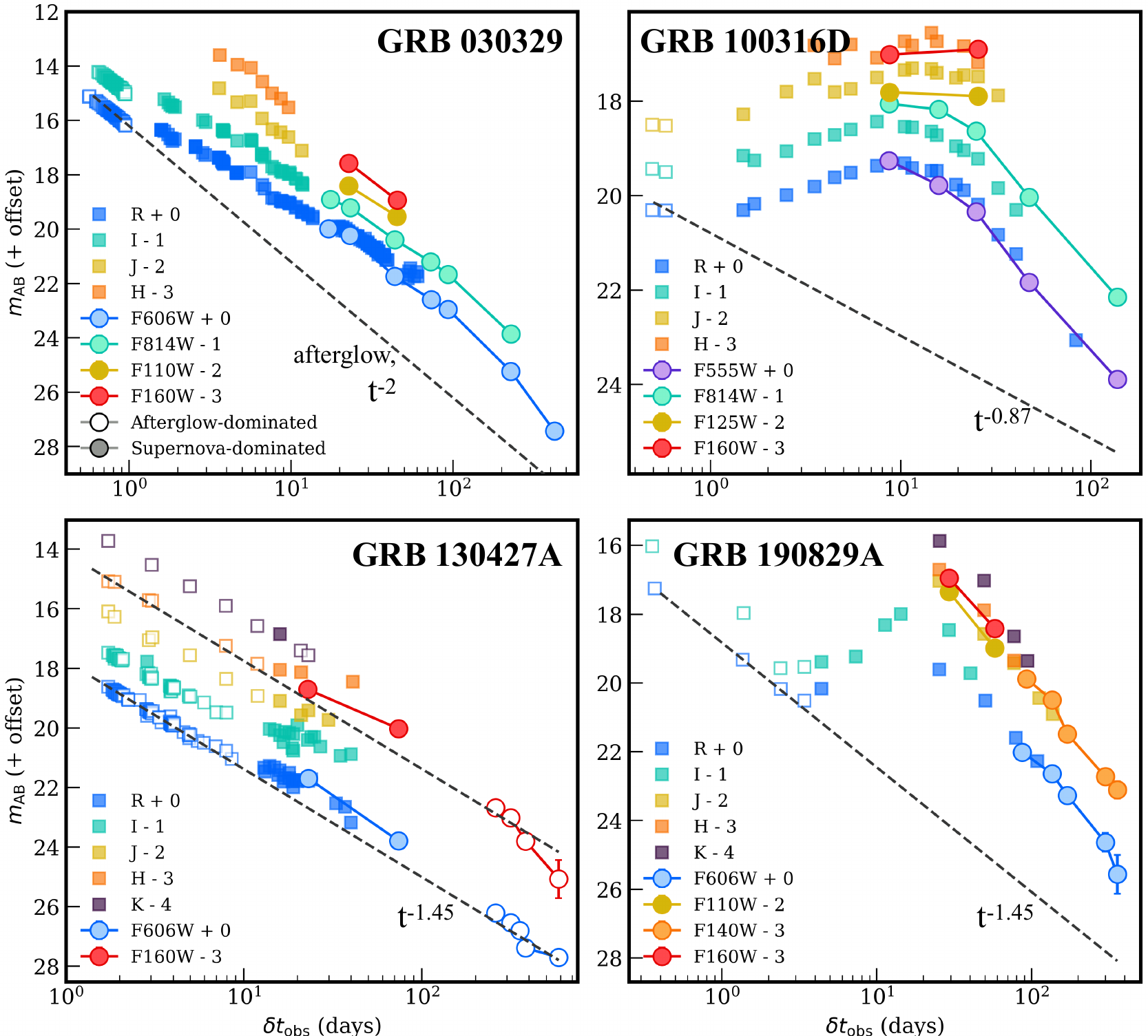}
\caption{All data used in this work for GRBs\,030329, 100316D, 130427A and 190829A in apparent magnitude ($m_{\rm AB}$) versus observed time after the GRB triggers ($\delta t_{\rm obs}$), corrected for Milky Way and local dust extinction. Ground-based observations are shown with squares while {\it HST} observations are represented with circles. In each panel we show the extrapolated $r$-band (and $H$-band for GRB\,130427A) afterglow decay with a dashed line (further described in Section~\ref{sec:ag_contrib}) and mark observations that are dominated or significantly ($> 0.3$~mag difference) contaminated by afterglow flux with open symbols. For instance, we expect that our {\it HST} observations of GRB\,130427A observed beyond 200~days are contaminated or dominated by afterglow flux.}
\label{fig:lc_hst_sn}
\end{figure*}

As we will show in Section~\ref{sec:ag_contrib}, for GRBs\,030329, 100316D and 190829A we expect the SN to dominate at late times, while for GRB\,130427A we expect the afterglow to be the main source of emission at the time of the template image (Figure~\ref{fig:lc_hst_sn}). To assess the SN contribution for the former GRB-SNe, we fit a simple $^{56}$Ni and $^{56}$Co decay model (e.g., \citealt{Arnett82,Tinyamont+22,Kilpatrick+23_20jfo}) to the late-time ($\delta t \gtrsim 110$~days) optical (typically, the $r$, $R$, F606W and/or F555W bands) light curve. We then extrapolate the expected magnitude to the time of the template image. We find that the SN contribution in the template images ranges between $m = 27.8 - 30.7$~mag with the greatest potential contribution for GRB\,100316D. However, as all of the GRB\,100316D photometry is significantly brighter than this limit ($m\leq24.6$~AB mag) we consider any template contribution negligible. To assess the afterglow contribution for GRB\,130427A we extrapolate the afterglow decay to late times using the parameters described in Section~\ref{sec:ag_contrib}, and predict any SN or afterglow contamination to be $m_{\rm F606W} \gtrsim 28.6$~mag and $m_{\rm F160W} \gtrsim 28.0$~mag. We note that this may indicate transient contamination that would impact later ($\delta t \gtrsim 300$~days) F160W observations. However, as we justify further in Section~\ref{sec:ag_contrib}, we do not use these late-time color measurements to probe SN $r$-process enrichment as they are likely dominated by afterglow emission. Further, as our earlier ($\delta t \approx 22, 74$~days) SN-dominated observations of GRB\,130427A are brighter than $23$~mag, we do not expect contamination in the template to affect our SN colors. 

\bigskip
\section{Determination of the Intrinsic Supernova Colors}
\label{sec:sn_corr}

\subsection{Correcting for Line-of-Sight Dust Extinction}
\label{sec:localdust}

Dust extinction along the GRB's line of sight may significantly contaminate the observed optical-NIR color of the SNe in our sample though, notably, it does not impact the relative color evolution. However, it is important to take into account because particularly high extinction may result in a reddened light curve that will affect any identification or constraints on $r$-process material. In addition to corrections for Galactic extinction, we correct for extinction from the local environment of the GRB (Table~\ref{tab:grbprops}), the assumptions of which we discuss below.

We utilize local or, when not available, host galaxy extinction values ($E(B-V)_{\rm loc}$) from the literature \citep{Matheson+03,Cano+11,Bufano+12,Levan+14,Chand+20,huang+23} and apply a Small Magellanic Cloud (SMC) extinction law \citep{CCM+89, Gordon+03} to determine corrections in each filter. We employ the SMC extinction law as it is a well-studied dust relation for a stellar population with a higher specific star formation rate compared to the Milky Way, appropriate for long GRB host galaxies \citep{Kann+06,Schady+12}. We note that the choice of extinction law will not significantly impact our optical-NIR corrections (as opposed to the ultraviolet regime).

For GRBs\,030329, 100316D and 130427A, the derived values are low ($E(B-V)_{\rm loc} < 0.15$~mag; Table~\ref{tab:grbprops}) and where multiple values exist in the literature, they are consistent within $\approx 0.10$~mag. For GRBs\,100316D and 130427A, we take values of $E(B-V)_{\rm loc}$ measured from spectroscopy of the afterglows covering the Na I D doublet ($\lambda\lambda 5890, 5896$) and calculated using a Milky Way gas-to-dust ratio (Table~\ref{tab:grbprops}; e.g., \citealt{Bufano+12,Xu+13}). For GRB\,130427A this value is consistent with $E(B-V)_{\rm loc}$ calculated from the multi-wavelength afterglow spectrum (\citealt{Perley+14}). For GRB\,030329 there is no measurement from afterglow spectroscopy or fitting, so we utilize a value derived from the H$\alpha$/H$\beta$ ratio in the host galaxy spectrum (\citealt{Matheson+03}). Literature values for local extinction of GRB\,190829A are significantly higher. Using the XRT and {\it Swift}/Ultra-Violet Optical Telescope (UVOT) spectrum, \citet{Chand+20} derive $E(B-V)_{\rm loc} =1.04$. We note there is also a measurement $E(B-V)_{\rm loc} = 0.64$ derived from the XRT light curve alone \citep{huang+23}, but we prefer the value that incorporates UVOT data as this encompasses the regime where dust is most pronounced.

\subsection{Characterizing the Afterglow Contribution}
\label{sec:ag_contrib}

The GRB afterglow, typically modeled as synchrotron emission from the interaction of the jet with the surrounding medium, is expected to dominate over any SN emission in the few days following a burst (though this timescale may be highly variable). On week- to month-long timescales, SN emission typically dominates in the optical band given the afterglow's steep power law decay (though there are notable exceptions in which the afterglow dominates for several months; e.g., GRB\,221009A; \citealt{Levan+23_221009a}). In the majority of cases, we expect that follow-up on timescales beyond a $\sim$month will probe the SN's color evolution. Therefore, the {\em HST} observations extending to very late times should be free from afterglow contamination. However, as using color evolution to determine $r$-process-enrichment in a GRB-SN is our primary objective, it is critical to determine which observations are clearly dominated by the SN emission.

Hence, we undertake a literature search to determine the temporal evolution ($F \propto t^\alpha$, where $F$ is the observed flux) for GRB afterglows to extrapolate as needed to the timescales of our observations. For GRB\,030329 we utilize the post jet-break (observed at $\approx 0.5$~day) afterglow decay of $\alpha = -2$ \citep{Lipkin+04,Moss+23}. We note that at early times this afterglow is complex and contaminated by flares (e.g., \citealt{Lipkin+04,Tiengo+04,Kamble+09}), but we expect its late-time behavior to decline smoothly. For GRB\,100316D, the optical afterglow was faint compared to the SN even at early times, and we use the late-time X-ray temporal slope of $\alpha = -0.87$ \citep{Margutti+13}. For GRB\,130427A we use $\alpha = -1.45$ measured from the multi-wavelength afterglow \citep{Perley+14}. We note that \citet{Maselli+14} observe a jet break in the X-ray and optical light curves at $\delta t \approx 10$~hours post-burst, followed by a decline of $\alpha = -1.36$. We prefer the values of \citet{Perley+14} which do not detect a jet break as their dataset includes greater temporal coverage (see also \citealt{dePasquale+16} who do not find evidence for a jet break in X-ray observations out to $\sim 2.5$~years). Finally, for GRB\,190829A we utilize the $r$-band temporal index of $\alpha = -1.45$ from early GTC imaging \citep{Hu+21}. Apart from GRB\,130427A, none of the GRBs in our sample have late-time jet breaks suggested in the literature. 

We employ the above values for $\alpha$ and early observations to extrapolate the afterglow decay in each band and determine in which observations the predicted afterglow flux dominates our observations or is within $0.3$~mag of the observed value. We plot the expected $r$-band afterglow contributions in Figure~\ref{fig:lc_hst_sn} as grey dashed lines.  In Figure~\ref{fig:lc_hst_sn} and Table~\ref{tab:hstobs} we denote observations that are likely afterglow-dominated, which we ignore in the analysis that follows. We do not anticipate any significant afterglow contamination in the light curves of GRBs\,030329, 100316D and 190829A past $\delta t \approx 4$~days. However, under our assumption of no observed jet break, the afterglow significantly contributes to or dominates the SN in the light curve of GRB\,130427A at $\delta t \gtrsim 200$~days (Figure~\ref{fig:lc_hst_sn}).

\subsection{Supernova Dust Contribution}

We briefly consider the possibility that dust produced by the SN may contribute to any observed reddening. The timescales for the production of dust in CCSNe remain uncertain, though new observations with {\it JWST} are beginning to constrain the dust abundance in Type II SNe on $\sim$decade timescales (e.g., \citealt{Hosseinzadeh+23,Shahbandeh+23}). However, unlike these Type II SNe, our sample includes only relativistic, stripped-envelope GRB-SNe, as identified by the broad-lined features in their spectra. The mean absorption velocity measured for a large sample of GRB-SNe is $\sim 0.07 c$ (\citealt{Modjaz+16}; see also \citealt{Mazzali+21}), indicating a high shock speed that would destroy any precursor dust grains before they may amass significantly. In addition, with the exception of a few rare cases, stripped-envelope SNe like GRB-SNe are rarely observed to have significant CSM, making any contribution from pre-existing dust unlikely (e.g., \citealt{Prentice2019,Szalai+21}). 

In principle, newly-formed dust may contaminate our later observations. We do not find strong observational or theoretical evidence in the literature that significant dust is formed in GRB-SNe on $\lesssim 3$-year timescales. There is evidence for reddening due to dust formation in other stripped envelope (e.g., SN\,2013ge; \citealt{Drout+14}) or highly energetic Type I superluminous SNe (SN\,2017ens; e.g., \citealt{Sun+22}). However, these events cannot be directly compared to GRB-SNe due to either lower ejecta velocities (in the case of SN\,2013ge) or evidence for greater CSM interaction \citep{Margutti+23}. Notably, the early optical light curves of our sample of GRB-SNe resemble that of an afterglow with no excess luminosity due to reprocessing of emission in an ejecta-CSM shock interaction (Figure~\ref{fig:lc_hst_sn}).

\section{Exploring $r$-Process Enrichment in Our Observational Sample}
\label{sec:datacomp}

\subsection{Direct Comparison of the Sample to Models}

We compare our extinction-corrected SN observations to the semi-analytic radiation transport models of \citet{BarnesMetzger22} for a collapsar SN enriched with $r$-process material. In this collapsar $r$-process scenario, weak interactions within the dense and hot accretion disk feeding the black hole favor neutronization of the disk material (above a critical accretion rate $\sim 10^{-3}-10^{-2}M_{\odot}$ s$^{-1}$; \citealt{De&Siegel21}). A fraction of this neutron-rich midplane material is then ejected in disk winds, which undergoes $r$-process nucleosynthesis in the outflow on large scales. This process is similar to the disk outflows which contribute significantly to $r$-process production and kilonova emission in neutron star mergers \citep{SiegelBarnesMetzger2019,BarnesMetzger22}. However, unlike in a merger, $r$-process disk wind ejecta collide and subsequently mix with the outer (non-$r$-process enriched) layers of the SN. The degree and radial extent of this mixing are uncertain and may be enhanced in the presence of a jet, resulting in a viewing angle dependence \citep{BarnesDuffell23}. 

\citet{BarnesMetzger22} predict that $r$-process material mixed with typical SN Ic-BL ejecta will produce a distinguishable red photometric color (distinct from the natural blue to red evolution of $r$-process-free models) that becomes pronounced on timescales of a few weeks to months. Naturally, larger dynamic ranges in filters hold larger discriminating power between models. The existing suite of models assumes a spherical distribution of ejecta parameterized by total SN ejecta mass ($M_{\rm tot}$), $^{56}$Ni ejecta mass ($M_{\rm 56Ni}$), average ejecta expansion velocity as a fraction of the speed of light ($\beta_{\rm ej}$), mass of $r$-process material ($M_{\rm rp}$) and the mixing coordinate ($\psi_{\rm mix}$). $\psi_{\rm mix}$ describes the radial extent out through the SN ejecta to which the $r$-process enriched layers are mixed homogeneously \citep{BarnesMetzger22,Anand+24}. The limit $\psi_{\rm mix}=1$ corresponds to the $r$-process material being homogeneously mixed all the way to the ejecta surface), resulting in a more pronounced and prolonged red color (Figure~\ref{fig:lc_with_models}). The $R-H$ color difference between the $r$-process-free and enriched cases may be up to $\approx 3$~mag but is typically $\lesssim 0.1$~mag prior to $\delta t \sim 30$~days and in the cases $\psi_{\rm mix} \lesssim 0.2$ \citep{BarnesMetzger22}.

\begin{figure}
\centering
\includegraphics[angle=0,width=\linewidth]{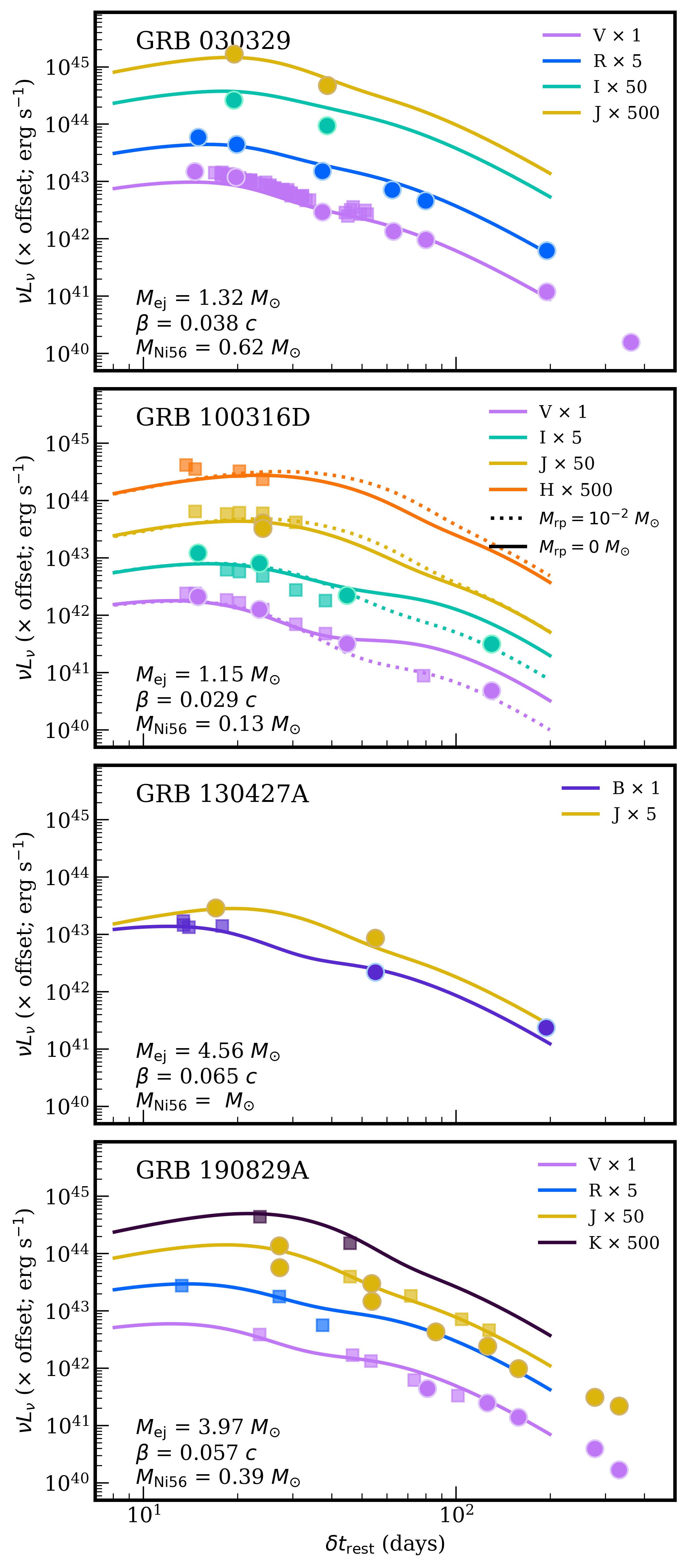}
\caption{Best-fit $r$-process-free (solid lines) models of four GRB-SNe in our sample and relevant rest-frame observations from our sample. We show the best-fit values of $M_{\rm ej}, \beta_{\rm ej}$, and $M_{\rm 56Ni}$ for each GRB-SNe. We also show the best-fit model for GRB\,100316D, the only event whose minimum $\chi^2$ model was enriched with $r$-process, when compared to the total (including $r$-process enriched) grid of models. We fix these values of $M_{\rm ej}, \beta_{\rm ej}$, and $M_{\rm 56Ni}$ for each GRB-SNe throughout our analysis.}
\label{fig:lc_lum}
\end{figure}

\begin{figure*}
\centering
\includegraphics[angle=0,width=\textwidth]{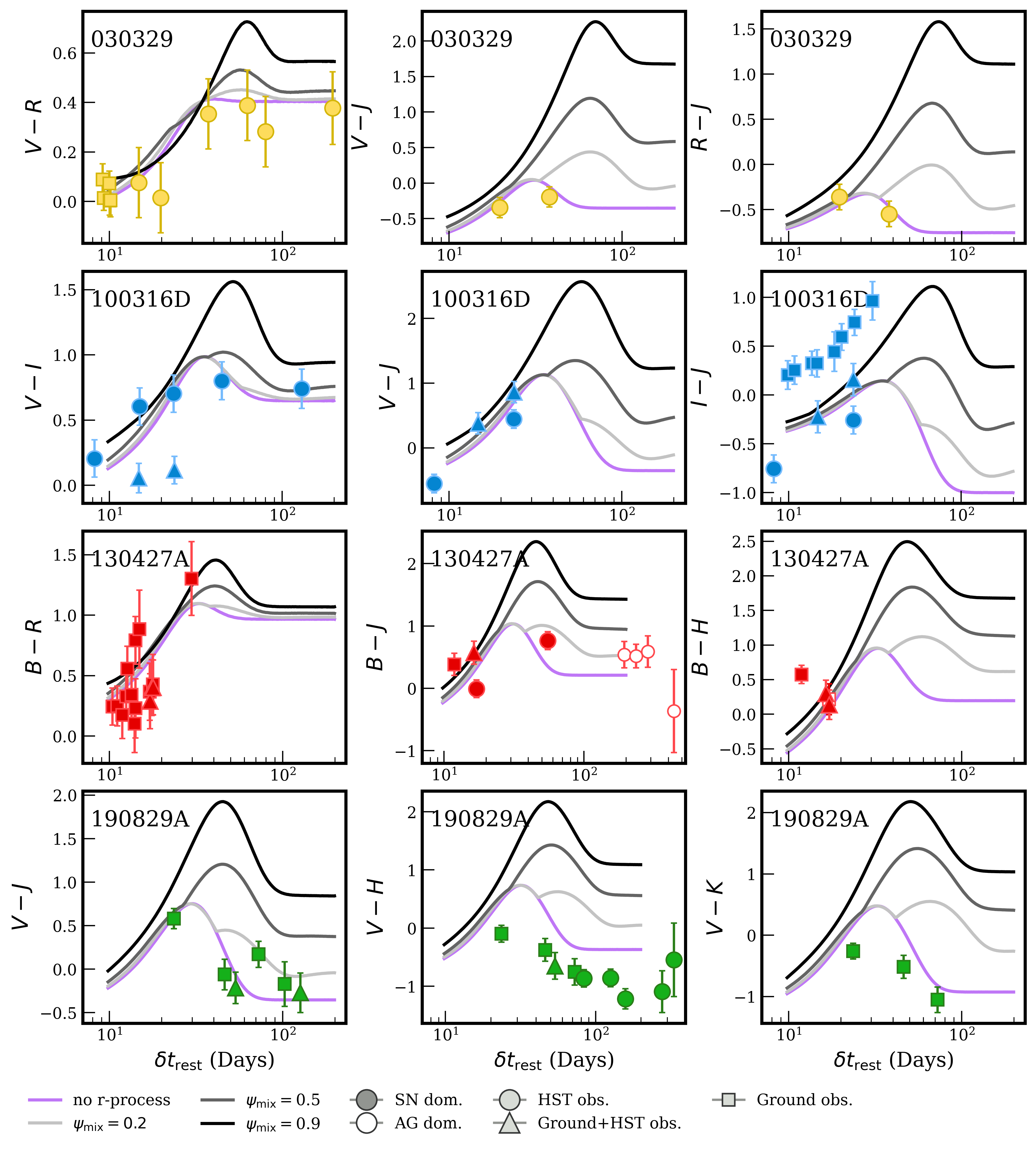}
\caption{Assorted colors of the four GRB-SNe in our sample versus rest-frame time compared to best-fit models from \citet{BarnesMetzger22} (Figure~\ref{fig:lc_lum} and Section~\ref{sec:datacomp}) with fixed $r$-process mass ($M_{\rm rp} = 0.03$) and varying values of $\psi_{\rm mix}$ (lines). We also plot models without $r$-process enrichment in purple. Afterglow-dominated observations are shown as open symbols, and measurements dominated by SN emission are filled symbols. Though color measurements between filters vary, none of our late-time measurements favor high values of $\psi_{\rm mix}$. Many measurements, particularly those that extend to late times, are unable to distinguish between models with $\psi_{\rm mix} \lesssim 0.2$ and those with no $r$-process. GRB\,190829A is, in the best-sampled $V-H$ filters, bluer than the unenriched models.}
\label{fig:lc_with_models}
\end{figure*}

We attempt to constrain the number of free parameters so we are left with only $M_{\rm rp}$ and $\psi_{\rm mix}$. To determine the values of $M_{\rm ej},~\beta_{\rm ej}$, and $M_{\rm 56Ni}$ that produce model light curves most comparable to our observations, we convert our observations to the rest-frame, and compare observations at $\delta t_{\rm rest} = 12-200$~days to the grid of $r$-process-free ($M_{\rm rp} = 0$) models spaced in values of  $M_{\rm ej}, \beta_{\rm ej}$, and $M_{\rm 56Ni}$ (described in \citealt{BarnesMetzger22}). We determine the best-fit parameters using $\chi^2$ minimization. We show the best-fit models and their parameters along with relevant rest-frame observations in Figure~\ref{fig:lc_lum}. Our best-fit parameters are comparable to those found in the literature for GRBs\,100316D, 130427A and 190829A (e.g., \citealt{Cano+17_review,Hu+21}) though we find a lower $M_{\rm ej}$ for GRB\,030329 compared to previous analyses \citep{Mazzali+03,Cano+17_review}.

To ensure that our initial choice of best-fit model based on a grid of $r$-process-free models does not bias our later conclusions about heavy element enrichment, we also run $\chi^2$ minimization over the full grid of models, including both enriched and unenriched cases. For GRBs\,030329, 130427A, and 190829A the best-fit model remains the $r$-process-free case shown in Figure~\ref{fig:lc_lum}. However, for GRB\,100316D the best-fit model is enriched with $M_{\rm rp} = 0.01 M_{\odot}$ and highly mixed ($\psi_{\rm mix} = 0.9$; dotted lines in Figure~\ref{fig:lc_lum}). The values of $M_{\rm ej}, \beta_{\rm ej}$, and $M_{\rm 56Ni}$ for this model only vary slightly from those found for the $r$-process-free best-fit model, and we do not expect the colors to vary significantly based on these parameters. We thus conclude that our use of SN parameters determined from fits to $r$-process-free models in subsequent analysis will not significantly affect our conclusions about enrichment.

We next plot these models against our observations in color space to determine if the color evolution of any GRB-SNe in our sample resembles that modeled for $r$-process enrichment. We use the models described above and shown in Figure~\ref{fig:lc_lum} enriched with $M_{\rm rp} = 0.03 M_{\odot}$, a moderate value consistent with theoretical yields (e.g., \citealt{SiegelBarnesMetzger2019}). In Figure~\ref{fig:lc_with_models} we show the three best-sampled colors for each burst. We calculate colors for observations taken within three days of each other across ground- and space-based facilities. We combine color errors in quadrature, and incorporate an additional 0.1~mag error term to account for differences between the model (output in Johnson filters) and HST bandpasses. Models are available in the Johnson $UBVRIJHK$-bands. We compare our observations to the nearest approximate rest-frame model band and correct for time dilation using the redshift of the GRB (Table~\ref{tab:grbprops}). We explore how the GRB-SN color evolution compares to models as a function of mixing fraction, and also compare to an $r$-process free model. We consider color measurements to be consistent with the model if they fall within $2\sigma$ errors.

For GRB\,030329, the latest available colors are rest-frame $V-R$, where model color differences are small between the $r$-process-free and enriched cases. Still, we find that half the $V-R$ measurements past 40~days are consistent with the highly-mixed case, while all are consistent with enrichment and $\psi_{\rm mix} < 0.5$.  GRB\,030329's optical-NIR ($V-J$ and $I-J$) colors are only consistent with the $r$-process-free model or low mixing $\psi_{\rm mix} \lesssim 0.2$ model.

For GRB\,100316D, our rest-frame $V-I$ HST measurements are also both consistent with $r$-process-enriched ($\psi_{\rm mix} \lesssim 0.5$) and unenriched models. For this GRB, the discrepancy between F814W and ground-based $i$-band measurements can likely be ascribed to their different bandpass coverages and the steepening of the observed spectrum around $7500-8000 \AA$ \citep{Chornock+10}. Optical-NIR observations of GRB\,100316D do not appear to favor either enriched or $r$-process-free models, though, as we note above, results are inconsistent between telescopes. Later NIR observations of this burst would be necessary to distinguish between enriched and unenriched models. 

In GRB\,130427A, the uncertainties and timing of rest-frame ground-based $B-R$ and $B-H$ data are not appropriate to distinguish between strong mixing or $r$-process-free models. The one optical-NIR color measurement at late times, ($B-J$ at $\delta t_{\rm rest} = 55$~days) is consistent with only the enriched model with $\psi_{\rm mix} = 0.1 - 0.2$. Unfortunately, observations at $\gtrsim 180$~days are likely contaminated by the afterglow and are thus not appropriate for this analysis (Section~\ref{sec:ag_contrib}; Figure~\ref{fig:lc_hst_sn}).

Finally, the $V-J$ colors of GRB\,190829A are reasonably matched to both the $r$-process-free and low $\psi_{\rm mix}$ models. On the other hand, the $V-H$ and $V-K$ colors are bluer than even the bluest $r$-process-free model. We note that this effect is also found for a number of SNe Ic-BL found without GRBs \citep{Anand+24}. Overall, we find that this GRB-SN provides a poor fit to the models, though the colors are more in line with the trends of the $r$-process-free case.

In summary, we find that, with a few exceptions, observations of GRBs\,030329, 100316D and 130427A favor no enrichment or low values of $\psi_{\rm mix}$ for $M_{\rm rp} = 0.03 M_{\odot}$ (Figure~\ref{fig:lc_with_models}), and inference between filters may vary. Most of these GRBs' color measurements are not on sufficient timescales to distinguish between $r$-process-free and low $\psi_{\rm mix}$ values. On the other hand, the well-sampled color measurements of GRB\,190829A provide a strong case for no $r$-process enrichment at the level of the models used in this section ($M_{\rm rp} = 0.03 M_{\odot}$). Future color measurements with the cadence and long baseline similar to GRB\,190829A would allow for more detailed population studies.

\subsection{Observed Color Diversity Amongst GRB-SNe Sample}
\label{sec:diversity}

\begin{figure}
\centering
\includegraphics[angle=0,width=0.46\textwidth]{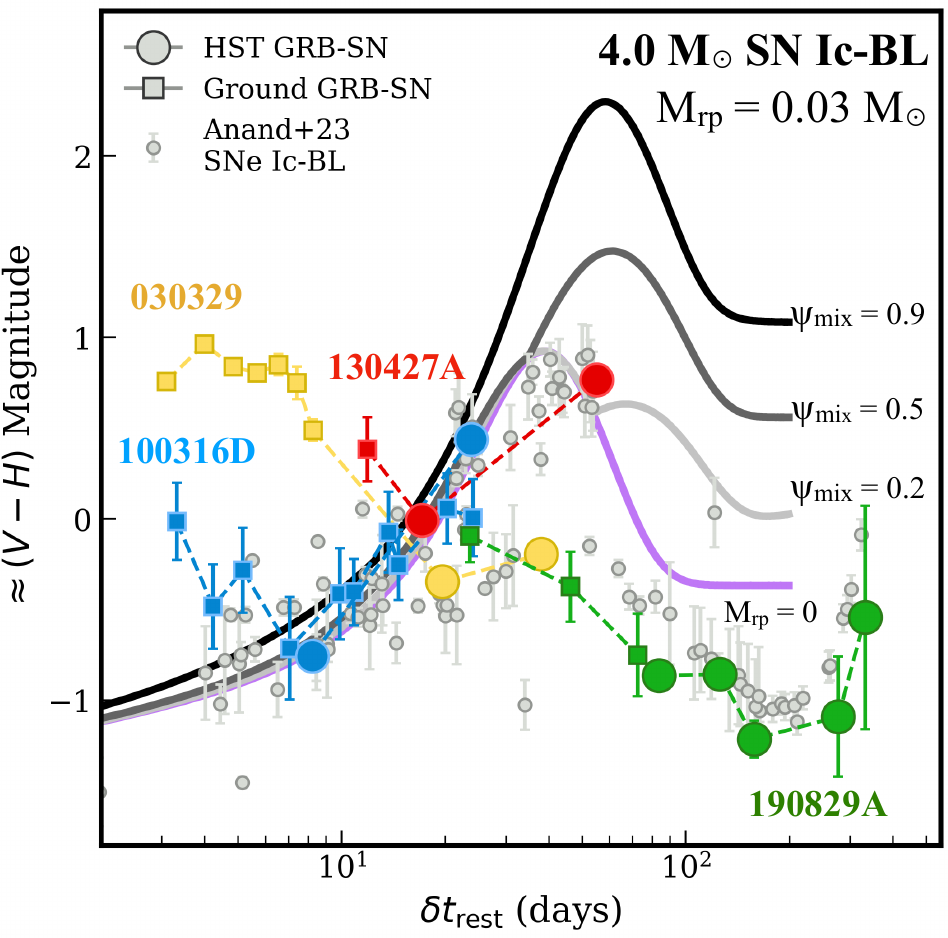}
\caption{The approximate rest-frame $V-H$ color evolution of SN-dominated observations of GRB\,030329 (yellow), GRB\,100316D (blue), GRB\,130427A (red), and GRB\,190829A (green), corrected for Galactic and local extinction. Ground-based observations are shown with squares, while observations obtained with {\it HST} are represented with circles, and extend the majority of the color evolution curves significantly. Against these observations we plot color evolution models for a moderate-mass SN Ic-BL enriched with 0.03 $M_{\odot}$ $r$-process material (greyscaled; color gradient corresponding to $\psi_{\rm mix}$) or $r$-process-free (purple; \citealt{BarnesMetzger22}). Broadly, the observations are consistent with both models free of $r$-process material and those with low mixing fractions. There is $\gtrsim 1$~magnitude of diversity in the color evolution of the bursts in our sample between $20 \lesssim \delta t \lesssim 80$~days.}
\label{fig:lc_diversity}
\end{figure}

Already, we observe diversity in color evolution within our sample of 4 GRB-SNe. In Figure~\ref{fig:lc_diversity} we plot SN-dominated color measurements for each GRB that are most closely matched to the $V-H$ filters (chosen due to late-time data availability). While for GRBs\,030329, 100316D and 190829A the rest-frame filters are reasonably comparable, the higher redshift of GRB\,130427A results in bluer rest-frame filters ($\approx B - J$) for the data plotted. Using the \citet{BarnesMetzger22} models we predict the k-corrections of low- and high-redshift rest-frame bands produce color differences up to $\approx 0.6$~mag (up to $\approx 0.3$~mag for the three lower-redshift GRBs). 

Figure~\ref{fig:lc_diversity} highlights the diversity within the GRB-SNe sample in terms of their color evolution.
At $20 \lesssim \delta t \lesssim 60$~days GRB\,100316D's SN is reddening rapidly, GRB\,030329's SN is more slowly reddening, and GRB\,190829A's SN is becoming bluer (Figure~\ref{fig:lc_diversity}). This behavior may be explained by differences in $M_{\rm ej}$ or $V_{\rm ej}$, or perhaps $M_{\rm rp}$ or $\psi_{\rm mix}$ between the GRBs.

In Figure ~\ref{fig:lc_diversity} we also compare the GRB-SNe to the large SNe Ic-BL compilation of \citet{Anand+24}. This sample includes color measurements for 25 nearby SNe Ic-BL, none of which show strong evidence for $r$-process enrichment when fit to models. Observations from this sample are corrected for Galactic extinction but not local extinction, though none of their SNe spectra indicate strong local dust. In general, the color evolution of the GRB-SNe in our sample is consistent with those of the \citet{Anand+24} sample, indicating there is no difference in $V-H$/$R-H$ color evolution and observed diversity between SNe Ic-BL discovered with and without GRB counterparts. This observed trend contrasts with theoretical expectations that GRB-SNe are more likely to produce larger $M_{\rm rp}$ and $\psi_{\rm mix}$ (and thus more pronounced red colors; \citealt{SiegelBarnesMetzger2019,BarnesDuffell23}) due to their high angular momentum and accretion disk sizes. Rather, we do not find evidence for differences in color evolution between GRB-SNe and SNe Ic-BL observed without jets. 

\begin{figure*}
\centering
\includegraphics[angle=0,width=\textwidth]{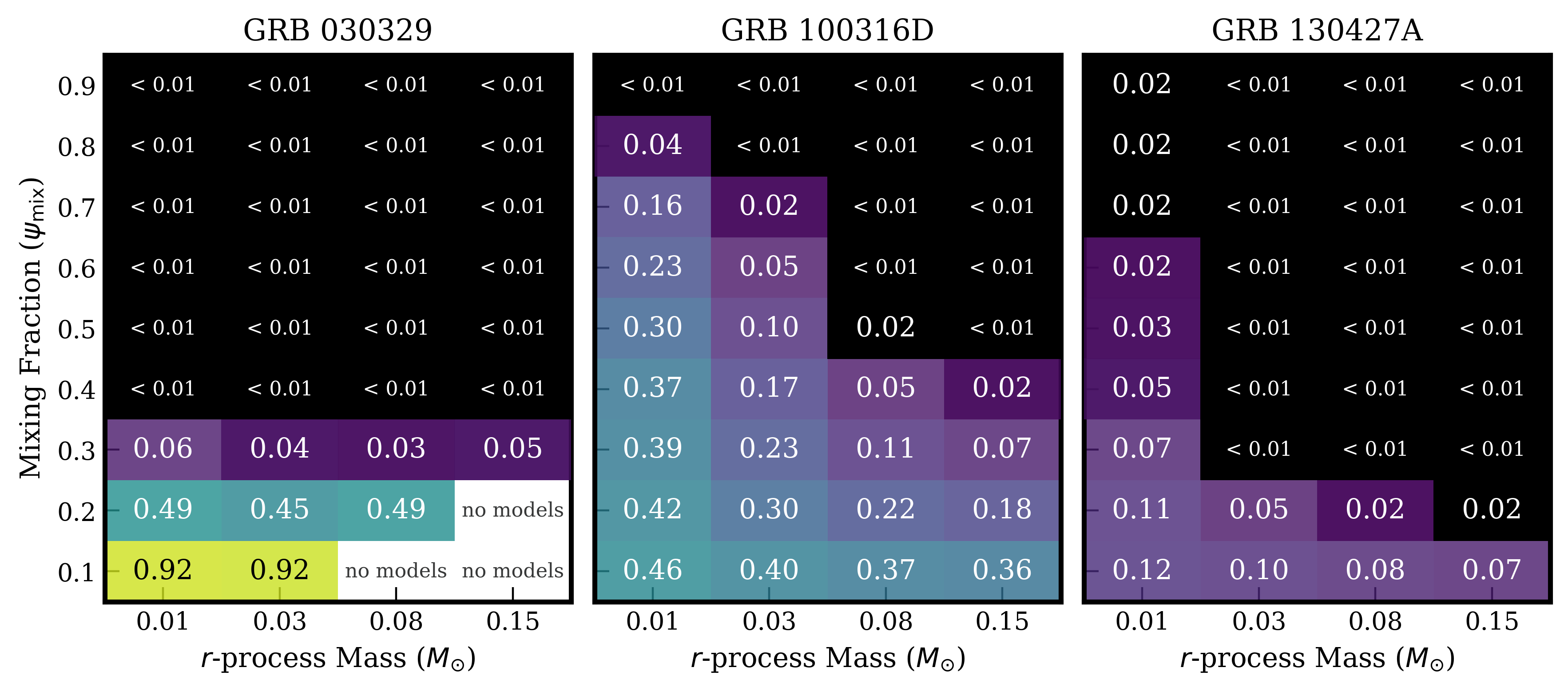}
\vspace{-.3cm}
\caption{Comparison of our datasets to large grids of models for a range of $r$-process masses and mixing fractions. Each cell in the grid represents a model from \citet{BarnesMetzger22} parameterized by the corresponding $M_{\rm rp}$ and $\psi_{\rm mix}$. Using the total color measurements per GRB meeting our criteria (Section~\ref{subsec:mrp_constraint}), we evaluate a $\chi^2$ value and report one-sided $p$-values for each model in the grid. The color shade and labeled number of each cell corresponds to the $p$-value of the model parameterized by that combination of $M_{\rm rp}$ and $\psi_{\rm mix}$. Models whose $p$-values are less than 0.02 are shaded black. Together, our observations favor low $\psi_{\rm mix}$ in nearly all cases but are consistent with the full range of $M_{\rm rp}$. We do not show a grid for GRB\,190829A as no models have $p$-values $>0.02$.}
\label{fig:mrp_xmix}
\end{figure*}

\subsection{Quantitative Constraints on $M_{\rm rp}$ and $\psi_{\rm mix}$}
\label{subsec:mrp_constraint}

We next consider a larger range of $M_{\rm rp}$ values and constrain $M_{\rm rp}$ and $\psi_{\rm mix}$ for each GRB-SN. To do this, we employ large grids of the best-fit models that are parameterized by combinations of $M_{\rm rp}$ and $\psi_{\rm mix}$ \citep{BarnesMetzger22}. We employ the same best-fit parameters for GRBs\,030329, 100316D, 130427A, and 190829A as in Section~\ref{sec:datacomp} (e.g., Figure~\ref{fig:lc_lum}) and hold these constant across the grid. The grids are linearly spaced in $\psi_{\rm mix}$ between 0.1 and 0.9, and at fixed values, $M_{\rm rp} = $ 0.01, 0.03, 0.08 and 0.15 $M_{\odot}$ \citep{BarnesMetzger22}. We consider only observations taken between $20 < \delta t < 200 \times (1 +z_{\rm GRB})$~days as the color differences on early timescales are negligible between models.  For each model, we determine a $\chi^2$ value using the $1\sigma$ color error. As in Section~\ref{sec:datacomp}, we account for bandpass differences between the models and observations with an additional 0.1~mag error. Finally, we convert the $\chi^2$ value to a $p$-value using the \texttt{scipy.chi2.sf} function.

In Figure~\ref{fig:mrp_xmix} we show the $M_{\rm rp}$-$\psi_{\rm mix}$ parameter space. We color code and label each cell according to the $p$-value of the model parameterized by the corresponding $M_{\rm rp}$-$\psi_{\rm mix}$ pair. Black space indicates that models for that $M_{\rm rp}$-$\psi_{\rm mix}$ pair are not consistent with observed colors ($p < 0.02$), and are ruled out in our analysis. White space indicates that no models were created for those coordinates.

For GRB\,030329, consistent models have $\psi_{\rm mix} \leq 0.3$. Our analysis does not provide strong constraints on $M_{\rm rp}$. For GRB\,100316D a wide portion of the parameter space is consistent with observations, although in general lower values of both $\psi_{\rm mix}$ and $M_{\rm rp}$ are favored. For GRB\,130427A we favor lower values for both parameters, though multiple combinations are consistent with observations. $M_{\rm rp} = 0.15 M_{\odot}$ is consistent only in the case where $\psi_{\rm mix} = 0.1$. Across the GRBs in our sample that are consistent with enriched models, observations disfavor high values of $\psi_{\rm mix}$ (except in some cases of $M_{\rm rp} = 0.01 M_{\odot}$ for GRB\,100316D). We do not find strong constraints on $M_{\rm rp}$ across $\psi_{\rm mix}$ values from our observations.

Finally, for GRB\,190829A none of the parameter combinations are consistent with observations, in line with our finding in Section~\ref{sec:datacomp} that enriched models are, in general, bluer than the observed colors. As GRB\,190829A has a high line-of-sight dust extinction that has not been measured with afterglow spectroscopy (Section~\ref{sec:localdust}), we re-run our analysis (including finding a new best-fit model) for our data corrected for a lower extinction value of $E(B-V)_{\rm loc} = 0.64$. Though the lower extinction value produces redder overall colors, it still does not provide any models for the grid of $\psi_{\rm mix}-M_{\rm rp}$ combinations with $p > 0.02$.

\section{Discussion}
\label{sec:discussion}

\subsection{Connection Between $\psi_{\rm mix}$, $\gamma$-ray and SN Properties}

We consider any potential connections between the $\gamma$-ray, SN and $r$-process enrichment properties and compare them against theoretical predictions from the literature. From 2D hydrodynamical simulations, \citet{BarnesDuffell23} propose that GRBs with the longest rest-frame $\gamma$-ray durations will be accompanied by SNe with higher $\psi_{\rm mix}$ values, as both properties are correlated with a long-lived, more massive disk wind. In addition, they find that a higher initial SN explosion energy is associated with lower values of $\psi_{\rm mix}$. 

In Section~\ref{sec:datacomp}, we concluded that, within the sample, GRB\,100316D observations are consistent with the highest mixing values (up to $\psi_{\rm mix} \approx 0.7$) while the observed colors of GRBs\,030329 and 130427A prefer $\psi_{\rm mix} \lesssim 0.3$ (Figure~\ref{fig:mrp_xmix}).  Comparing the results of these events to the $T_{90}$ rest-frame $\gamma$-ray durations listed in Table~\ref{tab:grbprops} and literature values of the SN explosion energy (e.g., \citealt{Mazzali+03,Cano+17_review,Cano+17,Hu+21}), we explore trends in our sample. GRB\,100316D (for which observations are consistent with $\psi_{\rm mix} \approx 0.4$) has the longest rest-frame $T_{90}$ $\gamma$-ray duration (a lower limit of 261~s as seen by \textit{Swift}) and the lowest SN explosion energy \citep{Cano+17_review} in our sample. We note that this GRB belongs to the low-luminosity class, which may be indicative of a less energetic central engine or shock breakout. GRBs\,030329 and 130427A (for which we deduce $\psi_{\rm mix} \leq 0.3$) have somewhat shorter $\gamma$-ray duration (rest-frame 18 and 182~s as seen by \textit{HETE-2} and \textit{Swift}) and higher estimated SN explosion energy \citep{Cano+17_review}. Though these are just three examples, they align with the trends predicted by \citet{BarnesDuffell23}. An expansion of $M_{\rm rp}$-$\psi_{\rm mix}$ constraints for GRB-SNe is necessary to determine if these trends hold within a statistically significant sample.

The findings of \citet{BarnesDuffell23} suggest that ultra-long GRBs (ULGRBs; \citealt{Levan+14_ULGRBs}) may be the ideal candidates for production of $r$-process elements.\footnote{On the other hand, if the accretion rate onto the black hole is too low, the disk may not neutronize in the first place (e.g., \citealt{SiegelBarnesMetzger2019,De&Siegel21}), precluding the production of $r$-process elements.} We consider but ultimately do not include several ULGRBs in our sample due to their high redshifts ($z\gtrsim0.6$; e.g., \citealt{Levan+14_ULGRBs,Greiner+15}) or an absence of confirmed SN counterpart (e.g., GRB\,130925A; \citealt{Evans+14,Piro+14}). However, these factors may also limit future rates of nearby ULGRBs suitable for deep NIR follow-up.

\subsection{Universal $r$-Process Enrichment Implications}
\label{sec:rp_universe}

As an exercise, we use our constraints to infer an average $r$-process mass from GRB-SNe, quantify the contribution of these events to the Universe's $r$-process budget, and compare our estimate to the Milky Way $r$-process enrichment. We caution that these results are highly model dependent and there are high uncertainties in these constraints. For this exercise, we separately consider the derived yields of the three GRBs in our sample consistent with $M_{\rm rp} \approx 0.01 - 0.15 M_{\odot}$ for low values of $\psi_{\rm mix}$. While a range of $r$-process yields are expected, the median is unknown at this time \citep{SiegelBarnesMetzger2019,BarnesMetzger22}. Specifically, we employ a yield of $M_{\rm rp} = 0.05 M_{\odot}$ based on the average of consistent $M_{\rm rp}$ values with $p > 0.05$ (Figure~\ref{fig:mrp_xmix}). We note that, with the exception of GRB\,100316D, all of our events were best-fit with an $r$-process-free model, though they return $p>0.05$ values for some enriched models (e.g., Section~\ref{sec:datacomp}).

We modify the equation of \citet{Rosswog+18} for events that produce $r$-process to determine the Milky Way contribution:
\begin{equation} \small
    M_r \sim 17\,000 M_{\odot}\left[\frac{\mathcal{R}_{\rm rp}}{500{\rm Gpc^{-3} yr ^{-1}}}\right]
    \left[\frac{\bar{m}_{\rm ej}}{0.03 M_{\odot}}\right]\left[\frac{\tau_{\rm gal}}{1.3 \times 10^{10} {\rm yr}}\right]
\end{equation}
\noindent where $\mathcal{R}_{\rm rp}$ is the event rate, $\bar{m}_{\rm ej}$ is the average $r$-process ejecta per event, and $\tau_{\rm gal}$ is the Milky Way age. We fix $\tau_{\rm gal} = 1.3 \times 10^{10}~{\rm yr}$ for all calculations. We first consider the case in which only CCSNe associated with long-duration GRBs (LGRBs) produce $r$-process such that $\mathcal{R}_{\rm rp} = \mathcal{R}_{\rm LGRB}$($z=0$) $= 79^{+57}_{-33}$ Gpc$^{-3}$ yr$^{-1}$ \citep{Ghirlanda+22}. This rate is modeled using the distributions of observed parameters such as fluence, $T_{90}$, and jet opening angle from \textit{Fermi}, the \textit{Compton Gamma Ray Observatory} and \textit{Swift} \citep{Ghirlanda+07,Ghirlanda+22}. For a yield of $M_{\rm rp} = 0.05 M_{\odot}$ this results in a total contribution of $M_{r} \sim 4500^{+2600}_{-7700} M_{\odot}$. This is significantly below the calculated total $r$-process of the Milky Way of $M_{\rm r, MW} \approx 23\,000\,M_{\odot}$ (elements of nucleon number $A \geq 69$; \citealt{Hotokezaka+18} measured from Europium abundances of local stars from \citealt{Venn+04,BattistiniBensby16}). We note that other works find the total $M_{\rm r, MW}$ may vary by several thousand $M_{\odot}$ (e.g., \citealt{Bauswein+14,Rosswog+18}).  However, taking $M_{\rm r, MW} = 23\,000 M_{\rm \odot}$ and considering the yields of both GRBs\,100316D and 190829A, we determine that GRB-SNe produce $11-34 \%$ of the Milky Way's $r$-process abundance. If only GRB-SNe produce $r$-process amongst CCSNe, either their average $M_{\rm rp}$ yields must be significantly higher than what we derive for GRB\,100316D and GRB\,190829A, the rates of GRB-SNe are higher than used in our estimate (indeed, the LGRB rate of \citealt{Ghirlanda+22} is based on bright GRBs and the rate of low-luminosity GRBs such as GRB\,100316D may be higher) or GRB-SNe are a subdominant $r$-process production channel compared to BNS mergers.

As our above analysis indicates that GRB-SNe alone cannot account for $M_{\rm r, MW}$, we consider the case in which collapsars, identified by SN Ic-BL alone, synthesize $r$-process elements. From the ZTF Bright Transient Survey, the total CCSNe rate is ($10.1^{+5.0}_{-3.5}$) $\times~10^4$ Gpc$^{-3}$ yr$^{-1}$ of which SNe Ic-BL represent $\approx 4$\% \citep{Perley+20}. Combined, this results in a rate $\mathcal{R}_{\rm rp} = \mathcal{R}_{\rm Ic-BL}$($z=0$) $= 3\,030$ Gpc$^{-3}$ yr$^{-1}$. Combined with our yield estimate of $M_{\rm rp} = 0.05 M_{\odot}$, this gives $M_{r} \sim 170\,000~M_{\odot}$ which significantly overpredicts Milky Way $r$-process enrichment. Thus, either only a fraction of SNe Ic-BL without GRBs produce $r$-process, or they are not a significant contributor, deduced from the significant mismatch in total $r$-process mass. This finding is consistent with the analysis of \citet{Anand+24}.

Our above calculations are dependent on model assumptions of $r$-process observables in GRB-SNe. Notably, the abundance pattern of elements produced by CCSNe are highly uncertain which likely affects the observed spectral energy distribution (SED) and, thus, the resulting model colors. On the modeling side, abundances depend on observationally unconstrained parameters such as the black hole accretion rate in the collapsar scenario. Typical CCSNe are not expected to produce the heaviest $r$-process elements (but may produce up to the first peak; e.g., \citealt{Wang+23}) while collapsars may produce up to the third peak elements under high accretion rates or strong magnetic fields (but may still struggle to produce actinides; \citealt{SiegelBarnesMetzger2019}). We note that the models used in our analysis assume a $r$-process SED based on the kilonova AT\,2017gfo, which likely produced up to and beyond the lanthanide elements ($A \geq 140$; e.g., \citealt{kasen+17}). Future models that account for mixing of only lighter $r$-process elements combined with {\it JWST} NIR spectroscopy (see Section~\ref{sec:futureobs}) may help to identify and delineate the abundance pattern. Notably, unlike BNS mergers, long GRBs are localized to galaxies with young ages and high specific star formation rates. These locations lend themselves to enriching early generations of stars, especially compared to BNS mergers which more frequently occur on the outskirts of galaxies \citep{Fong+22,Mandhai+22,O'Connor+22,vandeVoort+22}.

\begin{figure*}
\centering
\includegraphics[angle=0,width=\textwidth]{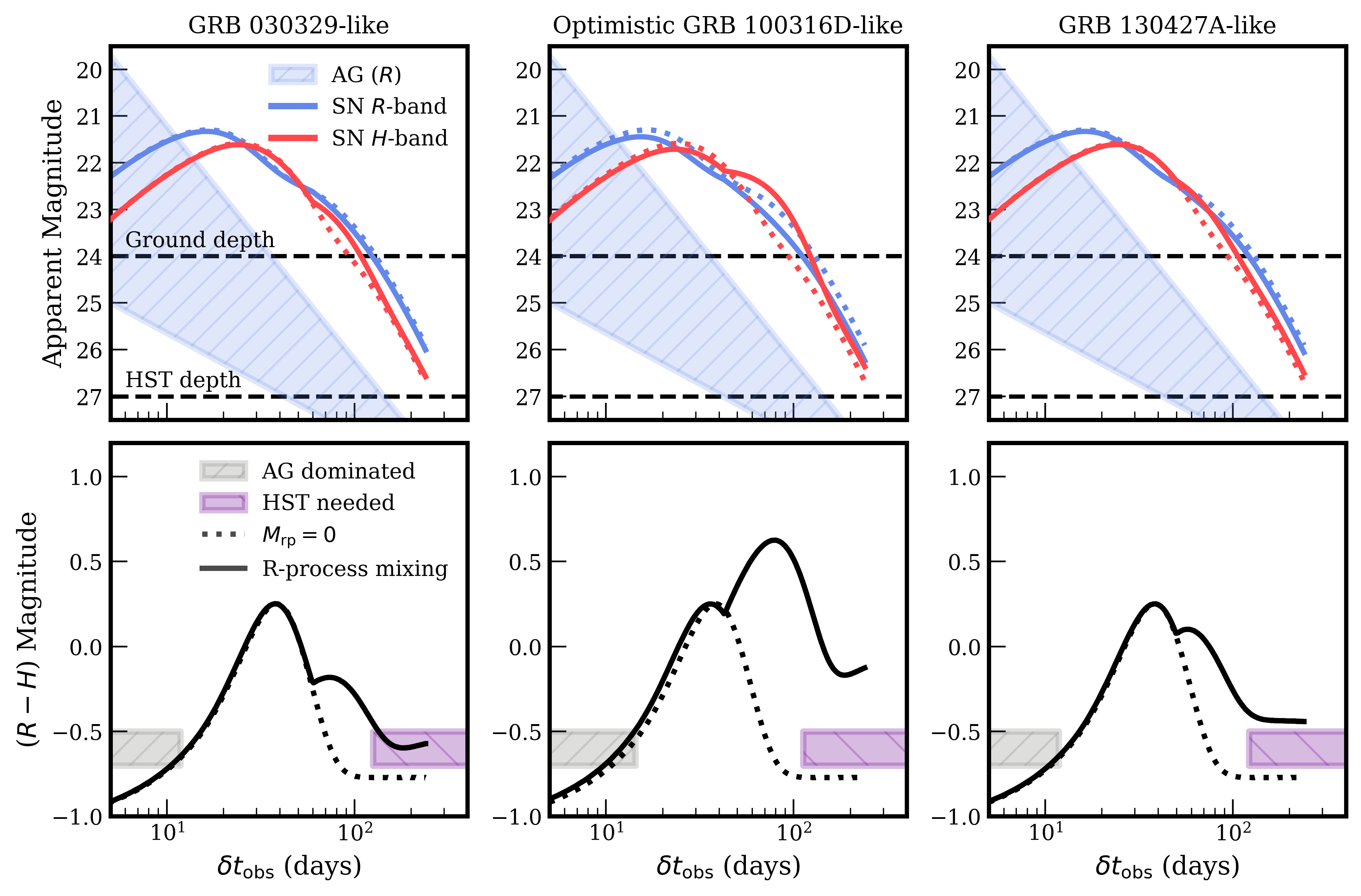}
\caption{\textit{Top:} expected $R$- and $H$-band light curves of a SN Ic-BL that is r-process-free (dotted lines) and enriched with $r$-process material (solid lines) set at $z=0.2$ \citep{BarnesMetzger22}. We show the expected afterglow brightness at $z=0.2$ and the depths of large ground and space-based facilities at which $\approx 5 \sigma$ detections should be possible (y-coordinate chose arbitrarily; more details in Section~\ref{sec:futureobs}). We show three models based on expected or potential outcomes of our analysis in Section~\ref{sec:datacomp}: our favored parameters for GRB\,190829A (left), our favored parameters for GRB\,100316D (middle) and an observationally optimistic model that is consistent with observations of GRB\,130427A (right). In the bottom panel, we plot $R-H$ over time for corresponding models, shading in grey the timespans at which the afterglow may dominate and in purple the timescales at which {\it HST} is necessary. To distinguish and monitor the color evolution of GRB\,190829A-like and GRB\,100316D-like models from the $r$-process-free case, space-based observations on the timescales of $\sim 60$~days are required. A SN Ic-BL like the optimistic GRB\,130427A-like case would be distinguishable by large-aperture ground-based observatories.}
\label{fig:future}
\end{figure*}

Our above derived rates are ballpark estimates and would benefit from additional modeling of $r$-process enriched collapsar light curves. Further modeling of $r$-process-enriched MR SNe light curves could also provide benchmarks for assessing heavy element production through a second mechanism. Similar to BNS mergers and kilonovae (e.g., \citealt{Bauswein+14,Shen+15,Rosswog+18,Hotokezaka+18} and references therein), our deduced $M_{r}$ values will be improved with more constrained rate estimates and characterization of the observed diversity of GRB-SNe $r$-process yields.

\subsection{Future Observations}
\label{sec:futureobs}

The observational sample presented in this work was not fine-tuned for comparison to the recent models of \citet{BarnesMetzger22}. Here, we consider if and how future observational strategies of GRB-SNe can be designed to best observe and constrain $M_{\rm rp}$ and $\psi_{\rm mix}$. In Figure~\ref{fig:future} we consider the observability of enriched models consistent with our analysis in Section~\ref{sec:datacomp} and \ref{sec:templates}. We plot GRBs\,030329 ($M_{\rm rp} = 0.03 M_{\odot}$, $\psi_{\rm mix} = 0.1$), 100316D ($M_{\rm rp} = 0.08 M_{\odot}$, $\psi_{\rm mix} = 0.3$) and 130427A ($M_{\rm rp} = 0.01 M_{\odot}$, $\psi_{\rm mix} = 0.3$) in both luminosity and color space, scaled to $z=0.2$. We also plot our $r$-band afterglow extrapolations of GRBs\,030329, 100316D and 190829A shifted to $z=0.2$ (spanning $m_r \approx 21-26$~AB mag at $\delta t \approx 1$~day; Section~\ref{sec:ag_contrib} and Figure~\ref{fig:lc_with_models}) to understand how significantly this component will contaminate future low-redshift GRB-SNe at all times. We do not include GRB\,130427A's afterglow in the range as it was known to be superlative in its brightness (e.g., \citealt{Perley+14,Laskar+13}). At peak brightness, many nearby GRB afterglows and SNe are observable with 1-2m-class telescopes. However, on the timescales that $r$-process color signatures may emerge ($\delta t \approx 40-200$~days), the expected GRB-SNe brightness spans $m = 22-27$~AB mag, thus requiring large-aperture ground-based telescopes or sensitive space facilities. In considering the sensitivity of large-aperture ground-based telescopes, we account for the fact that most GRB-SNe are embedded in their host galaxies, thus requiring image subtraction which reduces the source's signal-to-noise. 

From Figure~\ref{fig:future} we show that for a typical afterglow, we do not expect significant contamination on the timescales of the models ($\delta t \lesssim 300$~days). Our models demonstrates that, to observe similar events at $z=0.2$, the SN color evolution could be monitored at peak and distinguished using large-aperture ground-based telescopes, but {\it HST} sensitivity is necessary to capture the full color evolution and distinguish between mixing fractions. Overall, we find a wide diversity in expected luminosity and color evolution, but find that both ground- and space-based facilities can play an important role. Looking at upcoming facilities, serendipitous or target-of-opportunity observations with the \textit{Nancy Grace Roman Telescope} can provide complementary near-IR colors, while \textit{JWST} is poised to cover a larger dynamic range in optical-near-IR colors and better constrain models.
As discussed in Section~\ref{sec:datacomp}, an observational cadence similar to that of GRB\,190829A is critical to constraining the $r$-process enrichment.

Beyond photometric searches, spectroscopy of GRB-SNe may definitively identify or place deep limits on observed emission lines from transitions of $r$-process elements. Spectroscopic observations of kilonovae have identified absorption and emission lines from individual elements, most notably Sr II ($\lambda 10500$; e.g., \citealt{Watson+19}) and [Te III] ($\lambda 21500$; e.g., \citealt{Gillanders+23,Hotokezaka+23,Levan+23_230307a}). Caution should be taken when interpreting individual lanthanide line features in GRB-SNe, most notably the Sr II feature as it is coincident with known He lines \citep{Tarumi+23}.  Further lines may be identified by future high-resolution optical-NIR spectroscopy of kilonovae, most notably by JWST. Future late-time spectroscopy during the nebular phase of GRB-SNe can then be used to measure or place deep limits on the ejecta mass produced of each element.

\section{Conclusions}
\label{sec:conclusion}

We have presented optical-NIR observations of four GRB-SNe extending out to $\delta t = 588$~days and search for signs of $r$-process enrichment in the context of the \citet{BarnesMetzger22} models. Our dataset primarily consists of newly-analyzed observations from HST, with additional data from the VLT and the MMT. This analysis provides a template for considerations (e.g., afterglow, local extinction contamination) that may affect observational designs aimed at obtaining future color measurements of GRB-SNe. Moreover, combined with literature data, these observations can be used in comparison to future, enhanced models of $r$-process-enriched GRB-SNe, including those that account for viewing angle. Our analysis presents the first GRB-SNe sample of color measurements extending to the NIR at late times and allows us to make constraints on $M_{\rm rp}$ and $\psi_{\rm mix}$ for individual objects. Based on our analysis we make the following conclusions: 

\begin{itemize}
    \item Comparing our observed colors to those from the models of \citet{BarnesMetzger22}, we find that GRB\,190829A is consistent with no $r$-process enrichment, while GRBs\,030329, 100316D and 130427A are consistent with models for some enrichment, especially at low mixing fractions ($\psi_{\rm mix} \lesssim 0.5$). However, we note these latter three bursts are also consistent with models for no $r$-process. In nearly all cases, we find that models parameterized by high mixing fractions ($\psi_{\rm mix} \gtrsim 0.3$) are inconsistent with observations. We caution that our analysis is based on a fiducial set of semi-analytic models \citep{BarnesMetzger22} that is unable to account for $\gtrsim 30$\% of our observations. Future modeling will improve on these conclusions.
    \item We do not observe differences in $R-H$ GRB-SNe color measurements compared to those of SN Ic-BL without GRBs \citep{Anand+24}, contrasting with predictions that the presence of a jet and a pole-on viewing angle result in larger observed $\psi_{\rm mix}$ \citep{BarnesDuffell23}.
    \item We calculate the $r$-process enrichment from GRB-SNe and compare it to the observed Milky Way value from the literature. Considering our derived yields for GRBs\,100315D and 190829A and the total Milky Way yield of \citet{Hotokezaka+18}, GRB-SNe may produce $11-33 \%$ of the Milky Way's $r$-process abundance. However, this fraction is highly uncertain, and will be improved with further understanding of $r$-process yields from GRB-SNe and rate estimates that consider low-luminosity GRBs.
    \item Building a statistically significant sample of inferences on ejecta mass from GRB-SNe requires large-aperture ground-based and space-based telescopes to monitor events at $\delta t \gtrsim 70$~days. The cadence and long baseline of GRB\,190829A observations is the archetype for future GRB-SN monitoring.
\end{itemize}

Despite the landmark discovery of $r$-process nucleosynthesis in a neutron star merger, it remains an open question as to whether other heavy element formation channels exist (e.g., \citealt{Rosswog+22}). The rates of BNS mergers (e.g., \citealt{GWTC-3,MandelBroekgaarden22,Rouco+23}) as well as the mass and composition yields of kilonovae remain highly uncertain but likely vary widely (e.g., \citealt{Gompertz+18,metzger19,Kawaguchi+20b,Rastinejad+21}). The uncertainties in cosmological and Galactic BNS merger rates and their heavy element yields leave room for the existence of a second $r$-process formation channel (e.g., \citealt{HolmbeckAndrews23}). At the same time, emerging observations of $r$-process-enriched metal-poor stellar populations indicate a source of heavy elements closely tracing star formation (e.g., \citealt{JiFrebel18,Naidu+22,Ji+23,Simon+23,Kirby+23}). 

Moving forward, to identify or place deep constraints on $r$-process in rare classes of CCSNe requires significant effort and resources on both the theoretical and observational end. The development of observational predictions for $r$-process-enriched collapsar and MR SNe (or newly developed theories) is critical to establishing these sources as sites from photometric color measurements alone. Further late-time color and/or spectroscopic observations of GRB-SNe, SNe Ic-BL and, potentially, superluminous SNe (e.g., \citealt{REichert+23}) will provide additional measurements or upper limits of their $r$-process yields. Though spectroscopic observations, especially with {\it JWST}, are critical to definitively establishing these events as sites of $r$-process element production, color measurements are possible for a greater volume of GRB-SNe, improving our understanding of the distribution of ejecta masses. Dedicated programs on space-based facilities such as {\it HST} and {\it JWST} are necessary for this late-time NIR follow-up. With unprecedented sensitivity in the NIR bands, the upcoming {\it Nancy Grace Roman} Space Telescope will be a critical facility for this field. Finally, these studies of some of the most promising candidates for $r$-process enrichment are not possible without the continued and ensured detection of well-localized GRBs by satellites such as {\it Swift} and its successors.

\facilities{{\it HST} (ACS, NICMOS, WFC3 and WFPC2), MMT (Binospec), VLT (HAWK-I, X-shooter)}

\software{astropy \citep{Astropy2013,Astropy2018,astropy22}, IRAF \citep{Tody86,Tody93}, matplotlib \citep{Hunter+07}, Numpy \citep{numpy11,numpy}, SExtractor \citep{Bertin1996}}

\section{Acknowledgements}

We gratefully acknowledge Jennifer Barnes, Steve Schulze, Huei Sears, Peter Blanchard and Alessandra Corsi for valuable conversations regarding this manuscript. 

The Fong Group at Northwestern acknowledges support by the National Science Foundation under grant Nos. AST-1909358, AST-2206494, AST-2308182, and CAREER grant No. AST-2047919. W.F. gratefully acknowledges support by the David and Lucile Packard Foundation, the Alfred P. Sloan Foundation, and the Research Corporation for Science Advancement through Cottrell Scholar Award 28284.
Support for this work was provided by the National Aeronautics and Space Administration through Chandra Award Number DD3-24139X issued by the Chandra X-ray Center, which is operated by the Smithsonian Astrophysical Observatory for and on behalf of the National Aeronautics Space Administration under contract NAS8-03060. G.P.L. is supported by a Royal Society Dorothy Hodgkin Fellowship (grant Nos. DHF-R1-221175 and DHF-ERE-221005).
G.S. and P.S. acknowledge the support by the State of Hesse within the Research Cluster ELEMENTS (Project ID 500/10.006).
E.P. acknowledges financial support through INAF Fundamental Research Grant 2022.
L.I. was supported by an INAF Fundamental Research Grant 2023.
S.~Anand acknowledges support from the National Science Foundation GROWTH PIRE grant No. 1545949.

Based on observations made with the NASA/ESA Hubble Space Telescope, obtained from the data archive at the Space Telescope Science Institute. STScI is operated by the Association of Universities for Research in Astronomy, Inc. under NASA contract NAS 5-26555. 
All of the Hubble Space Telescope data presented in this article were obtained from the Mikulski Archive for Space Telescopes (MAST) at the Space Telescope Science Institute. The specific observations analyzed can be accessed via \dataset[DOI: 10.17909/fnr6-cq77]{https://doi.org/10.17909/fnr6-cq77}. 
Observations reported here were obtained at the MMT Observatory, a joint facility of the University of Arizona and the Smithsonian Institution.  MMT Observatory access was supported by Northwestern University and the Center for Interdisciplinary Exploration and Research in Astrophysics (CIERA).
Based on observations collected at the European Organisation for Astronomical Research in the Southern Hemisphere. This research has made use of NASA’s Astrophysics Data System. 

\bibliographystyle{aasjournal}
\bibliography{refs}

\section{Supplementary Information} \label{appendix}
\input{phot_hst}

\end{document}

%% file: affiliation.tex
\newcommand{\NU}{\affiliation{Center for Interdisciplinary Exploration and Research in Astrophysics (CIERA) and Department of Physics and Astronomy, Northwestern University, Evanston, IL 60208, USA}}

\newcommand{\Radboud}{\affiliation{Department of Astrophysics/IMAPP, Radboud University, 6525 AJ Nijmegen, The Netherlands}}

\newcommand{\Leicester}{\affiliation{School of Physics and Astronomy, University of Leicester, University Road, Leicester, LE1 7RH, UK}}

%% file: authors.tex
\author[0000-0002-9267-6213]{J. C.~Rastinejad}
\NU

\author[0000-0002-7374-935X]{W.~Fong}
\NU

\author[0000-0001-7821-9369]{A. J. Levan}
\Radboud
\affil{Department of Physics, University of Warwick, Coventry, CV4 7AL, UK}

\author[0000-0003-3274-6336]{N. R. Tanvir}
\Leicester

\author[0000-0002-5740-7747]{C.~D.~Kilpatrick}
\NU

\author [0000-0002-6652-9279]{A. S. Fruchter}
\affil{Space Telescope Science Institute, 3700 San Martin Drive, Baltimore, MD 21218, USA}

\author[0000-0003-3768-7515]{S. Anand}
\affil{Cahill Center for Astrophysics, California Institute of Technology, Pasadena CA 91125, USA}

\author[0000-0003-0136-1281]{K. Bhirombhakdi}
\affil{Space Telescope Science Institute, 3700 San Martin Drive, Baltimore, MD 21218, USA}

\author[0000-0001-9078-5507]{S. Covino}
\affil{INAF-Osservatorio Astronomico di Brera, Via Bianchi 46, I-23807, Merate (LC), Italy}

\author[0000-0002-8149-8298]{J. P. U. Fynbo}
\affil{Cosmic DAWN Center, Denmark}
\affil{Niels Bohr Institute, University of Copenhagen, Jagtvej 155, 2200 Copenhagen N, Denmark}

\author[0000-0002-7232-101X]{G. Halevi}
\NU

\author[0000-0002-8028-0991]{D. H. Hartmann}
\affil{Department of Physics and Astronomy, Clemson University, Clemson, SC 29634-0978, USA}

\author[0000-0002-9389-7413]{K. E. Heintz}
\affil{Cosmic DAWN Center, Denmark}
\affil{Niels Bohr Institute, University of Copenhagen, Jagtvej 155, 2200 Copenhagen N, Denmark}

\author[0000-0001-9695-8472]{L. Izzo}
\affil{INAF-Osservatorio Astronomico di Capodimonte, Salita Moiariello 16, I-80131, Napoli, Italy}
\affil{DARK, Niels Bohr Institute, University of Copenhagen, Jagtvej 155, 2200 Copenhagen N, Denmark}

\author[0000-0002-9404-5650]{P. Jakobsson}
\affil{Centre for Astrophysics and Cosmology, Science Institute, University of Iceland, Dunhagi 5, 107, Reykjavik, Iceland}
\author[0000-0002-5477-0217]{T. Kangas}
\affil{Finnish Centre for Astronomy with ESO (FINCA), FI-20014 University of Turku, Finland}
\affil{Tuorla Observatory, Department of Physics and Astronomy, FI-20014 University of Turku, Finland}

\author[0000-0001-5169-4143]{G. P. Lamb}
\affil{Astrophysics Research Institute, Liverpool John Moores University, IC2 Liverpool Science Park, 146 Brownlow Hill, Liverpool, L3 5RF, UK}

\author[0000-0002-7517-326X]{D. B. Malesani}
\affiliation{Cosmic DAWN Center, Denmark}
\affiliation{Niels Bohr Institute, University of Copenhagen, Jagtvej 155, 2200 Copenhagen N, Denmark}
\Radboud

\author[0000-0002-2810-2143]{A. Melandri}
\affil{NAF-Osservatorio Astronomico di Roma, Via di Frascati 33, I-00078, Monte Porzio Catone (RM), Italy}

\author[0000-0002-4670-7509]{B. D. Metzger}
\affil{Department of Physics and Columbia Astrophysics Laboratory, Columbia University, Pupin Hall, New York, NY 10027, USA}
\affil{Center for Computational Astrophysics, Flatiron Institute, 162 5th Ave, New York, NY 10010, USA}

\author[0000-0002-2281-2785]{B. Milvang-Jensen}
\affil{Cosmic DAWN Center, Denmark}
\affil{Niels Bohr Institute, University of Copenhagen, Jagtvej 155, 2200 Copenhagen N, Denmark}

\author[0000-0001-8646-4858]{E. Pian}
\affil{INAF, Astrophysics and Space Science Observatory, Via P. Gobetti 101, 40129 Bologna, Italy}

\author[0000-0003-3457-9375]{G. Pugliese}
\affiliation{Astronomical Institute Anton Pannekoek, University of Amsterdam, 1090 GE Amsterdam, The Netherlands}

\author[0000-0002-8860-6538]{A. Rossi}
\affiliation{INAF - Osservatorio di Astrofisica e Scienza dello Spazio, via Piero Gobetti 93/3, 40129 Bologna, Italy}

\author[0000-0001-6374-6465]{D. M.~Siegel}
\affil{Institute of Physics, University of Greifswald, D-17489 Greifswald, Germany}
\affil{Department of Physics, University of Guelph, Guelph, Ontario N1G 2W1, Canada}

\author[0000-0003-1006-6970]{P. Singh}
\affil{Institut f\"ur Theoretische Physik, Goethe Universit\"at, Max-von-Laue-Str. 1, 60438 Frankfurt am Main, Germany}

\author[0000-0003-1055-7980]{G. Stratta}
\affil{Institut f\"ur Theoretische Physik, Goethe Universit\"at, Max-von-Laue-Str. 1, 60438 Frankfurt am Main, Germany}
\affil{INAF - Istituto di Astrofisica e Planetologia Spaziali, via Fosso del Cavaliere 100, I-00133 Roma, Italy}
\affil{INAF - Osservatorio di Astrofisica e Scienza dello Spazio, via Piero Gobetti 93/3, 40129 Bologna, Italy}
\affil{INFN - Roma 1, Piazzale Aldo Moro 2, 00185, Roma}

%% file: grbs.tex
\begin{deluxetable*}{ccccccccc}
\savetablenum{1}
\tabletypesize{\small}
\centering
\tablecolumns{8}
\tabcolsep0.06in
\tablecaption{GRB Sample \& Properties
\label{tab:grbprops}}
\tablehead {
\colhead {GRB}		&
\colhead {SN}	&
\colhead {$T_{\rm 90, rest}^\dagger$}		& 
\colhead {$z$}	&
\colhead {A$_{V, {\rm MW}}$}	&
\colhead {E(B-V)$_{\rm loc}$}	&
\colhead {HST Observed Filters$^\ddagger$} &
\colhead {Program ID(s)} 
\\
\colhead {}		&
\colhead {}		&
\colhead {(s)}		& 
\colhead {}		&
\colhead {(mag)}	&
\colhead {(mag)}	& 
\colhead {}	&
\colhead {}}
\startdata
030329 & 2003dh & 18 & 0.1685 & 0.069 & 0.04$^1$ & F606W, F814W, F110W, F160W & 9405 \\
100316D$^*$ & 2010bh & 261 & 0.0591 & 0.319 & 0.14$^2$ &F555W, F814W, F125W, F160W & 11709, 12323 \\
130427A & 2013cq & 182 & 0.3399 & 0.055 & 0.05$^3$ & F606W, F160W & 13110, 13117, 13230,13951 \\
190829A & 2019oyw & 49 & 0.0785 & 0.133 & 1.04$^4$ & F606W, F110W, F140W, F160W & 15089, 15510, 16042, 16320 \\
\enddata
\tablecomments{
$\dagger$ $T_{90}$ duration as seen by \textit{Swift}, except for GRB\,030329 which was observed by HETE-II, converted to the rest frame using the redshifts listed here. \\
$\ddagger$ Filters selected based on data availability on timescales relevant for observing $r$-process-enriched component (contemporaneous colors observed at $\delta t \gtrsim 30$~days).\\
$^*$ Low-luminosity GRB. \\
Milky Way extinction values are taken from \citet{SchlaflyFinkbeiner11}. \\ References: (1) \cite{Matheson+03}; (2) \cite{Bufano+12}; (3) \cite{Levan+14}; (4) \cite{Chand+20}.}
\vspace{-0.3in}
\end{deluxetable*}
\vspace{-0.3in}

%% file: phot_hst.tex
\startlongtable
\begin{deluxetable*}{ccccCccc}
\savetablenum{2}
\tabletypesize{\footnotesize}
\centering
\tablecolumns{9}
\tabcolsep0.06in
\tablecaption{Sample of Observations
\label{tab:hstobs}}
\tablehead {
\colhead {GRB}		&
\colhead {$z$}		&
\colhead {Tel./Instum.}		&
\colhead {$\delta t_{\rm rest}$}	&
\colhead {Filter}		& 
\colhead {Magnitude$^{\dagger}$}	&
\colhead {Error} &
\colhead {Reference}  \\
\colhead {} &
\colhead {} &
\colhead {} &
\colhead {(days)}		& 
\colhead {} &
\colhead {(AB mag)}	&
\colhead {(AB mag)} &
\colhead {} 
}
\startdata
030329 & 0.1685 & Clay & 0.6 & R & 15.23$^*$ & 0.02 & \citet{Matheson+03} \\
 &  & Clay & 0.6 & R & 15.2$^*$ & 0.02 &  \\
 &  & FLWO & 0.6 & R & 15.39$^*$ & 0.01 &  \\
 &  & FLWO & 0.6 & R & 15.38$^*$ & 0.01 &  \\
 &  & FLWO & 0.6 & R & 15.39$^*$ & 0.01 &  \\
 &  & FLWO & 0.6 & R & 15.44$^*$ & 0.01 &  \\
 &  & FLWO & 0.6 & I & 15.27$^*$ & 0.01 &  \\
 &  & FLWO & 0.7 & R & 15.48$^*$ & 0.01 &  \\
 &  & FLWO & 0.7 & I & 15.3$^*$ & 0.01 &  \\
 &  & FLWO & 0.7 & R & 15.52$^*$ & 0.01 &  \\
 &  & FLWO & 0.7 & I & 15.35$^*$ & 0.01 &  \\
 &  & KAIT & 0.7 & R & 15.61$^*$ & 0.03 &  \\
 &  & FLWO & 0.7 & R & 15.59$^*$ & 0.01 &  \\
 &  & FLWO & 0.7 & I & 15.39$^*$ & 0.01 &  \\
 &  & KAIT & 0.7 & I & 15.42$^*$ & 0.03 &  \\
 &  & FLWO & 0.7 & R & 15.6$^*$ & 0.01 &  \\
 &  & FLWO & 0.7 & I & 15.43$^*$ & 0.01 &  \\
 &  & KAIT & 0.7 & R & 15.63$^*$ & 0.02 &  \\
 &  & FLWO & 0.7 & R & 15.63$^*$ & 0.01 &  \\
 &  & KAIT & 0.7 & I & 15.44$^*$ & 0.03 &  \\
 &  & FLWO & 0.7 & I & 15.5$^*$ & 0.01 &  \\
 &  & FLWO & 0.7 & R & 15.67$^*$ & 0.01 &  \\
 &  & FLWO & 0.7 & I & 15.5$^*$ & 0.01 &  \\
 &  & KAIT & 0.7 & R & 15.67$^*$ & 0.01 &  \\
 &  & KAIT & 0.7 & I & 15.48$^*$ & 0.02 &  \\
 &  & FLWO & 0.7 & R & 15.71 & 0.01 &  \\
 &  & FLWO & 0.7 & I & 15.49 & 0.01 &  \\
 &  & FLWO & 0.7 & R & 15.71 & 0.01 &  \\
 &  & FLWO & 0.7 & I & 15.55 & 0.01 &  \\
 &  & FLWO & 0.8 & R & 15.74 & 0.01 &  \\
 &  & FLWO & 0.8 & I & 15.58 & 0.01 &  \\
 &  & FLWO & 0.8 & R & 15.79 & 0.01 &  \\
 &  & KAIT & 0.8 & R & 15.8 & 0.02 &  \\
 &  & FLWO & 0.8 & I & 15.62 & 0.01 &  \\
 &  & KAIT & 0.8 & I & 15.58 & 0.03 &  \\
 &  & FLWO & 0.8 & R & 15.79 & 0.01 &  \\
 &  & FLWO & 0.8 & I & 15.63 & 0.01 &  \\
 &  & KAIT & 0.8 & R & 15.84 & 0.02 &  \\
 &  & KAIT & 0.8 & I & 15.67 & 0.03 &  \\
 &  & FLWO & 0.8 & R & 15.83 & 0.01 &  \\
 &  & FLWO & 0.8 & I & 15.68 & 0.01 &  \\
 &  & FLWO & 0.8 & R & 15.84 & 0.01 &  \\
 &  & FLWO & 0.8 & I & 15.67 & 0.02 &  \\
 &  & FLWO & 0.8 & R & 15.87 & 0.01 &  \\
 &  & FLWO & 0.8 & I & 15.68 & 0.02 &  \\
 &  & FLWO & 0.8 & R & 15.91 & 0.01 &  \\
 &  & FLWO & 0.8 & I & 15.74 & 0.01 &  \\
 &  & KAIT & 0.8 & R & 15.91 & 0.01 &  \\
 &  & KAIT & 0.8 & I & 15.74 & 0.02 &  \\
 &  & FLWO & 0.8 & R & 15.96$^*$ & 0.01 &  \\
 &  & FLWO & 0.8 & I & 15.78$^*$ & 0.01 &  \\
 &  & FLWO & 0.8 & R & 15.95$^*$ & 0.01 &  \\
 &  & FLWO & 0.8 & I & 15.8$^*$ & 0.02 &  \\
 &  & KAIT & 0.8 & R & 16.0$^*$ & 0.08 &  \\
 &  & KAIT & 0.8 & I & 15.75 & 0.03 &  \\
 &  & FLWO & 0.8 & R & 15.98$^*$ & 0.01 &  \\
 &  & FLWO & 0.9 & I & 15.83$^*$ & 0.01 &  \\
 &  & FLWO & 0.9 & R & 16.01$^*$ & 0.01 &  \\
 &  & FLWO & 0.9 & I & 15.87$^*$ & 0.02 &  \\
 &  & KAIT & 0.9 & R & 16.01$^*$ & 0.01 &  \\
 &  & KAIT & 0.9 & I & 15.85$^*$ & 0.02 &  \\
 &  & FLWO & 0.9 & R & 16.06$^*$ & 0.01 &  \\
 &  & FLWO & 0.9 & I & 15.86$^*$ & 0.02 &  \\
 &  & KAIT & 0.9 & R & 16.08$^*$ & 0.01 &  \\
 &  & KAIT & 0.9 & I & 15.87$^*$ & 0.02 &  \\
 &  & FLWO & 0.9 & R & 16.08$^*$ & 0.01 &  \\
 &  & FLWO & 0.9 & I & 15.86$^*$ & 0.02 &  \\
 &  & FLWO & 0.9 & R & 16.1$^*$ & 0.01 &  \\
 &  & FLWO & 0.9 & I & 15.87$^*$ & 0.02 &  \\
 &  & KAIT & 0.9 & R & 16.09$^*$ & 0.02 &  \\
 &  & KAIT & 0.9 & I & 15.92$^*$ & 0.02 &  \\
 &  & FLWO & 0.9 & R & 16.12$^*$ & 0.01 &  \\
 &  & FLWO & 0.9 & I & 15.89$^*$ & 0.02 &  \\
 &  & FLWO & 0.9 & R & 16.16$^*$ & 0.01 &  \\
 &  & KAIT & 0.9 & R & 16.13$^*$ & 0.02 &  \\
 &  & FLWO & 0.9 & I & 15.96$^*$ & 0.02 &  \\
 &  & KAIT & 0.9 & I & 15.98$^*$ & 0.02 &  \\
 &  & FLWO & 0.9 & R & 16.19$^*$ & 0.02 &  \\
 &  & FLWO & 0.9 & I & 16.06$^*$ & 0.02 &  \\
 &  & KAIT & 0.9 & R & 16.15$^*$ & 0.02 &  \\
 &  & KAIT & 0.9 & I & 16.02$^*$ & 0.04 &  \\
 &  & FLWO & 0.9 & R & 16.21$^*$ & 0.02 &  \\
 &  & FLWO & 0.9 & I & 16.1$^*$ & 0.02 &  \\
 &  & FLWO & 0.9 & R & 16.24$^*$ & 0.02 &  \\
 &  & FLWO & 0.9 & I & 16.1$^*$ & 0.02 &  \\
 &  & KAIT & 0.9 & R & 16.29$^*$ & 0.06 &  \\
 &  & KAIT & 0.9 & I & 16.02$^*$ & 0.03 &  \\
 &  & FLWO & 1.0 & R & 16.28$^*$ & 0.02 &  \\
 &  & FLWO & 1.0 & I & 16.1$^*$ & 0.03 &  \\
 &  & Clay & 1.6 & R & 16.44 & 0.02 &  \\
 &  & Clay & 1.6 & R & 16.44 & 0.02 &  \\
 &  & Clay & 1.6 & R & 16.43 & 0.02 &  \\
 &  & Clay & 1.6 & R & 16.44 & 0.02 &  \\
 &  & FLWO & 1.7 & R & 16.48 & 0.01 &  \\
 &  & FLWO & 1.7 & I & 16.28 & 0.01 &  \\
 &  & FLWO & 1.7 & R & 16.61 & 0.01 &  \\
 &  & FLWO & 1.7 & I & 16.4 & 0.01 &  \\
 &  & KAIT & 1.8 & R & 16.75 & 0.02 &  \\
 &  & KAIT & 1.8 & I & 16.49 & 0.04 &  \\
 &  & FLWO & 1.8 & R & 16.75 & 0.01 &  \\
 &  & FLWO & 1.8 & I & 16.54 & 0.01 &  \\
 &  & KAIT & 1.9 & R & 16.86 & 0.07 &  \\
 &  & KAIT & 1.9 & I & 16.51 & 0.07 &  \\
 &  & FLWO & 1.9 & R & 16.78 & 0.01 &  \\
 &  & FLWO & 1.9 & I & 16.57 & 0.01 &  \\
 &  & Clay & 2.6 & R & 17.02 & 0.02 &  \\
 &  & Clay & 2.6 & R & 17.06 & 0.02 &  \\
 &  & Clay & 2.6 & R & 17.07 & 0.02 &  \\
 &  & Clay & 2.6 & R & 17.07 & 0.02 &  \\
 &  & FLWO & 2.9 & R & 17.28 & 0.01 &  \\
 &  & FLWO & 2.9 & I & 17.04 & 0.01 &  \\
 &  & FLWO & 3.0 & R & 17.36 & 0.01 &  \\
 &  & FLWO & 3.0 & I & 17.13 & 0.02 &  \\
 &  & Clay & 3.6 & R & 17.45 & 0.02 &  \\
 &  & Clay & 3.6 & R & 17.45 & 0.02 &  \\
 &  & Clay & 3.6 & R & 17.46 & 0.02 &  \\
 &  & Clay & 3.6 & R & 17.45 & 0.02 &  \\
 &  & Clay & 3.6 & R & 17.46 & 0.02 &  \\
 &  & LCO40 & 3.6 & J & 16.84 & 0.02 &  \\
 &  & Clay & 3.6 & R & 17.44 & 0.02 &  \\
 &  & LCO40 & 3.6 & H & 16.61 & 0.02 &  \\
 &  & FLWO & 3.7 & R & 17.57 & 0.03 &  \\
 &  & FLWO & 3.8 & R & 17.6 & 0.02 &  \\
 &  & FLWO & 3.8 & I & 17.41 & 0.02 &  \\
 &  & FLWO & 3.8 & R & 17.62 & 0.01 &  \\
 &  & FLWO & 3.8 & I & 17.43 & 0.02 &  \\
 &  & FLWO & 3.8 & R & 17.64 & 0.01 &  \\
 &  & FLWO & 3.8 & I & 17.42 & 0.02 &  \\
 &  & FLWO & 3.8 & R & 17.67 & 0.01 &  \\
 &  & FLWO & 3.8 & I & 17.46 & 0.02 &  \\
 &  & FLWO & 3.9 & R & 17.7 & 0.02 &  \\
 &  & FLWO & 3.9 & I & 17.48 & 0.05 &  \\
 &  & Clay & 4.6 & R & 18.01 & 0.02 &  \\
 &  & Clay & 4.6 & R & 18.0 & 0.02 &  \\
 &  & Clay & 4.6 & R & 18.0 & 0.02 &  \\
 &  & Clay & 4.6 & R & 18.0 & 0.02 &  \\
 &  & Clay & 4.6 & R & 18.01 & 0.02 &  \\
 &  & Clay & 4.6 & R & 18.01 & 0.02 &  \\
 &  & Clay & 4.6 & R & 18.02 & 0.02 &  \\
 &  & Clay & 4.6 & R & 18.01 & 0.02 &  \\
 &  & Clay & 4.6 & R & 18.04 & 0.02 &  \\
 &  & LCO40 & 4.6 & J & 17.36 & 0.03 &  \\
 &  & FLWO & 4.7 & R & 18.0 & 0.02 &  \\
 &  & FLWO & 4.7 & I & 17.81 & 0.02 &  \\
 &  & LCO40 & 4.7 & H & 16.96 & 0.03 &  \\
 &  & FLWO & 5.6 & R & 17.99 & 0.02 &  \\
 &  & FLWO & 5.6 & I & 17.77 & 0.02 &  \\
 &  & LCO40 & 5.7 & J & 17.33 & 0.02 &  \\
 &  & LCO40 & 5.7 & H & 17.08 & 0.03 &  \\
 &  & LCO100 & 5.7 & I & 17.82 & 0.02 &  \\
 &  & LCO100 & 5.7 & I & 17.82 & 0.02 &  \\
 &  & LCO100 & 6.6 & I & 18.28 & 0.02 &  \\
 &  & LCO100 & 6.6 & I & 18.33 & 0.03 &  \\
 &  & LCO40 & 6.6 & J & 17.96 & 0.04 &  \\
 &  & LCO40 & 6.6 & H & 17.59 & 0.04 &  \\
 &  & FLWO & 6.7 & R & 18.46 & 0.02 &  \\
 &  & FLWO & 6.7 & I & 18.35 & 0.03 &  \\
 &  & FLWO & 6.9 & R & 18.6 & 0.02 &  \\
 &  & FLWO & 6.9 & I & 18.42 & 0.03 &  \\
 &  & FLWO & 7.6 & R & 18.94 & 0.03 &  \\
 &  & FLWO & 7.7 & I & 18.76 & 0.04 &  \\
 &  & LCO40 & 7.7 & H & 18.02 & 0.05 &  \\
 &  & LCO40 & 7.7 & J & 18.35 & 0.03 &  \\
 &  & FLWO & 7.9 & R & 18.96 & 0.04 &  \\
 &  & FLWO & 7.9 & I & 18.84 & 0.05 &  \\
 &  & LCO40 & 8.6 & J & 18.47 & 0.04 &  \\
 &  & FLWO & 8.6 & R & 19.06 & 0.03 &  \\
 &  & FLWO & 8.6 & I & 18.92 & 0.04 &  \\
 &  & LCO40 & 8.7 & H & 18.22 & 0.08 &  \\
 &  & FLWO & 8.7 & R & 19.04 & 0.04 &  \\
 &  & FLWO & 8.8 & R & 19.06 & 0.03 &  \\
 &  & FLWO & 8.8 & I & 18.92 & 0.04 &  \\
 &  & FLWO & 8.9 & R & 19.09 & 0.03 &  \\
 &  & FLWO & 8.9 & I & 18.97 & 0.05 &  \\
 &  & FLWO & 9.6 & R & 19.1 & 0.03 &  \\
 &  & LCO40 & 9.6 & J & 18.65 & 0.04 &  \\
 &  & FLWO & 9.6 & I & 19.05 & 0.05 &  \\
 &  & LCO40 & 9.7 & H & 18.54 & 0.05 &  \\
 &  & FLWO & 9.8 & R & 19.14 & 0.03 &  \\
 &  & FLWO & 9.8 & I & 18.98 & 0.04 &  \\
 &  & FLWO & 9.9 & R & 19.15 & 0.04 &  \\
 &  & FLWO & 9.9 & I & 19.03 & 0.06 &  \\
 &  & FLWO & 10.7 & R & 19.31 & 0.04 &  \\
 &  & FLWO & 10.7 & I & 19.19 & 0.05 &  \\
 &  & FLWO & 10.8 & I & 19.22 & 0.04 &  \\
 &  & FLWO & 10.8 & R & 19.26 & 0.03 &  \\
 &  & LCO40 & 11.6 & J & 19.14 & 0.07 &  \\
 &  & FLWO & 11.7 & R & 19.46 & 0.03 &  \\
 &  & FLWO & 11.7 & I & 19.36 & 0.04 &  \\
 &  & FLWO & 11.7 & R & 19.42 & 0.03 &  \\
 &  & FLWO & 11.7 & I & 19.39 & 0.05 &  \\
 &  & FLWO & 11.8 & R & 19.47 & 0.03 &  \\
 &  & FLWO & 11.8 & I & 19.43 & 0.06 &  \\
 &  & FLWO & 12.7 & R & 19.52 & 0.03 &  \\
 &  & FLWO & 12.8 & R & 19.57 & 0.04 &  \\
 &  & FLWO & 13.6 & R & 19.62 & 0.08 &  \\
 &  & FLWO & 13.7 & R & 19.69 & 0.08 &  \\
 &  & HST/WFPC2 & 17.0 & F606W & 20.14 & 0.002 & This work \\
 &  & HST/WFPC2 & 17.5 & F814W & 20.01 & 0.0 & This work \\
 &  & FLWO & 19.7 & R & 19.99 & 0.04 & \citet{Matheson+03} \\
 &  & FLWO & 19.7 & R & 20.02 & 0.04 &  \\
 &  & FLWO & 19.8 & R & 20.02 & 0.06 &  \\
 &  & FLWO & 20.7 & R & 20.07 & 0.04 &  \\
 &  & FLWO & 20.8 & R & 20.15 & 0.05 &  \\
 &  & FLWO & 20.8 & R & 20.0 & 0.05 &  \\
 &  & FLWO & 21.7 & R & 20.15 & 0.03 &  \\
 &  & FLWO & 21.8 & R & 20.16 & 0.04 &  \\
 &  & FLWO & 21.9 & R & 20.08 & 0.06 &  \\
 &  & LCO40 & 22.6 & R & 20.16 & 0.05 &  \\
 &  & FLWO & 22.6 & R & 20.23 & 0.04 &  \\
 &  & HST/NICMOS/NIC2 & 22.7 & F110W & 20.47 & 0.0 & This work \\
 &  & FLWO & 22.7 & R & 20.22 & 0.04 & \citet{Matheson+03} \\
 &  & HST/NICMOS/NIC2 & 22.8 & F160W & 20.6 & 0.0 & This work \\
 &  & HST/WFPC2 & 23.1 & F606W & 20.39 & 0.0 & This work \\
 &  & HST/WFPC2 & 23.2 & F814W & 20.31 & 0.0 & This work \\
 &  & FLWO & 23.7 & R & 20.29 & 0.03 & \citet{Matheson+03} \\
 &  & FLWO & 23.8 & R & 20.25 & 0.04 &  \\
 &  & FLWO & 24.7 & R & 20.36 & 0.04 &  \\
 &  & FLWO & 24.8 & R & 20.33 & 0.03 &  \\
 &  & KPNO4m & 25.7 & R & 20.37 & 0.02 &  \\
 &  & FLWO & 25.7 & R & 20.35 & 0.04 &  \\
 &  & KPNO4m & 25.7 & R & 20.36 & 0.02 &  \\
 &  & FLWO & 25.7 & R & 20.39 & 0.04 &  \\
 &  & FLWO & 26.7 & R & 20.46 & 0.04 &  \\
 &  & FLWO & 27.7 & R & 20.53 & 0.05 &  \\
 &  & FLWO & 28.7 & R & 20.44 & 0.05 &  \\
 &  & FLWO & 28.7 & R & 20.56 & 0.06 &  \\
 &  & FLWO & 29.7 & R & 20.63 & 0.04 &  \\
 &  & FLWO & 29.7 & R & 20.56 & 0.04 &  \\
 &  & FLWO & 29.7 & R & 20.63 & 0.05 &  \\
 &  & FLWO & 30.7 & R & 20.69 & 0.04 &  \\
 &  & FLWO & 30.7 & R & 20.7 & 0.04 &  \\
 &  & FLWO & 31.7 & R & 20.78 & 0.04 &  \\
 &  & LCO40 & 32.6 & R & 20.76 & 0.07 &  \\
 &  & FLWO & 32.7 & R & 20.8 & 0.04 &  \\
 &  & KPNO4m & 33.6 & R & 20.88 & 0.03 &  \\
 &  & FLWO & 33.8 & R & 20.77 & 0.06 &  \\
 &  & FLWO & 33.8 & R & 20.87 & 0.06 &  \\
 &  & KPNO4m & 34.6 & R & 20.91 & 0.04 &  \\
 &  & FLWO & 34.7 & R & 21.02 & 0.07 &  \\
 &  & FLWO & 34.7 & R & 20.92 & 0.05 &  \\
 &  & FLWO & 36.8 & R & 21.08 & 0.08 &  \\
 &  & FLWO & 37.7 & R & 21.03 & 0.07 &  \\
 &  & FLWO & 37.7 & R & 21.11 & 0.05 &  \\
 &  & FLWO & 37.7 & R & 21.03 & 0.06 &  \\
 &  & FLWO & 38.7 & R & 21.23 & 0.06 &  \\
 &  & FLWO & 39.7 & R & 21.22 & 0.07 &  \\
 &  & HST/WFPC2 & 43.7 & F814W & 21.49 & 0.01 & This work \\
 &  & HST/WFPC2 & 43.7 & F606W & 21.9 & 0.01 & This work \\
 &  & HST/NICMOS/NIC2 & 45.1 & F110W & 21.58 & 0.01 & This work \\
 &  & HST/NICMOS/NIC2 & 45.2 & F160W & 21.97 & 0.01 & This work \\
 &  & FLWO & 51.7 & R & 21.76 & 0.06 & \citet{Matheson+03} \\
 &  & FLWO & 52.7 & R & 21.91 & 0.11 &  \\
 &  & FLWO & 53.7 & R & 21.64 & 0.08 &  \\
 &  & FLWO & 54.7 & R & 21.52 & 0.09 &  \\
 &  & FLWO & 57.7 & R & 21.82 & 0.06 &  \\
 &  & FLWO & 59.7 & R & 21.67 & 0.07 &  \\
 &  & FLWO & 60.7 & R & 21.83 & 0.09 &  \\
 &  & HST/WFPC2 & 72.9 & F814W & 22.3 & 0.01 & This work \\
 &  & HST/WFPC2 & 73.6 & F606W & 22.75 & 0.01 & This work \\
 &  & HST/WFPC2 & 93.3 & F814W & 22.77 & 0.01 & This work \\
 &  & HST/WFPC2 & 93.4 & F606W & 23.11 & 0.01 & This work \\
 &  & HST/WFPC2 & 227.7 & F606W & 25.39 & 0.02 & This work \\
 &  & HST/WFPC2 & 227.8 & F814W & 24.95 & 0.04 & This work \\
 &  & HST/WFPC2 & 422.6 & F606W & 27.59 & 0.1 & This work \\
100316D & 0.0591 & GROND & 0.5 & r & 20.9$^*$ & 0.04 & \citet{Olivares+12} \\
 &  & GROND & 0.5 & J & 20.69$^*$ & 0.14 &  \\
 &  & GROND & 0.5 & i & 20.87$^*$ & 0.05 &  \\
 &  & GROND & 0.6 & J & 20.71$^*$ & 0.26 &  \\
 &  & GROND & 0.6 & r & 20.91$^*$ & 0.03 &  \\
 &  & GROND & 0.6 & i & 20.94$^*$ & 0.04 &  \\
 &  & GROND & 1.5 & r & 20.9 & 0.04 &  \\
 &  & GROND & 1.5 & i & 20.59 & 0.05 &  \\
 &  & GROND & 1.5 & J & 20.47 & 0.22 &  \\
 &  & GROND & 1.7 & i & 20.69 & 0.05 &  \\
 &  & GROND & 1.7 & r & 20.77 & 0.03 &  \\
 &  & GROND & 2.5 & i & 20.5 & 0.05 &  \\
 &  & GROND & 2.5 & r & 20.58 & 0.04 &  \\
 &  & GROND & 2.5 & J & 19.99 & 0.17 &  \\
 &  & GROND & 3.5 & r & 20.4 & 0.03 &  \\
 &  & GROND & 3.5 & i & 20.24 & 0.03 &  \\
 &  & GROND & 3.5 & J & 19.72 & 0.17 &  \\
 &  & GROND & 3.5 & H & 19.94 & 0.21 &  \\
 &  & GROND & 4.5 & J & 20.0 & 0.19 &  \\
 &  & GROND & 4.5 & r & 20.21 & 0.03 &  \\
 &  & GROND & 4.5 & i & 20.15 & 0.04 &  \\
 &  & GROND & 4.5 & H & 20.22 & 0.23 &  \\
 &  & GROND & 5.5 & H & 19.92 & 0.23 &  \\
 &  & GROND & 5.5 & r & 20.11 & 0.04 &  \\
 &  & GROND & 5.5 & i & 20.04 & 0.04 &  \\
 &  & GROND & 5.5 & J & 19.93 & 0.19 &  \\
 &  & GROND & 7.5 & H & 20.2 & 0.28 &  \\
 &  & GROND & 7.5 & r & 19.96 & 0.04 &  \\
 &  & GROND & 7.5 & J & 19.69 & 0.19 &  \\
 &  & GROND & 7.5 & i & 19.87 & 0.06 &  \\
 &  & HST/WFC3/UVIS & 8.7 & F555W & 19.99 & 0.014 & This work \\
 &  & HST/WFC3/UVIS & 8.7 & F814W & 19.45 & 0.01 & This work \\
 &  & HST/WFC3/IR & 8.7 & F125W & 20.0 & 0.01 & This work \\
 &  & HST/WFC3/IR & 8.7 & F160W & 20.15 & 0.01 & This work \\
 &  & GROND & 10.5 & H & 19.85 & 0.25 & \citet{Olivares+12} \\
 &  & GROND & 10.5 & i & 19.98 & 0.04 &  \\
 &  & GROND & 10.5 & r & 19.91 & 0.03 &  \\
 &  & GROND & 10.5 & J & 19.53 & 0.14 &  \\
 &  & GROND & 11.5 & r & 20.01 & 0.03 &  \\
 &  & GROND & 11.5 & H & 19.94 & 0.18 &  \\
 &  & GROND & 11.5 & J & 19.49 & 0.14 &  \\
 &  & GROND & 11.5 & i & 19.99 & 0.04 &  \\
 &  & GROND & 14.5 & r & 20.07 & 0.03 &  \\
 &  & GROND & 14.5 & i & 20.08 & 0.03 &  \\
 &  & GROND & 14.5 & J & 19.51 & 0.12 &  \\
 &  & GROND & 14.5 & H & 19.67 & 0.22 &  \\
 &  & GROND & 15.5 & i & 20.16 & 0.05 &  \\
 &  & GROND & 15.5 & H & 19.85 & 0.19 &  \\
 &  & GROND & 15.5 & r & 20.07 & 0.04 &  \\
 &  & GROND & 15.5 & J & 19.59 & 0.13 &  \\
 &  & HST/WFC3/UVIS & 15.8 & F555W & 20.51 & 0.02 & This work \\
 &  & HST/WFC3/UVIS & 15.8 & F814W & 19.57 & 0.01 & This work \\
 &  & GROND & 19.5 & J & 19.7 & 0.19 & \citet{Olivares+12} \\
 &  & GROND & 19.5 & r & 20.36 & 0.05 &  \\
 &  & GROND & 19.5 & i & 20.39 & 0.07 &  \\
 &  & GROND & 21.5 & J & 19.64 & 0.13 &  \\
 &  & GROND & 21.5 & i & 20.48 & 0.04 &  \\
 &  & GROND & 21.5 & H & 19.95 & 0.19 &  \\
 &  & GROND & 21.5 & r & 20.48 & 0.04 &  \\
 &  & HST/WFC3/UVIS & 24.9 & F555W & 21.07 & 0.02 & This work \\
 &  & HST/WFC3/UVIS & 24.9 & F814W & 20.03 & 0.01 & This work \\
 &  & HST/WFC3/IR & 25.5 & F125W & 20.08 & 0.01 & This work \\
 &  & GROND & 25.5 & J & 19.67 & 0.13 & \citet{Olivares+12} \\
 &  & GROND & 25.5 & H & 20.3 & 0.21 &  \\
 &  & GROND & 25.5 & i & 20.66 & 0.03 &  \\
 &  & GROND & 25.5 & r & 20.78 & 0.03 &  \\
 &  & HST/WFC3/IR & 25.5 & F160W & 20.03 & 0.01 & This work \\
 &  & GROND & 32.5 & r & 21.42 & 0.04 & \citet{Olivares+12} \\
 &  & GROND & 32.5 & J & 20.07 & 0.19 &  \\
 &  & GROND & 32.5 & i & 21.28 & 0.05 &  \\
 &  & GROND & 40.5 & i & 21.74 & 0.07 &  \\
 &  & GROND & 40.5 & r & 21.83 & 0.06 &  \\
 &  & HST/WFC3/UVIS & 47.4 & F555W & 22.57 & 0.03 & This work \\
 &  & HST/WFC3/UVIS & 47.4 & F814W & 21.43 & 0.02 & This work \\
 &  & GROND & 83.4 & r & 23.66 & 0.23 & \citet{Olivares+12} \\
 &  & HST/WFC3/UVIS & 137.5 & F555W & 24.62 & 0.03 & This work \\
 &  & HST/WFC3/UVIS & 137.5 & F814W & 23.55 & 0.04 & This work \\
130427A & 0.3399 & GROND & 1.7 & r & 18.61$^*$ & 0.04 & \citet{Perley+14} \\
 &  & GROND & 1.7 & i & 18.47$^*$ & 0.05 &  \\
 &  & GROND & 1.7 & J & 18.09$^*$ & 0.05 &  \\
 &  & GROND & 1.7 & H & 18.07$^*$ & 0.07 &  \\
 &  & GROND & 1.7 & K & 17.72$^*$ & 0.07 &  \\
 &  & P60 & 1.8 & r & 18.7$^*$ & 0.04 &  \\
 &  & P60 & 1.8 & i & 18.56 & 0.05 &  \\
 &  & P60 & 1.8 & r & 18.73$^*$ & 0.02 &  \\
 &  & P60 & 1.9 & i & 18.54$^*$ & 0.04 &  \\
 &  & P60 & 1.9 & r & 18.74$^*$ & 0.02 &  \\
 &  & P60 & 1.9 & i & 18.59$^*$ & 0.05 &  \\
 &  & KAIT & 1.9 & R & 18.81 & 0.05 &  \\
 &  & P60 & 1.9 & r & 18.75$^*$ & 0.03 &  \\
 &  & KAIT & 1.9 & I & 18.68 & 0.08 &  \\
 &  & RATIR & 1.9 & H & 18.09$^*$ & 0.05 &  \\
 &  & RATIR & 1.9 & J & 18.27$^*$ & 0.05 &  \\
 &  & P60 & 1.9 & i & 18.58$^*$ & 0.05 &  \\
 &  & P60 & 1.9 & r & 18.77$^*$ & 0.03 &  \\
 &  & P60 & 1.9 & i & 18.6$^*$ & 0.05 &  \\
 &  & P60 & 1.9 & r & 18.8$^*$ & 0.02 &  \\
 &  & P60 & 1.9 & i & 18.55$^*$ & 0.05 &  \\
 &  & P60 & 1.9 & r & 18.8$^*$ & 0.03 &  \\
 &  & P60 & 1.9 & i & 18.59$^*$ & 0.05 &  \\
 &  & P60 & 1.9 & r & 18.82$^*$ & 0.03 &  \\
 &  & P60 & 1.9 & i & 18.59$^*$ & 0.06 &  \\
 &  & P60 & 2.0 & r & 18.81$^*$ & 0.03 &  \\
 &  & P60 & 2.0 & r & 18.85 & 0.04 &  \\
 &  & P60 & 2.0 & i & 18.66$^*$ & 0.05 &  \\
 &  & P60 & 2.0 & r & 18.88 & 0.03 &  \\
 &  & P60 & 2.0 & i & 18.71$^*$ & 0.05 &  \\
 &  & P60 & 2.0 & r & 18.85 & 0.03 &  \\
 &  & P60 & 2.0 & i & 18.67$^*$ & 0.05 &  \\
 &  & P60 & 2.0 & r & 18.89$^*$ & 0.03 &  \\
 &  & P60 & 2.0 & i & 18.68$^*$ & 0.05 &  \\
 &  & P60 & 2.0 & r & 18.85$^*$ & 0.03 &  \\
 &  & P60 & 2.0 & i & 18.67$^*$ & 0.05 &  \\
 &  & P60 & 2.0 & r & 18.88$^*$ & 0.04 &  \\
 &  & P60 & 2.0 & i & 18.72$^*$ & 0.05 &  \\
 &  & P60 & 2.0 & r & 18.91$^*$ & 0.04 &  \\
 &  & P60 & 2.1 & i & 18.74$^*$ & 0.05 &  \\
 &  & P60 & 2.1 & r & 18.91$^*$ & 0.04 &  \\
 &  & P60 & 2.1 & i & 18.73$^*$ & 0.07 &  \\
 &  & P60 & 2.1 & r & 18.93$^*$ & 0.05 &  \\
 &  & P60 & 2.1 & i & 18.68$^*$ & 0.07 &  \\
 &  & P60 & 2.1 & r & 18.89$^*$ & 0.05 &  \\
 &  & GMG & 2.2 & r & 19.02$^*$ & 0.03 &  \\
 &  & GMG & 2.2 & r & 19.05$^*$ & 0.03 &  \\
 &  & GMG & 2.2 & r & 19.06$^*$ & 0.03 &  \\
 &  & T100 & 2.6 & R & 19.07 & 0.05 &  \\
 &  & P60 & 2.8 & r & 19.37$^*$ & 0.03 &  \\
 &  & P60 & 2.8 & i & 19.2$^*$ & 0.06 &  \\
 &  & P60 & 2.8 & r & 19.37$^*$ & 0.04 &  \\
 &  & KAIT & 2.9 & R & 19.59 & 0.12 &  \\
 &  & KAIT & 2.9 & I & 18.77 & 0.15 &  \\
 &  & P60 & 2.9 & r & 19.39$^*$ & 0.03 &  \\
 &  & P60 & 2.9 & i & 19.15$^*$ & 0.09 &  \\
 &  & P60 & 2.9 & r & 19.44$^*$ & 0.03 &  \\
 &  & P60 & 2.9 & i & 19.29$^*$ & 0.07 &  \\
 &  & RATIR & 2.9 & H & 18.71$^*$ & 0.06 &  \\
 &  & RATIR & 2.9 & J & 19.05$^*$ & 0.06 &  \\
 &  & P60 & 2.9 & r & 19.44$^*$ & 0.03 &  \\
 &  & P60 & 3.0 & i & 19.3$^*$ & 0.07 &  \\
 &  & P60 & 3.0 & r & 19.48$^*$ & 0.03 &  \\
 &  & P60 & 3.0 & i & 19.26$^*$ & 0.07 &  \\
 &  & P60 & 3.0 & r & 19.43$^*$ & 0.04 &  \\
 &  & P60 & 3.0 & i & 19.35$^*$ & 0.08 &  \\
 &  & P60 & 3.0 & r & 19.49$^*$ & 0.03 &  \\
 &  & P60 & 3.0 & i & 19.17$^*$ & 0.06 &  \\
 &  & P60 & 3.0 & r & 19.48$^*$ & 0.05 &  \\
 &  & UKIRT & 3.0 & K & 18.53$^*$ & 0.03 &  \\
 &  & P60 & 3.0 & i & 19.3$^*$ & 0.07 &  \\
 &  & UKIRT & 3.0 & H & 18.73$^*$ & 0.03 &  \\
 &  & P60 & 3.1 & r & 19.52$^*$ & 0.04 &  \\
 &  & P60 & 3.1 & i & 19.34$^*$ & 0.07 &  \\
 &  & UKIRT & 3.1 & J & 18.96$^*$ & 0.03 &  \\
 &  & P60 & 3.1 & r & 19.43$^*$ & 0.05 &  \\
 &  & GMG & 3.3 & r & 19.63$^*$ & 0.04 &  \\
 &  & GMG & 3.3 & r & 19.66$^*$ & 0.04 &  \\
 &  & T100 & 3.4 & R & 19.81 & 0.07 &  \\
 &  & P60 & 3.8 & r & 19.78$^*$ & 0.05 &  \\
 &  & P60 & 3.8 & i & 19.58$^*$ & 0.08 &  \\
 &  & P60 & 3.8 & r & 19.86$^*$ & 0.04 &  \\
 &  & P60 & 3.9 & i & 19.59$^*$ & 0.09 &  \\
 &  & Nickel & 3.9 & R & 19.61 & 0.12 &  \\
 &  & Nickel & 3.9 & I & 19.78 & 0.12 &  \\
 &  & P60 & 3.9 & r & 19.83$^*$ & 0.05 &  \\
 &  & P60 & 3.9 & i & 19.71$^*$ & 0.1 &  \\
 &  & P60 & 3.9 & r & 19.85$^*$ & 0.05 &  \\
 &  & P60 & 3.9 & i & 19.62$^*$ & 0.09 &  \\
 &  & P60 & 3.9 & r & 19.88$^*$ & 0.05 &  \\
 &  & P60 & 3.9 & i & 19.78$^*$ & 0.1 &  \\
 &  & P60 & 3.9 & r & 19.87$^*$ & 0.07 &  \\
 &  & P60 & 3.9 & r & 19.89$^*$ & 0.06 &  \\
 &  & P60 & 3.9 & i & 19.66$^*$ & 0.1 &  \\
 &  & P60 & 4.0 & r & 19.92$^*$ & 0.05 &  \\
 &  & P60 & 4.0 & i & 19.61$^*$ & 0.09 &  \\
 &  & P60 & 4.0 & r & 19.92$^*$ & 0.05 &  \\
 &  & P60 & 4.0 & i & 19.64$^*$ & 0.09 &  \\
 &  & P60 & 4.0 & r & 19.88$^*$ & 0.05 &  \\
 &  & P60 & 4.0 & i & 19.66$^*$ & 0.08 &  \\
 &  & P60 & 4.0 & r & 19.82$^*$ & 0.05 &  \\
 &  & P60 & 4.0 & i & 19.64$^*$ & 0.08 &  \\
 &  & P60 & 4.0 & r & 19.92$^*$ & 0.04 &  \\
 &  & P60 & 4.0 & i & 19.68$^*$ & 0.1 &  \\
 &  & P60 & 4.1 & r & 19.86$^*$ & 0.05 &  \\
 &  & P60 & 4.1 & i & 19.64$^*$ & 0.1 &  \\
 &  & T100 & 4.6 & R & 20.14 & 0.07 &  \\
 &  & P60 & 4.9 & r & 20.24$^*$ & 0.06 &  \\
 &  & P60 & 4.9 & i & 19.91$^*$ & 0.1 &  \\
 &  & P60 & 4.9 & r & 20.19$^*$ & 0.04 &  \\
 &  & P60 & 4.9 & i & 19.96$^*$ & 0.07 &  \\
 &  & KAST & 4.9 & R & 20.35 & 0.14 &  \\
 &  & KAST & 4.9 & R & 20.26 & 0.03 &  \\
 &  & UKIRT & 5.0 & K & 19.24$^*$ & 0.04 &  \\
 &  & UKIRT & 5.0 & J & 19.56$^*$ & 0.05 &  \\
 &  & P60 & 5.0 & r & 20.24$^*$ & 0.05 &  \\
 &  & P60 & 5.0 & i & 19.94$^*$ & 0.1 &  \\
 &  & T100 & 5.5 & R & 20.43 & 0.08 &  \\
 &  & P60 & 6.0 & r & 20.5$^*$ & 0.05 &  \\
 &  & P60 & 6.0 & i & 20.14$^*$ & 0.09 &  \\
 &  & P60 & 7.0 & r & 20.62$^*$ & 0.05 &  \\
 &  & P60 & 7.0 & i & 20.47$^*$ & 0.1 &  \\
 &  & UKIRT & 7.9 & K & 19.9$^*$ & 0.06 &  \\
 &  & UKIRT & 7.9 & H & 20.24$^*$ & 0.07 &  \\
 &  & UKIRT & 7.9 & J & 20.35$^*$ & 0.08 &  \\
 &  & P60 & 8.0 & r & 20.76$^*$ & 0.07 &  \\
 &  & P60 & 8.0 & i & 20.48$^*$ & 0.12 &  \\
 &  & Tautenburg & 8.6 & R & 21.04 & 0.15 &  \\
 &  & UKIRT & 11.9 & K & 20.57$^*$ & 0.09 &  \\
 &  & UKIRT & 11.9 & H & 20.84$^*$ & 0.12 &  \\
 &  & UKIRT & 11.9 & J & 20.93$^*$ & 0.13 &  \\
 &  & P60 & 13.0 & r & 21.33$^*$ & 0.1 &  \\
 &  & Keck & 13.1 & R & 21.47 & 0.07 &  \\
 &  & P60 & 13.9 & r & 21.28$^*$ & 0.05 &  \\
 &  & P60 & 14.0 & i & 21.03$^*$ & 0.14 &  \\
 &  & P60 & 14.8 & r & 21.33$^*$ & 0.07 &  \\
 &  & P60 & 14.9 & i & 21.08$^*$ & 0.15 &  \\
 &  & Tautenburg & 15.5 & R & 21.58 & 0.18 &  \\
 &  & P60 & 15.9 & r & 21.42$^*$ & 0.08 &  \\
 &  & P60 & 15.9 & i & 21.25$^*$ & 0.18 &  \\
 &  & UKIRT & 15.9 & K & 20.85 & 0.1 &  \\
 &  & UKIRT & 16.0 & H & 21.04 & 0.16 &  \\
 &  & UKIRT & 16.0 & J & 21.09 & 0.14 &  \\
 &  & GROND & 16.7 & r & 21.81$^*$ & 0.09 &  \\
 &  & GROND & 16.7 & i & 21.48 & 0.17 &  \\
 &  & P60 & 17.0 & i & 21.13 & 0.15 &  \\
 &  & P60 & 17.0 & r & 21.69$^*$ & 0.1 &  \\
 &  & P60 & 17.9 & i & 21.16 & 0.16 &  \\
 &  & P60 & 18.0 & r & 21.5$^*$ & 0.09 &  \\
 &  & P200 & 18.0 & r & 21.69$^*$ & 0.24 &  \\
 &  & GROND & 18.7 & i & 21.68 & 0.21 &  \\
 &  & GROND & 18.7 & r & 21.79$^*$ & 0.11 &  \\
 &  & Gemini-N & 18.9 & i & 21.76 & 0.21 &  \\
 &  & Gemini-N & 18.9 & r & 21.99$^*$ & 0.12 &  \\
 &  & P60 & 18.9 & i & 21.2 & 0.16 &  \\
 &  & P60 & 19.0 & r & 21.75$^*$ & 0.1 &  \\
 &  & P60 & 20.0 & i & 20.91 & 0.24 &  \\
 &  & Gemini-N & 20.0 & r & 21.79$^*$ & 0.21 &  \\
 &  & P60 & 20.0 & r & 21.75$^*$ & 0.11 &  \\
 &  & UKIRT & 20.9 & K & 21.4$^*$ & 0.17 &  \\
 &  & UKIRT & 20.9 & H & 21.13 & 0.16 &  \\
 &  & UKIRT & 21.0 & J & 21.57 & 0.23 &  \\
 &  & HST/WFC3/IR & 22.8 & F160W & 21.74 & 0.014 & This work \\
 &  & P60 & 22.9 & i & 21.41 & 0.2 & \citet{Perley+14} \\
 &  & P60 & 22.9 & r & 21.78 & 0.12 &  \\
 &  & UKIRT & 23.0 & K & 21.56$^*$ & 0.19 &  \\
 &  & HST/WFC3/UVIS & 23.0 & F606W & 21.87 & 0.01 & This work \\
 &  & UKIRT & 23.1 & J & 21.41 & 0.18 & \citet{Perley+14} \\
 &  & P60 & 23.9 & i & 21.3 & 0.21 &  \\
 &  & P60 & 23.9 & r & 21.72 & 0.14 &  \\
 &  & P60 & 24.9 & i & 21.3 & 0.2 &  \\
 &  & P60 & 26.9 & i & 21.63 & 0.26 &  \\
 &  & UKIRT & 30.0 & J & 21.73 & 0.26 &  \\
 &  & P60 & 32.9 & r & 22.53 & 0.17 &  \\
 &  & P60 & 34.9 & i & 21.94 & 0.27 &  \\
 &  & P60 & 36.9 & r & 22.65 & 0.17 &  \\
 &  & P60 & 39.9 & i & 21.88 & 0.21 &  \\
 &  & Keck & 40.0 & R & 23.18 & 0.22 &  \\
 &  & UKIRT & 40.9 & H & 21.45 & 0.18 &  \\
 &  & HST/WFC3/IR & 73.9 & F160W & 23.06 & 0.02 & This work \\
 &  & HST/WFC3/UVIS & 73.9 & F606W & 23.96 & 0.02 & This work \\
 &  & HST/WFC3/IR & 259.8 & F160W & 25.71$^*$ & 0.14 & This work \\
 &  & HST/WFC3/UVIS & 259.9 & F606W & 26.39$^*$ & 0.07 & This work \\
 &  & HST/WFC3/UVIS & 315.9 & F606W & 26.71$^*$ & 0.06 & This work \\
 &  & HST/WFC3/IR & 316.0 & F160W & 26.05$^*$ & 0.11 & This work \\
 &  & HST/WFC3/UVIS & 355.9 & F606W & 26.98$^*$ & 0.19 & This work \\
 &  & HST/WFC3/IR & 382.8 & F160W & 26.84$^*$ & 0.17 & This work \\
 &  & HST/WFC3/UVIS & 382.9 & F606W & 27.56$^*$ & 0.12 & This work \\
 &  & HST/WFC3/IR & 587.5 & F160W & 28.11$^*$ & 0.64 & This work \\
 &  & HST/WFC3/UVIS & 587.6 & F606W & 27.88$^*$ & 0.12 & This work \\
190829A & 0.0785 & GTC & 0.4 & i & 18.87 & 0.01 & \citet{Hu+21} \\
 &  & GTC & 0.4 & r & 19.79$^*$ & 0.01 &  \\
 &  & GTC & 1.4 & r & 21.86$^*$ & 0.06 &  \\
 &  & GTC & 1.4 & i & 20.81$^*$ & 0.05 &  \\
 &  & GTC & 2.4 & i & 22.42$^*$ & 0.06 &  \\
 &  & GTC & 2.4 & r & 22.71$^*$ & 0.03 &  \\
 &  & GTC & 3.4 & r & 23.05$^*$ & 0.03 &  \\
 &  & GTC & 3.4 & i & 22.38$^*$ & 0.02 &  \\
 &  & GTC & 4.4 & i & 22.23$^*$ & 0.08 &  \\
 &  & GTC & 4.4 & r & 22.7 & 0.06 &  \\
 &  & GTC & 7.3 & i & 22.08 & 0.04 &  \\
 &  & GTC & 11.4 & i & 21.16 & 0.03 &  \\
 &  & GTC & 14.3 & i & 20.84 & 0.09 &  \\
 &  & VLT/HAWKI & 25.4 & J & 19.86 & 0.021 & This work \\
 &  & VLT/XSHOOTER & 25.4 & r & 22.15 & 0.12 & This work \\
 &  & VLT/HAWKI & 25.4 & H & 20.22 & 0.09 & This work \\
 &  & VLT/HAWKI & 25.4 & K_s & 20.19 & 0.06 & This work \\
 &  & HST/WFC3/IR & 29.3 & F110W & 20.28 & 0.05 & This work \\
 &  & GTC & 29.3 & i & 21.31 & 0.02 & \citet{Hu+21} \\
 &  & HST/WFC3/IR & 29.4 & F160W & 20.51 & 0.09 & This work \\
 &  & GTC & 40.3 & i & 22.56 & 0.01 & \citet{Hu+21} \\
 &  & VLT/HAWKI & 49.4 & J & 21.4 & 0.02 & This work \\
 &  & VLT/HAWKI & 49.4 & H & 21.4 & 0.08 & This work \\
 &  & VLT/HAWKI & 49.4 & K_s & 21.35 & 0.07 & This work \\
 &  & VLT/XSHOOTER & 50.3 & r & 23.05 & 0.17 & This work \\
 &  & HST/WFC3/IR & 57.8 & F110W & 21.91 & 0.12 & This work \\
 &  & HST/WFC3/IR & 58.0 & F160W & 21.98 & 0.18 & This work \\
 &  & VLT/HAWKI & 77.2 & J & 22.25 & 0.04 & This work \\
 &  & VLT/HAWKI & 77.2 & H & 22.86 & 0.18 & This work \\
 &  & VLT/HAWKI & 77.2 & K_s & 22.97 & 0.15 & This work \\
 &  & VLT/XSHOOTER & 79.3 & r & 24.13 & 0.14 & This work \\
 &  & HST/WFC3/UVIS & 87.2 & F606W & 24.71 & 0.03 & This work \\
 &  & HST/WFC3/IR & 92.9 & F140W & 23.55 & 0.04 & This work \\
 &  & VLT/HAWKI & 94.4 & K_s & 23.68 & 0.28 & This work \\
 &  & VLT/XSHOOTER & 109.2 & r & 24.81 & 0.19 & This work \\
 &  & VLT/HAWKI & 112.2 & J & 23.27 & 0.18 & This work \\
 &  & HST/WFC3/UVIS & 135.6 & F606W & 25.34 & 0.03 & This work \\
 &  & HST/WFC3/IR & 135.8 & F140W & 24.17 & 0.05 & This work \\
 &  & VLT/HAWKI & 137.3 & J & 23.74 & 0.2 & This work \\
 &  & HST/WFC3/IR & 170.4 & F140W & 25.16 & 0.08 & This work \\
 &  & HST/WFC3/UVIS & 170.5 & F606W & 25.96 & 0.06 & This work \\
 &  & HST/WFC3/IR & 299.3 & F140W & 26.4 & 0.19 & This work \\
 &  & HST/WFC3/UVIS & 299.4 & F606W & 27.33 & 0.27 & This work \\
 &  & HST/WFC3/IR & 358.0 & F140W & 26.78 & 0.25 & This work \\
 &  & HST/WFC3/UVIS & 358.0 & F606W & 28.26 & 0.56 & This work \\
\enddata
\tablecomments{
($*$) Indicates observation is dominated or significantly affected by afterglow emission. \\
($\dagger$) Observations are not corrected for Galactic nor local extinction.}
\end{deluxetable*}

%% file: main.bbl
\begin{thebibliography}{}
\expandafter\ifx\csname natexlab\endcsname\relax\def\natexlab#1{#1}\fi
\providecommand{\url}[1]{\href{#1}{#1}}
\providecommand{\dodoi}[1]{doi:~\href{http://doi.org/#1}{\nolinkurl{#1}}}
\providecommand{\doeprint}[1]{\href{http://ascl.net/#1}{\nolinkurl{http://ascl.net/#1}}}
\providecommand{\doarXiv}[1]{\href{https://arxiv.org/abs/#1}{\nolinkurl{https://arxiv.org/abs/#1}}}

\bibitem[{{Abbott} {et~al.}(2017{\natexlab{a}}){Abbott}, {Abbott}, {Abbott},
  {Acernese}, {Ackley}, {Adams}, {Adams}, {Addesso}, {Adhikari}, {Adya},
  {Affeldt}, {Afrough}, {Agarwal}, {Agathos}, {Agatsuma}, {Aggarwal}, {Aguiar},
  {Aiello}, {Ain}, {Ajith}, {Allen}, {Allen}, {Allocca}, {Altin}, {Amato},
  {Ananyeva}, {Anderson}, \& {Anderson}}]{gw170817mma}
{Abbott}, B.~P., {Abbott}, R., {Abbott}, T.~D., {et~al.} 2017{\natexlab{a}},
  \apjl, 848, L12, \dodoi{10.3847/2041-8213/aa91c9}

\bibitem[{{Abbott} {et~al.}(2017{\natexlab{b}}){Abbott}, {Abbott}, {Abbott},
  {Acernese}, {Ackley}, {Adams}, {Adams}, {Addesso}, {Adhikari}, {Adya},
  {Affeldt}, {Afrough}, {Agarwal}, {Agathos}, {Agatsuma}, {Aggarwal}, {Aguiar},
  {Aiello}, {Ain}, {Ajith}, {Allen}, {Allen}, {Allocca}, {Aloy}, {Altin},
  {Amato}, {Ananyeva}, {Anderson}, {Anderson}, {Angelova}, {Antier}, {Appert},
  {Arai}, {Araya}, {Areeda}, {Arnaud}, {Arun}, {Ascenzi}, {Ashton}, {Ast},
  {Aston}, {Astone}, {Atallah}, {Aufmuth}, {Aulbert}, {AultONeal}, {Austin},
  {Avila-Alvarez}, {Babak}, {Bacon}, {Bader}, {Bae}, {Baker}, {Baldaccini},
  {Ballardin}, {Ballmer}, {Banagiri}, {Barayoga}, {Barclay}, {Barish},
  {Barker}, {Barkett}, {Barone}, {Barr}, {Barsotti}, {Barsuglia}, {Barta},
  {Bartlett}, {Bartos}, {Bassiri}, {Basti}, {Batch}, {Bawaj}, {Bayley},
  {Bazzan}, {B{\'e}csy}, {Beer}, {Bejger}, {Belahcene}, {Bell}, {Berger},
  {Bergmann}, {Bero}, {Berry}, {Bersanetti}, {Bertolini}, {Betzwieser},
  {Bhagwat}, {Bhandare}, {Bilenko}, {Billingsley}, {Billman}, {Birch},
  {Birney}, {Birnholtz}, {Biscans}, {Biscoveanu}, {Bisht}, {Bitossi}, {Biwer},
  {Bizouard}, {Blackburn}, {Blackman}, {Blair}, {Blair}, {Blair}, {Bloemen},
  {Bock}, {Bode}, {Boer}, {Bogaert}, {Bohe}, {Bondu}, {Bonilla}, {Bonnand},
  {Boom}, {Bork}, {Boschi}, {Bose}, {Bossie}, {Bouffanais}, {Bozzi},
  {Bradaschia}, {Brady}, {Branchesi}, {Brau}, {Briant}, {Brillet}, {Brinkmann},
  {Brisson}, {Brockill}, {Broida}, {Brooks}, {Brown}, {Brown}, {Brunett},
  {Buchanan}, {Buikema}, {Bulik}, {Bulten}, {Buonanno}, {Buskulic}, {Buy},
  {Byer}, {Cabero}, {Cadonati}, {Cagnoli}, {Cahillane}, {Calder{\'o}n
  Bustillo}, {Callister}, {Calloni}, {Camp}, {Canepa}, {Canizares}, {Cannon},
  {Cao}, {Cao}, {Capano}, {Capocasa}, {Carbognani}, {Caride}, {Carney},
  {Casanueva Diaz}, {Casentini}, {Caudill}, {Cavagli{\`a}}, {Cavalier},
  {Cavalieri}, {Cella}, {Cepeda}, {Cerd{\'a}-Dur{\'a}n}, {Cerretani},
  {Cesarini}, {Chamberlin}, {Chan}, {Chao}, {Charlton}, {Chase},
  {Chassande-Mottin}, {Chatterjee}, {Chatziioannou}, {Cheeseboro}, {Chen},
  {Chen}, {Chen}, {Cheng}, {Chia}, {Chincarini}, {Chiummo}, {Chmiel}, {Cho},
  {Cho}, {Chow}, {Christensen}, {Chu}, {Chua}, {Chua}, {Chung}, {Chung},
  {Ciani}, {Ciolfi}, {Cirelli}, {Cirone}, {Clara}, {Clark}, {Clearwater},
  {Cleva}, {Cocchieri}, {Coccia}, {Cohadon}, {Cohen}, {Colla}, {Collette},
  {Cominsky}, {Constancio}, {Conti}, {Cooper}, {Corban}, {Corbitt},
  {Cordero-Carri{\'o}n}, {Corley}, {Cornish}, {Corsi}, {Cortese}, {Costa},
  {Coughlin}, {Coughlin}, {Coulon}, {Countryman}, {Couvares}, {Covas}, {Cowan},
  {Coward}, {Cowart}, {Coyne}, {Coyne}, {Creighton}, {Creighton}, {Cripe},
  {Crowder}, {Cullen}, {Cumming}, {Cunningham}, {Cuoco}, {Dal Canton},
  {D{\'a}lya}, {Danilishin}, {D'Antonio}, {Danzmann}, {Dasgupta}, {Da Silva
  Costa}, {Dattilo}, {Dave}, {Davier}, {Davis}, {Daw}, {Day}, {De}, {DeBra},
  {Degallaix}, {De Laurentis}, {Del{\'e}glise}, {Del Pozzo}, {Demos}, {Denker},
  {Dent}, {De Pietri}, {Dergachev}, {De Rosa}, {DeRosa}, {De Rossi}, {DeSalvo},
  {de Varona}, {Devenson}, {Dhurandhar}, {D{\'\i}az}, {Di Fiore}, {Di
  Giovanni}, {Di Girolamo}, {Di Lieto}, {Di Pace}, {Di Palma}, {Di Renzo},
  {Doctor}, {Dolique}, {Donovan}, {Dooley}, {Doravari}, {Dorrington},
  {Douglas}, {Dovale {\'A}lvarez}, {Downes}, {Drago}, {Dreissigacker},
  {Driggers}, {Du}, {Ducrot}, {Dupej}, {Dwyer}, {Edo}, {Edwards}, {Effler},
  {Eggenstein}, {Ehrens}, {Eichholz}, {Eikenberry}, {Eisenstein}, {Essick},
  {Estevez}, {Etienne}, {Etzel}, {Evans}, {Evans}, {Factourovich}, {Fafone},
  {Fair}, {Fairhurst}, {Fan}, {Farinon}, {Farr}, {Farr}, {Fauchon-Jones},
  {Favata}, {Fays}, {Fee}, {Fehrmann}, {Feicht}, {Fejer}, {Fernandez-Galiana},
  {Ferrante}, {Ferreira}, {Ferrini}, {Fidecaro}, {Finstad}, {Fiori},
  {Fiorucci}, {Fishbach}, {Fisher}, {Fitz-Axen}, {Flaminio}, {Fletcher},
  {Fong}, {Font}, {Forsyth}, {Forsyth}, {Fournier}, {Frasca}, {Frasconi},
  {Frei}, {Freise}, {Frey}, {Frey}, {Fries}, {Fritschel}, {Frolov}, {Fulda},
  {Fyffe}, {Gabbard}, {Gadre}, {Gaebel}, {Gair}, {Gammaitoni}, {Ganija},
  {Gaonkar}, {Garcia-Quiros}, {Garufi}, {Gateley}, {Gaudio}, {Gaur},
  {Gayathri}, {Gehrels}, {Gemme}, {Genin}, {Gennai}, {George}, {George},
  {Gergely}, {Germain}, {Ghonge}, {Ghosh}, {Ghosh}, {Ghosh}, {Giaime},
  {Giardina}, {Giazotto}, {Gill}, {Glover}, {Goetz}, {Goetz}, {Gomes},
  {Goncharov}, {Gonz{\'a}lez}, {Gonzalez Castro}, {Gopakumar}, {Gorodetsky},
  {Gossan}, {Gosselin}, {Gouaty}, {Grado}, {Graef}, {Granata}, {Grant}, {Gras},
  {Gray}, {Greco}, {Green}, {Gretarsson}, {Groot}, {Grote}, {Grunewald},
  {Gruning}, {Guidi}, {Guo}, {Gupta}, {Gupta}, {Gushwa}, {Gustafson},
  {Gustafson}, {Halim}, {Hall}, {Hall}, {Hamilton}, {Hammond}, {Haney},
  {Hanke}, {Hanks}, {Hanna}, {Hannam}, {Hannuksela}, {Hanson}, {Hardwick},
  {Harms}, {Harry}, {Harry}, {Hart}, {Haster}, {Haughian}, {Healy}, {Heidmann},
  {Heintze}, {Heitmann}, {Hello}, {Hemming}, {Hendry}, {Heng}, {Hennig},
  {Heptonstall}, {Heurs}, {Hild}, {Hinderer}, {Hoak}, {Hofman}, {Holt}, {Holz},
  {Hopkins}, {Horst}, {Hough}, {Houston}, {Howell}, {Hreibi}, {Hu}, {Huerta},
  {Huet}, {Hughey}, {Husa}, {Huttner}, {Huynh-Dinh}, {Indik}, {Inta}, {Intini},
  {Isa}, {Isac}, {Isi}, {Iyer}, {Izumi}, {Jacqmin}, {Jani}, {Jaranowski},
  {Jawahar}, {Jim{\'e}nez-Forteza}, {Johnson}, {Johnson-McDaniel}, {Jones},
  {Jones}, {Jonker}, {Ju}, {Junker}, {Kalaghatgi}, {Kalogera}, {Kamai},
  {Kandhasamy}, {Kang}, {Kanner}, {Kapadia}, {Karki}, {Karvinen}, {Kasprzack},
  {Kastaun}, {Katolik}, {Katsavounidis}, {Katzman}, {Kaufer}, {Kawabe},
  {K{\'e}f{\'e}lian}, {Keitel}, {Kemball}, {Kennedy}, {Kent}, {Key}, {Khalili},
  {Khan}, {Khan}, {Khan}, {Khazanov}, {Kijbunchoo}, {Kim}, {Kim}, {Kim}, {Kim},
  {Kim}, {Kim}, {Kimbrell}, {King}, {King}, {Kinley-Hanlon}, {Kirchhoff},
  {Kissel}, {Kleybolte}, {Klimenko}, {Knowles}, {Koch}, {Koehlenbeck}, {Koley},
  {Kondrashov}, {Kontos}, {Korobko}, {Korth}, {Kowalska}, {Kozak},
  {Kr{\"a}mer}, {Kringel}, {Krishnan}, {Kr{\'o}lak}, {Kuehn}, {Kumar}, {Kumar},
  {Kumar}, {Kuo}, {Kutynia}, {Kwang}, {Lackey}, {Lai}, {Landry}, {Lang},
  {Lange}, {Lantz}, {Lanza}, {Lartaux-Vollard}, {Lasky}, {Laxen}, {Lazzarini},
  {Lazzaro}, {Leaci}, {Leavey}, {Lee}, {Lee}, {Lee}, {Lee}, {Lee}, {Lehmann},
  {Lenon}, {Leonardi}, {Leroy}, {Letendre}, {Levin}, {Li}, {Linker},
  {Littenberg}, {Liu}, {Lo}, {Lockerbie}, {London}, {Lord}, {Lorenzini},
  {Loriette}, {Lormand}, {Losurdo}, {Lough}, {Lousto}, {Lovelace}, {L{\"u}ck},
  {Lumaca}, {Lundgren}, {Lynch}, {Ma}, {Macas}, {Macfoy}, {Machenschalk},
  {MacInnis}, {Macleod}, {Maga{\~n}a Hernandez}, {Maga{\~n}a-Sandoval},
  {Maga{\~n}a Zertuche}, {Magee}, {Majorana}, {Maksimovic}, {Man}, {Mandic},
  {Mangano}, {Mansell}, {Manske}, {Mantovani}, {Marchesoni}, {Marion},
  {M{\'a}rka}, {M{\'a}rka}, {Markakis}, {Markosyan}, {Markowitz}, {Maros},
  {Marquina}, {Martelli}, {Martellini}, {Martin}, {Martin}, {Martynov},
  {Mason}, {Massera}, {Masserot}, {Massinger}, {Masso-Reid}, {Mastrogiovanni},
  {Matas}, {Matichard}, {Matone}, {Mavalvala}, {Mazumder}, {McCarthy},
  {McClelland}, {McCormick}, {McCuller}, {McGuire}, {McIntyre}, {McIver},
  {McManus}, {McNeill}, {McRae}, {McWilliams}, {Meacher}, {Meadors}, {Mehmet},
  {Meidam}, {Mejuto-Villa}, {Melatos}, {Mendell}, {Mercer}, {Merilh},
  {Merzougui}, {Meshkov}, {Messenger}, {Messick}, {Metzdorff}, {Meyers},
  {Miao}, {Michel}, {Middleton}, {Mikhailov}, {Milano}, {Miller}, {Miller},
  {Miller}, {Millhouse}, {Milovich-Goff}, {Minazzoli}, {Minenkov}, {Ming},
  {Mishra}, {Mitra}, {Mitrofanov}, {Mitselmakher}, {Mittleman}, {Moffa},
  {Moggi}, {Mogushi}, {Mohan}, {Mohapatra}, {Montani}, {Moore}, {Moraru},
  {Moreno}, {Morriss}, {Mours}, {Mow-Lowry}, {Mueller}, {Muir}, {Mukherjee},
  {Mukherjee}, {Mukherjee}, {Mukund}, {Mullavey}, {Munch}, {Mu{\~n}iz},
  {Muratore}, {Murray}, {Napier}, {Nardecchia}, {Naticchioni}, {Nayak},
  {Neilson}, {Nelemans}, {Nelson}, {Nery}, {Neunzert}, {Nevin}, {Newport},
  {Newton}, {Ng}, {Nguyen}, {Nichols}, {Nielsen}, {Nissanke}, {Nitz}, {Noack},
  {Nocera}, {Nolting}, {North}, {Nuttall}, {Oberling}, {O'Dea}, {Ogin}, {Oh},
  {Oh}, {Ohme}, {Okada}, {Oliver}, {Oppermann}, {Oram}, {O'Reilly}, {Ormiston},
  {Ortega}, {O'Shaughnessy}, {Ossokine}, {Ottaway}, {Overmier}, {Owen}, {Pace},
  {Page}, {Page}, {Pai}, {Pai}, {Palamos}, {Palashov}, {Palomba}, {Pal-Singh},
  {Pan}, {Pan}, {Pang}, {Pang}, {Pankow}, {Pannarale}, {Pant}, {Paoletti},
  {Paoli}, {Papa}, {Parida}, {Parker}, {Pascucci}, {Pasqualetti},
  {Passaquieti}, {Passuello}, {Patil}, {Patricelli}, {Pearlstone}, {Pedraza},
  {Pedurand}, {Pekowsky}, {Pele}, {Penn}, {Perez}, {Perreca}, {Perri},
  {Pfeiffer}, {Phelps}, {Piccinni}, {Pichot}, {Piergiovanni}, {Pierro},
  {Pillant}, {Pinard}, {Pinto}, {Pirello}, {Pitkin}, {Poe}, {Poggiani},
  {Popolizio}, {Porter}, {Post}, {Powell}, {Prasad}, {Pratt}, {Pratten},
  {Predoi}, {Prestegard}, {Prijatelj}, {Principe}, {Privitera}, {Prodi},
  {Prokhorov}, {Puncken}, {Punturo}, {Puppo}, {P{\"u}rrer}, {Qi}, {Quetschke},
  {Quintero}, {Quitzow-James}, {Raab}, {Rabeling}, {Radkins}, {Raffai}, {Raja},
  {Rajan}, {Rajbhandari}, {Rakhmanov}, {Ramirez}, {Ramos-Buades}, {Rapagnani},
  {Raymond}, {Razzano}, {Read}, {Regimbau}, {Rei}, {Reid}, {Reitze}, {Ren},
  {Reyes}, {Ricci}, {Ricker}, {Rieger}, {Riles}, {Rizzo}, {Robertson}, {Robie},
  {Robinet}, {Rocchi}, {Rolland}, {Rollins}, {Roma}, {Romano}, {Romel},
  {Romie}, {Rosi{\'n}ska}, {Ross}, {Rowan}, {R{\"u}diger}, {Ruggi}, {Rutins},
  {Ryan}, {Sachdev}, {Sadecki}, {Sadeghian}, {Sakellariadou}, {Salconi},
  {Saleem}, {Salemi}, {Samajdar}, {Sammut}, {Sampson}, {Sanchez}, {Sanchez},
  {Sanchis-Gual}, {Sandberg}, {Sanders}, {Sassolas}, {Sathyaprakash},
  {Saulson}, {Sauter}, {Savage}, {Sawadsky}, {Schale}, {Scheel}, {Scheuer},
  {Schmidt}, {Schmidt}, {Schnabel}, {Schofield}, {Sch{\"o}nbeck}, {Schreiber},
  {Schuette}, {Schulte}, {Schutz}, {Schwalbe}, {Scott}, {Scott}, {Seidel},
  {Sellers}, {Sengupta}, {Sentenac}, {Sequino}, {Sergeev}, {Shaddock},
  {Shaffer}, {Shah}, {Shahriar}, {Shaner}, {Shao}, {Shapiro}, {Shawhan},
  {Sheperd}, {Shoemaker}, {Shoemaker}, {Siellez}, {Siemens}, {Sieniawska},
  {Sigg}, {Silva}, {Singer}, {Singh}, {Singhal}, {Sintes}, {Slagmolen},
  {Smith}, {Smith}, {Smith}, {Somala}, {Son}, {Sonnenberg}, {Sorazu},
  {Sorrentino}, {Souradeep}, {Spencer}, {Srivastava}, {Staats}, {Staley},
  {Steinke}, {Steinlechner}, {Steinlechner}, {Steinmeyer}, {Stevenson},
  {Stone}, {Stops}, {Strain}, {Stratta}, {Strigin}, {Strunk}, {Sturani},
  {Stuver}, {Summerscales}, {Sun}, {Sunil}, {Suresh}, {Sutton}, {Swinkels},
  {Szczepa{\'n}czyk}, {Tacca}, {Tait}, {Talbot}, {Talukder}, {Tanner},
  {T{\'a}pai}, {Taracchini}, {Tasson}, {Taylor}, {Taylor}, {Tewari}, {Theeg},
  {Thies}, {Thomas}, {Thomas}, {Thomas}, {Thorne}, {Thorne}, {Thrane},
  {Tiwari}, {Tiwari}, {Tokmakov}, {Toland}, {Tonelli}, {Tornasi},
  {Torres-Forn{\'e}}, {Torrie}, {T{\"o}yr{\"a}}, {Travasso}, {Traylor},
  {Trinastic}, {Tringali}, {Trozzo}, {Tsang}, {Tse}, {Tso}, {Tsukada}, {Tsuna},
  {Tuyenbayev}, {Ueno}, {Ugolini}, {Unnikrishnan}, {Urban}, {Usman},
  {Vahlbruch}, {Vajente}, {Valdes}, {van Bakel}, {van Beuzekom}, {van den
  Brand}, {Van Den Broeck}, {Vander-Hyde}, {van der Schaaf}, {van Heijningen},
  {van Veggel}, {Vardaro}, {Varma}, {Vass}, {Vas{\'u}th}, {Vecchio},
  {Vedovato}, {Veitch}, {Veitch}, {Venkateswara}, {Venugopalan}, {Verkindt},
  {Vetrano}, {Vicer{\'e}}, {Viets}, {Vinciguerra}, {Vine}, {Vinet}, {Vitale},
  {Vo}, {Vocca}, {Vorvick}, {Vyatchanin}, {Wade}, {Wade}, {Wade}, {Walet},
  {Walker}, {Wallace}, {Walsh}, {Wang}, {Wang}, {Wang}, {Wang}, {Wang}, {Ward},
  {Warner}, {Was}, {Watchi}, {Weaver}, {Wei}, {Weinert}, {Weinstein}, {Weiss},
  {Wen}, {Wessel}, {We{\ss}els}, {Westerweck}, {Westphal}, {Wette}, {Whelan},
  {Whitcomb}, {Whiting}, {Whittle}, {Wilken}, {Williams}, {Williams},
  {Williamson}, {Willis}, {Willke}, {Wimmer}, {Winkler}, {Wipf}, {Wittel},
  {Woan}, {Woehler}, {Wofford}, {Wong}, {Worden}, {Wright}, {Wu}, {Wysocki},
  {Xiao}, {Yamamoto}, {Yancey}, {Yang}, {Yap}, {Yazback}, {Yu}, {Yu}, {Yvert},
  {Zadro{\.z}ny}, {Zanolin}, {Zelenova}, {Zendri}, {Zevin}, {Zhang}, {Zhang},
  {Zhang}, {Zhang}, {Zhao}, {Zhou}, {Zhou}, {Zhu}, {Zhu}, {Zimmerman},
  {Zucker}, {Zweizig}, {(LIGO Scientific Collaboration}, {Virgo Collaboration},
  {Burns}, {Veres}, {Kocevski}, {Racusin}, {Goldstein}, {Connaughton},
  {Briggs}, {Blackburn}, {Hamburg}, {Hui}, {von Kienlin}, {McEnery}, {Preece},
  {Wilson-Hodge}, {Bissaldi}, {Cleveland}, {Gibby}, {Giles}, {Kippen},
  {McBreen}, {Meegan}, {Paciesas}, {Poolakkil}, {Roberts}, {Stanbro},
  {Gamma-ray Burst Monitor}, {Savchenko}, {Ferrigno}, {Kuulkers}, {Bazzano},
  {Bozzo}, {Brandt}, {Chenevez}, {Courvoisier}, {Diehl}, {Domingo}, {Hanlon},
  {Jourdain}, {Laurent}, {Lebrun}, {Lutovinov}, {Mereghetti}, {Natalucci},
  {Rodi}, {Roques}, {Sunyaev}, {Ubertini}, \&
  {(INTEGRAL}}]{gw170817_grb170817a}
---. 2017{\natexlab{b}}, \apjl, 848, L13, \dodoi{10.3847/2041-8213/aa920c}

\bibitem[{{Ackermann} {et~al.}(2014){Ackermann}, {Ajello}, {Asano}, {Atwood},
  {Axelsson}, {Baldini}, {Ballet}, {Barbiellini}, {Baring}, {Bastieri},
  {Bechtol}, {Bellazzini}, {Bissaldi}, {Bonamente}, {Bregeon}, {Brigida},
  {Bruel}, {Buehler}, {Burgess}, {Buson}, {Caliandro}, {Cameron}, {Caraveo},
  {Cecchi}, {Chaplin}, {Charles}, {Chekhtman}, {Cheung}, {Chiang}, {Chiaro},
  {Ciprini}, {Claus}, {Cleveland}, {Cohen-Tanugi}, {Collazzi}, {Cominsky},
  {Connaughton}, {Conrad}, {Cutini}, {D'Ammando}, {de Angelis}, {DeKlotz}, {de
  Palma}, {Dermer}, {Desiante}, {Diekmann}, {Di Venere}, {Drell},
  {Drlica-Wagner}, {Favuzzi}, {Fegan}, {Ferrara}, {Finke}, {Fitzpatrick},
  {Focke}, {Franckowiak}, {Fukazawa}, {Funk}, {Fusco}, {Gargano}, {Gehrels},
  {Germani}, {Gibby}, {Giglietto}, {Giles}, {Giordano}, {Giroletti}, {Godfrey},
  {Granot}, {Grenier}, {Grove}, {Gruber}, {Guiriec}, {Hadasch}, {Hanabata},
  {Harding}, {Hayashida}, {Hays}, {Horan}, {Hughes}, {Inoue}, {Jogler},
  {J{\'o}hannesson}, {Johnson}, {Kawano}, {Kn{\"o}dlseder}, {Kocevski}, {Kuss},
  {Lande}, {Larsson}, {Latronico}, {Longo}, {Loparco}, {Lovellette}, {Lubrano},
  {Mayer}, {Mazziotta}, {McEnery}, {Michelson}, {Mizuno}, {Moiseev}, {Monzani},
  {Moretti}, {Morselli}, {Moskalenko}, {Murgia}, {Nemmen}, {Nuss}, {Ohno},
  {Ohsugi}, {Okumura}, {Omodei}, {Orienti}, {Paneque}, {Pelassa}, {Perkins},
  {Pesce-Rollins}, {Petrosian}, {Piron}, {Pivato}, {Porter}, {Racusin},
  {Rain{\`o}}, {Rando}, {Razzano}, {Razzaque}, {Reimer}, {Reimer}, {Ritz},
  {Roth}, {Ryde}, {Sartori}, {Parkinson}, {Scargle}, {Schulz}, {Sgr{\`o}},
  {Siskind}, {Sonbas}, {Spandre}, {Spinelli}, {Tajima}, {Takahashi}, {Thayer},
  {Thayer}, {Thompson}, {Tibaldo}, {Tinivella}, {Torres}, {Tosti}, {Troja},
  {Usher}, {Vandenbroucke}, {Vasileiou}, {Vianello}, {Vitale}, {Winer}, {Wood},
  {Yamazaki}, {Younes}, {Yu}, {Zhu}, {Bhat}, {Briggs}, {Byrne}, {Foley},
  {Goldstein}, {Jenke}, {Kippen}, {Kouveliotou}, {McBreen}, {Meegan},
  {Paciesas}, {Preece}, {Rau}, {Tierney}, {van der Horst}, {von Kienlin},
  {Wilson-Hodge}, {Xiong}, {Cusumano}, {La Parola}, \&
  {Cummings}}]{Ackerman+14}
{Ackermann}, M., {Ajello}, M., {Asano}, K., {et~al.} 2014, Science, 343, 42,
  \dodoi{10.1126/science.1242353}

\bibitem[{{Alam} {et~al.}(2015){Alam}, {Albareti}, {Allende Prieto}, {Anders},
  {Anderson}, {Anderton}, {Andrews}, {Armengaud}, {Aubourg}, {Bailey}, {Basu},
  {Bautista}, {Beaton}, {Beers}, {Bender}, {Berlind}, {Beutler}, {Bhardwaj},
  {Bird}, {Bizyaev}, {Blake}, {Blanton}, {Blomqvist}, {Bochanski}, {Bolton},
  {Bovy}, {Shelden Bradley}, {Brandt}, {Brauer}, {Brinkmann}, {Brown},
  {Brownstein}, {Burden}, {Burtin}, {Busca}, {Cai}, {Capozzi}, {Carnero
  Rosell}, {Carr}, {Carrera}, {Chambers}, {Chaplin}, {Chen}, {Chiappini},
  {Chojnowski}, {Chuang}, {Clerc}, {Comparat}, {Covey}, {Croft}, {Cuesta},
  {Cunha}, {da Costa}, {Da Rio}, {Davenport}, {Dawson}, {De Lee}, {Delubac},
  {Deshpande}, {Dhital}, {Dutra-Ferreira}, {Dwelly}, {Ealet}, {Ebelke},
  {Edmondson}, {Eisenstein}, {Ellsworth}, {Elsworth}, {Epstein}, {Eracleous},
  {Escoffier}, {Esposito}, {Evans}, {Fan}, {Fern{\'a}ndez-Alvar}, {Feuillet},
  {Filiz Ak}, {Finley}, {Finoguenov}, {Flaherty}, {Fleming}, {Font-Ribera},
  {Foster}, {Frinchaboy}, {Galbraith-Frew}, {Garc{\'\i}a},
  {Garc{\'\i}a-Hern{\'a}ndez}, {Garc{\'\i}a P{\'e}rez}, {Gaulme}, {Ge},
  {G{\'e}nova-Santos}, {Georgakakis}, {Ghezzi}, {Gillespie}, {Girardi},
  {Goddard}, {Gontcho}, {Gonz{\'a}lez Hern{\'a}ndez}, {Grebel}, {Green},
  {Grieb}, {Grieves}, {Gunn}, {Guo}, {Harding}, {Hasselquist}, {Hawley},
  {Hayden}, {Hearty}, {Hekker}, {Ho}, {Hogg}, {Holley-Bockelmann}, {Holtzman},
  {Honscheid}, {Huber}, {Huehnerhoff}, {Ivans}, {Jiang}, {Johnson},
  {Kinemuchi}, {Kirkby}, {Kitaura}, {Klaene}, {Knapp}, {Kneib}, {Koenig},
  {Lam}, {Lan}, {Lang}, {Laurent}, {Le Goff}, {Leauthaud}, {Lee}, {Lee},
  {Licquia}, {Liu}, {Long}, {L{\'o}pez-Corredoira}, {Lorenzo-Oliveira},
  {Lucatello}, {Lundgren}, {Lupton}, {Mack}, {Mahadevan}, {Maia}, {Majewski},
  {Malanushenko}, {Malanushenko}, {Manchado}, {Manera}, {Mao}, {Maraston},
  {Marchwinski}, {Margala}, {Martell}, {Martig}, {Masters}, {Mathur},
  {McBride}, {McGehee}, {McGreer}, {McMahon}, {M{\'e}nard}, {Menzel},
  {Merloni}, {M{\'e}sz{\'a}ros}, {Miller}, {Miralda-Escud{\'e}}, {Miyatake},
  {Montero-Dorta}, {More}, {Morganson}, {Morice-Atkinson}, {Morrison},
  {Mosser}, {Muna}, {Myers}, {Nandra}, {Newman}, {Neyrinck}, {Nguyen},
  {Nichol}, {Nidever}, {Noterdaeme}, {Nuza}, {O'Connell}, {O'Connell},
  {O'Connell}, {Ogando}, {Olmstead}, {Oravetz}, {Oravetz}, {Osumi}, {Owen},
  {Padgett}, {Padmanabhan}, {Paegert}, {Palanque-Delabrouille}, {Pan},
  {Parejko}, {P{\^a}ris}, {Park}, {Pattarakijwanich}, {Pellejero-Ibanez},
  {Pepper}, {Percival}, {P{\'e}rez-Fournon}, {P{\'e}rez-R{\`a}fols},
  {Petitjean}, {Pieri}, {Pinsonneault}, {Porto de Mello}, {Prada}, {Prakash},
  {Price-Whelan}, {Protopapas}, {Raddick}, {Rahman}, {Reid}, {Rich}, {Rix},
  {Robin}, {Rockosi}, {Rodrigues}, {Rodr{\'\i}guez-Torres}, {Roe}, {Ross},
  {Ross}, {Rossi}, {Ruan}, {Rubi{\~n}o-Mart{\'\i}n}, {Rykoff},
  {Salazar-Albornoz}, {Salvato}, {Samushia}, {S{\'a}nchez}, {Santiago},
  {Sayres}, {Schiavon}, {Schlegel}, {Schmidt}, {Schneider}, {Schultheis},
  {Schwope}, {Sc{\'o}ccola}, {Scott}, {Sellgren}, {Seo}, {Serenelli}, {Shane},
  {Shen}, {Shetrone}, {Shu}, {Silva Aguirre}, {Sivarani}, {Skrutskie},
  {Slosar}, {Smith}, {Sobreira}, {Souto}, {Stassun}, {Steinmetz}, {Stello},
  {Strauss}, {Streblyanska}, {Suzuki}, {Swanson}, {Tan}, {Tayar}, {Terrien},
  {Thakar}, {Thomas}, {Thomas}, {Thompson}, {Tinker}, {Tojeiro}, {Troup},
  {Vargas-Maga{\~n}a}, {Vazquez}, {Verde}, {Viel}, {Vogt}, {Wake}, {Wang},
  {Weaver}, {Weinberg}, {Weiner}, {White}, {Wilson}, {Wisniewski},
  {Wood-Vasey}, {Ye`che}, {York}, {Zakamska}, {Zamora}, {Zasowski}, {Zehavi},
  {Zhao}, {Zheng}, {Zhou}, {Zhou}, {Zou}, \& {Zhu}}]{Alam+15}
{Alam}, S., {Albareti}, F.~D., {Allende Prieto}, C., {et~al.} 2015, \apjs, 219,
  12, \dodoi{10.1088/0067-0049/219/1/12}

\bibitem[{{Anand} {et~al.}(2024){Anand}, {Barnes}, {Yang}, {Kasliwal},
  {Coughlin}, {Sollerman}, {De}, {Fremling}, {Corsi}, {Ho}, {Balasubramanian},
  {Omand}, {Srinivasaragavan}, {Cenko}, {Ahumada}, {Andreoni}, {Dahiwale},
  {Das}, {Jencson}, {Karambelkar}, {Kumar}, {Metzger}, {Perley}, {Sarin},
  {Schweyer}, {Schulze}, {Sharma}, {Sit}, {Stein}, {Tartaglia}, {Tinyanont},
  {Tzanidakis}, {van Roestel}, {Yao}, {Bloom}, {Cook}, {Dekany}, {Graham},
  {Groom}, {Kaplan}, {Masci}, {Medford}, {Riddle}, \& {Zhang}}]{Anand+24}
{Anand}, S., {Barnes}, J., {Yang}, S., {et~al.} 2024, \apj, 962, 68,
  \dodoi{10.3847/1538-4357/ad11df}

\bibitem[{{Arcavi} {et~al.}(2017){Arcavi}, {Hosseinzadeh}, {Howell}, {McCully},
  {Poznanski}, {Kasen}, {Barnes}, {Zaltzman}, {Vasylyev}, {Maoz}, \&
  {Valenti}}]{Arcavi+17}
{Arcavi}, I., {Hosseinzadeh}, G., {Howell}, D.~A., {et~al.} 2017, \nat, 551,
  64, \dodoi{10.1038/nature24291}

\bibitem[{{Arcones} {et~al.}(2007){Arcones}, {Janka}, \& {Scheck}}]{ARcones+07}
{Arcones}, A., {Janka}, H.~T., \& {Scheck}, L. 2007, \aap, 467, 1227,
  \dodoi{10.1051/0004-6361:20066983}

\bibitem[{{Arnett}(1982)}]{Arnett82}
{Arnett}, W.~D. 1982, \apj, 253, 785, \dodoi{10.1086/159681}

\bibitem[{{Astropy Collaboration} {et~al.}(2013){Astropy Collaboration},
  {Robitaille}, {Tollerud}, {Greenfield}, {Droettboom}, {Bray}, {Aldcroft},
  {Davis}, {Ginsburg}, {Price-Whelan}, {Kerzendorf}, {Conley}, {Crighton},
  {Barbary}, {Muna}, {Ferguson}, {Grollier}, {Parikh}, {Nair}, {Unther},
  {Deil}, {Woillez}, {Conseil}, {Kramer}, {Turner}, {Singer}, {Fox}, {Weaver},
  {Zabalza}, {Edwards}, {Azalee Bostroem}, {Burke}, {Casey}, {Crawford},
  {Dencheva}, {Ely}, {Jenness}, {Labrie}, {Lim}, {Pierfederici}, {Pontzen},
  {Ptak}, {Refsdal}, {Servillat}, \& {Streicher}}]{Astropy2013}
{Astropy Collaboration}, {Robitaille}, T.~P., {Tollerud}, E.~J., {et~al.} 2013,
  \aap, 558, A33, \dodoi{10.1051/0004-6361/201322068}

\bibitem[{{Astropy Collaboration} {et~al.}(2018){Astropy Collaboration},
  {Price-Whelan}, {Sip{\H{o}}cz}, {G{\"u}nther}, {Lim}, {Crawford}, {Conseil},
  {Shupe}, {Craig}, {Dencheva}, {Ginsburg}, {VanderPlas}, {Bradley},
  {P{\'e}rez-Su{\'a}rez}, {de Val-Borro}, {Aldcroft}, {Cruz}, {Robitaille},
  {Tollerud}, {Ardelean}, {Babej}, {Bach}, {Bachetti}, {Bakanov}, {Bamford},
  {Barentsen}, {Barmby}, {Baumbach}, {Berry}, {Biscani}, {Boquien}, {Bostroem},
  {Bouma}, {Brammer}, {Bray}, {Breytenbach}, {Buddelmeijer}, {Burke},
  {Calderone}, {Cano Rodr{\'\i}guez}, {Cara}, {Cardoso}, {Cheedella}, {Copin},
  {Corrales}, {Crichton}, {D'Avella}, {Deil}, {Depagne}, {Dietrich}, {Donath},
  {Droettboom}, {Earl}, {Erben}, {Fabbro}, {Ferreira}, {Finethy}, {Fox},
  {Garrison}, {Gibbons}, {Goldstein}, {Gommers}, {Greco}, {Greenfield},
  {Groener}, {Grollier}, {Hagen}, {Hirst}, {Homeier}, {Horton}, {Hosseinzadeh},
  {Hu}, {Hunkeler}, {Ivezi{\'c}}, {Jain}, {Jenness}, {Kanarek}, {Kendrew},
  {Kern}, {Kerzendorf}, {Khvalko}, {King}, {Kirkby}, {Kulkarni}, {Kumar},
  {Lee}, {Lenz}, {Littlefair}, {Ma}, {Macleod}, {Mastropietro}, {McCully},
  {Montagnac}, {Morris}, {Mueller}, {Mumford}, {Muna}, {Murphy}, {Nelson},
  {Nguyen}, {Ninan}, {N{\"o}the}, {Ogaz}, {Oh}, {Parejko}, {Parley}, {Pascual},
  {Patil}, {Patil}, {Plunkett}, {Prochaska}, {Rastogi}, {Reddy Janga},
  {Sabater}, {Sakurikar}, {Seifert}, {Sherbert}, {Sherwood-Taylor}, {Shih},
  {Sick}, {Silbiger}, {Singanamalla}, {Singer}, {Sladen}, {Sooley},
  {Sornarajah}, {Streicher}, {Teuben}, {Thomas}, {Tremblay}, {Turner},
  {Terr{\'o}n}, {van Kerkwijk}, {de la Vega}, {Watkins}, {Weaver}, {Whitmore},
  {Woillez}, {Zabalza}, \& {Astropy Contributors}}]{Astropy2018}
{Astropy Collaboration}, {Price-Whelan}, A.~M., {Sip{\H{o}}cz}, B.~M., {et~al.}
  2018, \aj, 156, 123, \dodoi{10.3847/1538-3881/aabc4f}

\bibitem[{{Astropy Collaboration} {et~al.}(2022){Astropy Collaboration},
  {Price-Whelan}, {Lim}, {Earl}, {Starkman}, {Bradley}, {Shupe}, {Patil},
  {Corrales}, {Brasseur}, {N{\"o}the}, {Donath}, {Tollerud}, {Morris},
  {Ginsburg}, {Vaher}, {Weaver}, {Tocknell}, {Jamieson}, {van Kerkwijk},
  {Robitaille}, {Merry}, {Bachetti}, {G{\"u}nther}, {Aldcroft},
  {Alvarado-Montes}, {Archibald}, {B{\'o}di}, {Bapat}, {Barentsen},
  {Baz{\'a}n}, {Biswas}, {Boquien}, {Burke}, {Cara}, {Cara}, {Conroy},
  {Conseil}, {Craig}, {Cross}, {Cruz}, {D'Eugenio}, {Dencheva}, {Devillepoix},
  {Dietrich}, {Eigenbrot}, {Erben}, {Ferreira}, {Foreman-Mackey}, {Fox},
  {Freij}, {Garg}, {Geda}, {Glattly}, {Gondhalekar}, {Gordon}, {Grant},
  {Greenfield}, {Groener}, {Guest}, {Gurovich}, {Handberg}, {Hart},
  {Hatfield-Dodds}, {Homeier}, {Hosseinzadeh}, {Jenness}, {Jones}, {Joseph},
  {Kalmbach}, {Karamehmetoglu}, {Ka{\l}uszy{\'n}ski}, {Kelley}, {Kern},
  {Kerzendorf}, {Koch}, {Kulumani}, {Lee}, {Ly}, {Ma}, {MacBride}, {Maljaars},
  {Muna}, {Murphy}, {Norman}, {O'Steen}, {Oman}, {Pacifici}, {Pascual},
  {Pascual-Granado}, {Patil}, {Perren}, {Pickering}, {Rastogi}, {Roulston},
  {Ryan}, {Rykoff}, {Sabater}, {Sakurikar}, {Salgado}, {Sanghi}, {Saunders},
  {Savchenko}, {Schwardt}, {Seifert-Eckert}, {Shih}, {Jain}, {Shukla}, {Sick},
  {Simpson}, {Singanamalla}, {Singer}, {Singhal}, {Sinha}, {Sip{\H{o}}cz},
  {Spitler}, {Stansby}, {Streicher}, {{\v{S}}umak}, {Swinbank}, {Taranu},
  {Tewary}, {Tremblay}, {de Val-Borro}, {Van Kooten}, {Vasovi{\'c}}, {Verma},
  {de Miranda Cardoso}, {Williams}, {Wilson}, {Winkel}, {Wood-Vasey}, {Xue},
  {Yoachim}, {Zhang}, {Zonca}, \& {Astropy Project Contributors}}]{astropy22}
{Astropy Collaboration}, {Price-Whelan}, A.~M., {Lim}, P.~L., {et~al.} 2022,
  \apj, 935, 167, \dodoi{10.3847/1538-4357/ac7c74}

\bibitem[{{Barnes} \& {Duffell}(2023)}]{BarnesDuffell23}
{Barnes}, J., \& {Duffell}, P.~C. 2023, arXiv e-prints, arXiv:2305.00056,
  \dodoi{10.48550/arXiv.2305.00056}

\bibitem[{{Barnes} \& {Metzger}(2022)}]{BarnesMetzger22}
{Barnes}, J., \& {Metzger}, B.~D. 2022, \apjl, 939, L29,
  \dodoi{10.3847/2041-8213/ac9b41}

\bibitem[{{Battistini} \& {Bensby}(2016)}]{BattistiniBensby16}
{Battistini}, C., \& {Bensby}, T. 2016, \aap, 586, A49,
  \dodoi{10.1051/0004-6361/201527385}

\bibitem[{{Bauswein} {et~al.}(2014){Bauswein}, {Ardevol Pulpillo}, {Janka}, \&
  {Goriely}}]{Bauswein+14}
{Bauswein}, A., {Ardevol Pulpillo}, R., {Janka}, H.~T., \& {Goriely}, S. 2014,
  \apjl, 795, L9, \dodoi{10.1088/2041-8205/795/1/L9}

\bibitem[{{Becker}(2015)}]{becker15}
{Becker}, A. 2015, {HOTPANTS: High Order Transform of PSF ANd Template
  Subtraction}.
\newblock \doeprint{1504.004}

\bibitem[{{Belczynski} {et~al.}(2002){Belczynski}, {Kalogera}, \&
  {Bulik}}]{Belczynski+02}
{Belczynski}, K., {Kalogera}, V., \& {Bulik}, T. 2002, \apj, 572, 407,
  \dodoi{10.1086/340304}

\bibitem[{{Bennett} {et~al.}(2014){Bennett}, {Larson}, {Weiland}, \&
  {Hinshaw}}]{Bennett+14}
{Bennett}, C.~L., {Larson}, D., {Weiland}, J.~L., \& {Hinshaw}, G. 2014, \apj,
  794, 135, \dodoi{10.1088/0004-637X/794/2/135}

\bibitem[{{Bertin} \& {Arnouts}(1996)}]{Bertin1996}
{Bertin}, E., \& {Arnouts}, S. 1996, \aaps, 117, 393,
  \dodoi{10.1051/aas:1996164}

\bibitem[{{Blanchard} {et~al.}(2016){Blanchard}, {Berger}, \&
  {Fong}}]{Blanchard+16}
{Blanchard}, P.~K., {Berger}, E., \& {Fong}, W.-f. 2016, \apj, 817, 144,
  \dodoi{10.3847/0004-637X/817/2/144}

\bibitem[{{Blanchard} {et~al.}(2023){Blanchard}, {Villar}, {Chornock},
  {Laskar}, {Li}, {Leja}, {Pierel}, {Berger}, {Margutti}, {Alexander},
  {Barnes}, {Cendes}, {Eftekhari}, {Kasen}, {LeBaron}, {Metzger}, {Muzerolle
  Page}, {Rest}, {Sears}, {Siegel}, \& {Karthik Yadavalli}}]{Blanchard+23}
{Blanchard}, P.~K., {Villar}, V.~A., {Chornock}, R., {et~al.} 2023, arXiv
  e-prints, arXiv:2308.14197, \dodoi{10.48550/arXiv.2308.14197}

\bibitem[{{Bohlin}(2016)}]{Bohlin16}
{Bohlin}, R.~C. 2016, \aj, 152, 60, \dodoi{10.3847/0004-6256/152/3/60}

\bibitem[{{Bufano} {et~al.}(2012){Bufano}, {Pian}, {Sollerman}, {Benetti},
  {Pignata}, {Valenti}, {Covino}, {D'Avanzo}, {Malesani}, {Cappellaro}, {Della
  Valle}, {Fynbo}, {Hjorth}, {Mazzali}, {Reichart}, {Starling}, {Turatto},
  {Vergani}, {Wiersema}, {Amati}, {Bersier}, {Campana}, {Cano},
  {Castro-Tirado}, {Chincarini}, {D'Elia}, {de Ugarte Postigo}, {Deng},
  {Ferrero}, {Filippenko}, {Goldoni}, {Gorosabel}, {Greiner}, {Hammer},
  {Jakobsson}, {Kaper}, {Kawabata}, {Klose}, {Levan}, {Maeda}, {Masetti},
  {Milvang-Jensen}, {Mirabel}, {M{\o}ller}, {Nomoto}, {Palazzi}, {Piranomonte},
  {Salvaterra}, {Stratta}, {Tagliaferri}, {Tanaka}, {Tanvir}, \&
  {Wijers}}]{Bufano+12}
{Bufano}, F., {Pian}, E., {Sollerman}, J., {et~al.} 2012, \apj, 753, 67,
  \dodoi{10.1088/0004-637X/753/1/67}

\bibitem[{{Burbidge} {et~al.}(1957){Burbidge}, {Burbidge}, {Fowler}, \&
  {Hoyle}}]{Burbidge+57}
{Burbidge}, E.~M., {Burbidge}, G.~R., {Fowler}, W.~A., \& {Hoyle}, F. 1957,
  Reviews of Modern Physics, 29, 547, \dodoi{10.1103/RevModPhys.29.547}

\bibitem[{{Burns} {et~al.}(2023){Burns}, {Svinkin}, {Fenimore}, {Kann},
  {Ag{\"u}{\'\i} Fern{\'a}ndez}, {Frederiks}, {Hamburg}, {Lesage}, {Temiraev},
  {Tsvetkova}, {Bissaldi}, {Briggs}, {Dalessi}, {Dunwoody}, {Fletcher},
  {Goldstein}, {Hui}, {Hristov}, {Kocevski}, {Lysenko}, {Mailyan}, {Mangan},
  {McBreen}, {Racusin}, {Ridnaia}, {Roberts}, {Ulanov}, {Veres},
  {Wilson-Hodge}, \& {Wood}}]{Burns+23}
{Burns}, E., {Svinkin}, D., {Fenimore}, E., {et~al.} 2023, \apjl, 946, L31,
  \dodoi{10.3847/2041-8213/acc39c}

\bibitem[{{Cameron}(1957)}]{Cameron57}
{Cameron}, A.~G.~W. 1957, \aj, 62, 9, \dodoi{10.1086/107435}

\bibitem[{{Cameron}(2003)}]{Cameron+03}
---. 2003, \apj, 587, 327, \dodoi{10.1086/368110}

\bibitem[{{Cano} {et~al.}(2017{\natexlab{a}}){Cano}, {Wang}, {Dai}, \&
  {Wu}}]{Cano+17_review}
{Cano}, Z., {Wang}, S.-Q., {Dai}, Z.-G., \& {Wu}, X.-F. 2017{\natexlab{a}},
  Advances in Astronomy, 2017, 8929054, \dodoi{10.1155/2017/8929054}

\bibitem[{{Cano} {et~al.}(2011){Cano}, {Bersier}, {Guidorzi}, {Kobayashi},
  {Levan}, {Tanvir}, {Wiersema}, {D'Avanzo}, {Fruchter}, {Garnavich}, {Gomboc},
  {Gorosabel}, {Kasen}, {Kopa{\v{c}}}, {Margutti}, {Mazzali}, {Melandri},
  {Mundell}, {Nugent}, {Pian}, {Smith}, {Steele}, {Wijers}, \&
  {Woosley}}]{Cano+11}
{Cano}, Z., {Bersier}, D., {Guidorzi}, C., {et~al.} 2011, \apj, 740, 41,
  \dodoi{10.1088/0004-637X/740/1/41}

\bibitem[{{Cano} {et~al.}(2017{\natexlab{b}}){Cano}, {Izzo}, {de Ugarte
  Postigo}, {Th{\"o}ne}, {Kr{\"u}hler}, {Heintz}, {Malesani}, {Geier},
  {Fuentes}, {Chen}, {Covino}, {D'Elia}, {Fynbo}, {Goldoni}, {Gomboc},
  {Hjorth}, {Jakobsson}, {Kann}, {Milvang-Jensen}, {Pugliese},
  {S{\'a}nchez-Ram{\'\i}rez}, {Schulze}, {Sollerman}, {Tanvir}, \&
  {Wiersema}}]{Cano+17}
{Cano}, Z., {Izzo}, L., {de Ugarte Postigo}, A., {et~al.} 2017{\natexlab{b}},
  \aap, 605, A107, \dodoi{10.1051/0004-6361/201731005}

\bibitem[{{Cardelli} {et~al.}(1989){Cardelli}, {Clayton}, \& {Mathis}}]{CCM+89}
{Cardelli}, J.~A., {Clayton}, G.~C., \& {Mathis}, J.~S. 1989, \apj, 345, 245,
  \dodoi{10.1086/167900}

\bibitem[{{Chand} {et~al.}(2020){Chand}, {Banerjee}, {Gupta}, {Dimple}, {Pal},
  {Joshi}, {Zhang}, {Basak}, {Tam}, {Sharma}, {Pandey}, {Kumar}, \&
  {Yang}}]{Chand+20}
{Chand}, V., {Banerjee}, A., {Gupta}, R., {et~al.} 2020, \apj, 898, 42,
  \dodoi{10.3847/1538-4357/ab9606}

\bibitem[{{Chornock} {et~al.}(2010){Chornock}, {Berger}, {Levesque},
  {Soderberg}, {Foley}, {Fox}, {Frebel}, {Simon}, {Bochanski}, {Challis},
  {Kirshner}, {Podsiadlowski}, {Roth}, {Rutledge}, {Schmidt}, {Sheppard}, \&
  {Simcoe}}]{Chornock+10}
{Chornock}, R., {Berger}, E., {Levesque}, E.~M., {et~al.} 2010, arXiv e-prints,
  arXiv:1004.2262, \dodoi{10.48550/arXiv.1004.2262}

\bibitem[{C{\^o}t{\'e} {et~al.}(2017)C{\^o}t{\'e}, Belczynski, Fryer, \& {et
  al.}}]{Cote+17}
C{\^o}t{\'e}, B., Belczynski, K., Fryer, C.~L., \& {et al.} 2017, {\apj}, 836,
  230, \dodoi{10.3847/1538-4357/aa5c8d}

\bibitem[{{Coulter} {et~al.}(2017){Coulter}, {Foley}, {Kilpatrick}, {Drout},
  {Piro}, {Shappee}, {Siebert}, {Simon}, {Ulloa}, {Kasen}, {Madore},
  {Murguia-Berthier}, {Pan}, {Prochaska}, {Ramirez-Ruiz}, {Rest}, \&
  {Rojas-Bravo}}]{Coulter+17}
{Coulter}, D.~A., {Foley}, R.~J., {Kilpatrick}, C.~D., {et~al.} 2017, Science,
  358, 1556, \dodoi{10.1126/science.aap9811}

\bibitem[{{Cowan} {et~al.}(2021){Cowan}, {Sneden}, {Lawler}, {Aprahamian},
  {Wiescher}, {Langanke}, {Mart{\'\i}nez-Pinedo}, \& {Thielemann}}]{Cowan+21}
{Cowan}, J.~J., {Sneden}, C., {Lawler}, J.~E., {et~al.} 2021, Reviews of Modern
  Physics, 93, 015002, \dodoi{10.1103/RevModPhys.93.015002}

\bibitem[{{Cowan} {et~al.}(1991){Cowan}, {Thielemann}, \& {Truran}}]{Cowan+91}
{Cowan}, J.~J., {Thielemann}, F.-K., \& {Truran}, J.~W. 1991, \physrep, 208,
  267, \dodoi{10.1016/0370-1573(91)90070-3}

\bibitem[{{Dainotti} {et~al.}(2022){Dainotti}, {De Simone}, {Islam},
  {Kawaguchi}, {Moriya}, {Takiwaki}, {Tominaga}, \&
  {Gangopadhyay}}]{Dainotti+22}
{Dainotti}, M.~G., {De Simone}, B., {Islam}, K.~M., {et~al.} 2022, \apj, 938,
  41, \dodoi{10.3847/1538-4357/ac8b77}

\bibitem[{{De} \& {Siegel}(2021)}]{De&Siegel21}
{De}, S., \& {Siegel}, D.~M. 2021, \apj, 921, 94,
  \dodoi{10.3847/1538-4357/ac110b}

\bibitem[{{De Pasquale} {et~al.}(2016){De Pasquale}, {Page}, {Kann}, {Oates},
  {Schulze}, {Zhang}, {Cano}, {Gendre}, {Malesani}, {Rossi}, {Troja}, {Piro},
  {Bo{\"e}r}, {Stratta}, \& {Gehrels}}]{dePasquale+16}
{De Pasquale}, M., {Page}, M.~J., {Kann}, D.~A., {et~al.} 2016, \mnras, 462,
  1111, \dodoi{10.1093/mnras/stw1704}

\bibitem[{{Dichiara} {et~al.}(2019){Dichiara}, {Bernardini}, {Burrows},
  {D'Avanzo}, {Gronwall}, {Gropp}, {Kennea}, {Klingler}, {Krimm}, {Kuin},
  {LaPorte}, {Melandri}, {Page}, {Palmer}, {Siegel}, {Simpson}, {Tohuvavohu},
  \& {Neil Gehrels Swift Observatory Team}}]{dichiara_190829a}
{Dichiara}, S., {Bernardini}, M.~G., {Burrows}, D.~N., {et~al.} 2019, GRB
  Coordinates Network, 25552, 1

\bibitem[{{Dichiara} {et~al.}(2022){Dichiara}, {Troja}, {Lipunov}, {Ricci},
  {Oates}, {Butler}, {Liuzzo}, {Ryan}, {O'Connor}, {Cenko}, {Cosentino},
  {Lien}, {Gorbovskoy}, {Tyurina}, {Balanutsa}, {Vlasenko}, {Gorbunov},
  {Podesta}, {Podesta}, {Rebolo}, {Serra}, \& {Buckley}}]{Dichiara+22}
{Dichiara}, S., {Troja}, E., {Lipunov}, V., {et~al.} 2022, \mnras, 512, 2337,
  \dodoi{10.1093/mnras/stac454}

\bibitem[{{Dominik} {et~al.}(2012){Dominik}, {Belczynski}, {Fryer}, {Holz},
  {Berti}, {Bulik}, {Mandel}, \& {O'Shaughnessy}}]{Dominik+12}
{Dominik}, M., {Belczynski}, K., {Fryer}, C., {et~al.} 2012, \apj, 759, 52,
  \dodoi{10.1088/0004-637X/759/1/52}

\bibitem[{{Drout} {et~al.}(2014){Drout}, {Chornock}, {Soderberg}, {Sanders},
  {McKinnon}, {Rest}, {Foley}, {Milisavljevic}, {Margutti}, {Berger},
  {Calkins}, {Fong}, {Gezari}, {Huber}, {Kankare}, {Kirshner}, {Leibler},
  {Lunnan}, {Mattila}, {Marion}, {Narayan}, {Riess}, {Roth}, {Scolnic},
  {Smartt}, {Tonry}, {Burgett}, {Chambers}, {Hodapp}, {Jedicke}, {Kaiser},
  {Magnier}, {Metcalfe}, {Morgan}, {Price}, \& {Waters}}]{Drout+14}
{Drout}, M.~R., {Chornock}, R., {Soderberg}, A.~M., {et~al.} 2014, \apj, 794,
  23, \dodoi{10.1088/0004-637X/794/1/23}

\bibitem[{{Evans} {et~al.}(2014){Evans}, {Willingale}, {Osborne}, {O'Brien},
  {Tanvir}, {Frederiks}, {Pal'shin}, {Svinkin}, {Lien}, {Cummings}, {Xiong},
  {Zhang}, {G{\"o}tz}, {Savchenko}, {Negoro}, {Nakahira}, {Suzuki}, {Wiersema},
  {Starling}, {Castro-Tirado}, {Beardmore}, {S{\'a}nchez-Ram{\'\i}rez},
  {Gorosabel}, {Jeong}, {Kennea}, {Burrows}, \& {Gehrels}}]{Evans+14}
{Evans}, P.~A., {Willingale}, R., {Osborne}, J.~P., {et~al.} 2014, \mnras, 444,
  250, \dodoi{10.1093/mnras/stu1459}

\bibitem[{{Fermi GBM Team}(2019)}]{fermi_190829a}
{Fermi GBM Team}. 2019, GRB Coordinates Network, 25551, 1

\bibitem[{{Fischer} {et~al.}(2010){Fischer}, {Whitehouse}, {Mezzacappa},
  {Thielemann}, \& {Liebend{\"o}rfer}}]{Fischer+10}
{Fischer}, T., {Whitehouse}, S.~C., {Mezzacappa}, A., {Thielemann}, F.~K., \&
  {Liebend{\"o}rfer}, M. 2010, \aap, 517, A80,
  \dodoi{10.1051/0004-6361/200913106}

\bibitem[{{Fong} {et~al.}(2022){Fong}, {Nugent}, {Dong}, {Berger}, {Paterson},
  {Chornock}, {Levan}, {Blanchard}, {Alexander}, {Andrews}, {Cobb},
  {Cucchiara}, {Fox}, {Fryer}, {Gordon}, {Kilpatrick}, {Lunnan}, {Margutti},
  {Miller}, {Milne}, {Nicholl}, {Perley}, {Rastinejad}, {Escorial},
  {Schroeder}, {Smith}, {Tanvir}, \& {Terreran}}]{Fong+22}
{Fong}, W.-f., {Nugent}, A.~E., {Dong}, Y., {et~al.} 2022, \apj, 940, 56,
  \dodoi{10.3847/1538-4357/ac91d0}

\bibitem[{{Frebel}(2018)}]{Frebel+18}
{Frebel}, A. 2018, Annual Review of Nuclear and Particle Science, 68, 237,
  \dodoi{10.1146/annurev-nucl-101917-021141}

\bibitem[{{Fujibayashi} {et~al.}(2022){Fujibayashi}, {Sekiguchi}, {Shibata}, \&
  {Wanajo}}]{Fujibayashi+22}
{Fujibayashi}, S., {Sekiguchi}, Y., {Shibata}, M., \& {Wanajo}, S. 2022, arXiv
  e-prints, arXiv:2212.03958, \dodoi{10.48550/arXiv.2212.03958}

\bibitem[{{Gehrels} {et~al.}(2004){Gehrels}, {Chincarini}, {Giommi}, {Mason},
  {Nousek}, {Wells}, {White}, {Barthelmy}, {Burrows}, {Cominsky}, {Hurley},
  {Marshall}, {M{\'e}sz{\'a}ros}, {Roming}, {Angelini}, {Barbier}, {Belloni},
  {Campana}, {Caraveo}, {Chester}, {Citterio}, {Cline}, {Cropper}, {Cummings},
  {Dean}, {Feigelson}, {Fenimore}, {Frail}, {Fruchter}, {Garmire}, {Gendreau},
  {Ghisellini}, {Greiner}, {Hill}, {Hunsberger}, {Krimm}, {Kulkarni}, {Kumar},
  {Lebrun}, {Lloyd-Ronning}, {Markwardt}, {Mattson}, {Mushotzky}, {Norris},
  {Osborne}, {Paczynski}, {Palmer}, {Park}, {Parsons}, {Paul}, {Rees},
  {Reynolds}, {Rhoads}, {Sasseen}, {Schaefer}, {Short}, {Smale}, {Smith},
  {Stella}, {Tagliaferri}, {Takahashi}, {Tashiro}, {Townsley}, {Tueller},
  {Turner}, {Vietri}, {Voges}, {Ward}, {Willingale}, {Zerbi}, \&
  {Zhang}}]{Gehrels+04}
{Gehrels}, N., {Chincarini}, G., {Giommi}, P., {et~al.} 2004, \apj, 611, 1005,
  \dodoi{10.1086/422091}

\bibitem[{{Ghirlanda} {et~al.}(2007){Ghirlanda}, {Nava}, {Ghisellini}, \&
  {Firmani}}]{Ghirlanda+07}
{Ghirlanda}, G., {Nava}, L., {Ghisellini}, G., \& {Firmani}, C. 2007, \aap,
  466, 127, \dodoi{10.1051/0004-6361:20077119}

\bibitem[{{Ghirlanda} \& {Salvaterra}(2022)}]{Ghirlanda+22}
{Ghirlanda}, G., \& {Salvaterra}, R. 2022, \apj, 932, 10,
  \dodoi{10.3847/1538-4357/ac6e43}

\bibitem[{{Gillanders} {et~al.}(2023){Gillanders}, {Troja}, {Fryer}, {Ristic},
  {O'Connor}, {Fontes}, {Yang}, {Domoto}, {Rahmouni}, {Tanaka}, {Fox}, \&
  {Dichiara}}]{Gillanders+23}
{Gillanders}, J.~H., {Troja}, E., {Fryer}, C.~L., {et~al.} 2023, arXiv
  e-prints, arXiv:2308.00633, \dodoi{10.48550/arXiv.2308.00633}

\bibitem[{Goldstein {et~al.}(2017)Goldstein, Veres, Burns, Briggs, Hamburg,
  Kocevski, Wilson-Hodge, Preece, Poolakkil, Roberts, \&
  et~al.}]{Goldstein+2017}
Goldstein, A., Veres, P., Burns, E., {et~al.} 2017, The Astrophysical Journal,
  848, L14, \dodoi{10.3847/2041-8213/aa8f41}

\bibitem[{{Gompertz} {et~al.}(2018){Gompertz}, {Levan}, {Tanvir}, {Hjorth},
  {Covino}, {Evans}, {Fruchter}, {Gonz{\'a}lez-Fern{\'a}ndez}, {Jin}, {Lyman},
  {Oates}, {O'Brien}, \& {Wiersema}}]{Gompertz+18}
{Gompertz}, B.~P., {Levan}, A.~J., {Tanvir}, N.~R., {et~al.} 2018, \apj, 860,
  62, \dodoi{10.3847/1538-4357/aac206}

\bibitem[{{Gonzaga} {et~al.}(2012){Gonzaga}, {Hack}, {Fruchter}, \&
  {Mack}}]{DrizzlePac+12}
{Gonzaga}, S., {Hack}, W., {Fruchter}, A., \& {Mack}, J. 2012, {The DrizzlePac
  Handbook}

\bibitem[{{Gordon} {et~al.}(2003){Gordon}, {Clayton}, {Misselt}, {Landolt}, \&
  {Wolff}}]{Gordon+03}
{Gordon}, K.~D., {Clayton}, G.~C., {Misselt}, K.~A., {Landolt}, A.~U., \&
  {Wolff}, M.~J. 2003, \apj, 594, 279, \dodoi{10.1086/376774}

\bibitem[{{Greiner} {et~al.}(2003){Greiner}, {Peimbert}, {Esteban}, {Kaufer},
  {Jaunsen}, {Smoke}, {Klose}, \& {Reimer}}]{Greiner+03}
{Greiner}, J., {Peimbert}, M., {Esteban}, C., {et~al.} 2003, GRB Coordinates
  Network, 2020, 1

\bibitem[{{Greiner} {et~al.}(2015){Greiner}, {Mazzali}, {Kann}, {Kr{\"u}hler},
  {Pian}, {Prentice}, {Olivares E.}, {Rossi}, {Klose}, {Taubenberger}, {Knust},
  {Afonso}, {Ashall}, {Bolmer}, {Delvaux}, {Diehl}, {Elliott}, {Filgas},
  {Fynbo}, {Graham}, {Guelbenzu}, {Kobayashi}, {Leloudas}, {Savaglio},
  {Schady}, {Schmidl}, {Schweyer}, {Sudilovsky}, {Tanga}, {Updike}, {van
  Eerten}, \& {Varela}}]{Greiner+15}
{Greiner}, J., {Mazzali}, P.~A., {Kann}, D.~A., {et~al.} 2015, \nat, 523, 189,
  \dodoi{10.1038/nature14579}

\bibitem[{{H.~E.~S.~S. Collaboration} {et~al.}(2021){H.~E.~S.~S.
  Collaboration}, {Abdalla}, {Aharonian}, {Ait Benkhali}, {Ang{\"u}ner},
  {Arcaro}, {Armand}, {Armstrong}, {Ashkar}, {Backes}, {Baghmanyan}, {Barbosa
  Martins}, {Barnacka}, {Barnard}, {Becherini}, {Berge}, {Bernl{\"o}hr}, {Bi},
  {Bissaldi}, {B{\"o}ttcher}, {Boisson}, {Bolmont}, {de Bony de Lavergne},
  {Breuhaus}, {Brun}, {Brun}, {Bryan}, {B{\"u}chele}, {Bulik}, {Bylund},
  {Caroff}, {Carosi}, {Casanova}, {Chand}, {Chandra}, {Chen}, {Cotter},
  {Cury{\l}o}, {Damascene Mbarubucyeye}, {Davids}, {Davies}, {Deil}, {Devin},
  {Dirson}, {Djannati-Ata{\"\i}}, {Dmytriiev}, {Donath}, {Doroshenko},
  {Dreyer}, {Duffy}, {Dyks}, {Egberts}, {Eichhorn}, {Einecke}, {Emery},
  {Ernenwein}, {Feijen}, {Fegan}, {Fiasson}, {Fichet de Clairfontaine},
  {Fontaine}, {Funk}, {F{\"u}{\ss}ling}, {Gabici}, {Gallant}, {Giavitto},
  {Giunti}, {Glawion}, {Glicenstein}, {Grondin}, {Hahn}, {Haupt}, {Hermann},
  {Hinton}, {Hofmann}, {Hoischen}, {Holch}, {Holler}, {H{\"o}rbe}, {Horns},
  {Huber}, {Jamrozy}, {Jankowsky}, {Jankowsky}, {Jardin-Blicq}, {Joshi},
  {Jung-Richardt}, {Kasai}, {Kastendieck}, {Katarzy{\'n}ski}, {Katz},
  {Khangulyan}, {Kh{\'e}lifi}, {Klepser}, {Klu{\'z}niak}, {Komin}, {Konno},
  {Kosack}, {Kostunin}, {Kreter}, {Lamanna}, {Lemi{\`e}re}, {Lemoine-Goumard},
  {Lenain}, {Leuschner}, {Levy}, {Lohse}, {Lypova}, {Mackey}, {Majumdar},
  {Malyshev}, {Malyshev}, {Marandon}, {Marchegiani}, {Marcowith}, {Mares},
  {Mart{\'\i}-Devesa}, {Marx}, {Maurin}, {Meintjes}, {Meyer}, {Mitchell},
  {Moderski}, {Mohrmann}, {Montanari}, {Moore}, {Morris}, {Moulin}, {Muller},
  {Murach}, {Nakashima}, {Nayerhoda}, {de Naurois}, {Ndiyavala}, {Niemiec},
  {Oakes}, {O'Brien}, {Odaka}, {Ohm}, {Olivera-Nieto}, {de Ona Wilhelmi},
  {Ostrowski}, {Panny}, {Panter}, {Parsons}, {Peron}, {Peyaud}, {Piel}, {Pita},
  {Poireau}, {Priyana Noel}, {Prokhorov}, {Prokoph}, {P{\"u}hlhofer}, {Punch},
  {Quirrenbach}, {Raab}, {Rauth}, {Reichherzer}, {Reimer}, {Reimer}, {Remy},
  {Renaud}, {Rieger}, {Rinchiuso}, {Romoli}, {Rowell}, {Rudak}, {Ruiz-Velasco},
  {Sahakian}, {Sailer}, {Salzmann}, {Sanchez}, {Santangelo}, {Sasaki},
  {Scalici}, {Sch{\"a}fer}, {Sch{\"u}ssler}, {Schutte}, {Schwanke},
  {Seglar-Arroyo}, {Senniappan}, {Seyffert}, {Shafi}, {Shapopi},
  {Shiningayamwe}, {Simoni}, {Sinha}, {Sol}, {Specovius}, {Spencer},
  {Spir-Jacob}, {Stawarz}, {Sun}, {Steenkamp}, {Stegmann}, {Steinmassl},
  {Steppa}, {Takahashi}, {Tam}, {Tavernier}, {Taylor}, {Terrier}, {Thiersen},
  {Tiziani}, {Tluczykont}, {Tomankova}, {Tsirou}, {Tuffs}, {Uchiyama}, {van der
  Walt}, {van Eldik}, {van Rensburg}, {van Soelen}, {Vasileiadis}, {Veh},
  {Venter}, {Vincent}, {Vink}, {V{\"o}lk}, {Wadiasingh}, {Wagner}, {Watson},
  {Werner}, {White}, {Wierzcholska}, {Wong}, {Yusafzai}, {Zacharias}, {Zanin},
  {Zargaryan}, {Zdziarski}, {Zech}, {Zhu}, {Zorn}, {Zouari}, {{\.Z}ywucka},
  {Evans}, \& {Page}}]{HESS+21}
{H.~E.~S.~S. Collaboration}, {Abdalla}, H., {Aharonian}, F., {et~al.} 2021,
  Science, 372, 1081, \dodoi{10.1126/science.abe8560}

\bibitem[{{Halevi} \& {M{\"o}sta}(2018)}]{HaleviMosta18}
{Halevi}, G., \& {M{\"o}sta}, P. 2018, \mnras, 477, 2366,
  \dodoi{10.1093/mnras/sty797}

\bibitem[{{Hansen} {et~al.}(2017){Hansen}, {Simon}, {Marshall}, {Li},
  {Carollo}, {DePoy}, {Nagasawa}, {Bernstein}, {Drlica-Wagner}, {Abdalla},
  {Allam}, {Annis}, {Bechtol}, {Benoit-L{\'e}vy}, {Brooks}, {Buckley-Geer},
  {Carnero Rosell}, {Carrasco Kind}, {Carretero}, {Cunha}, {da Costa}, {Desai},
  {Eifler}, {Fausti Neto}, {Flaugher}, {Frieman}, {Garc{\'\i}a-Bellido},
  {Gaztanaga}, {Gerdes}, {Gruen}, {Gruendl}, {Gschwend}, {Gutierrez}, {James},
  {Krause}, {Kuehn}, {Kuropatkin}, {Lahav}, {Miquel}, {Plazas}, {Romer},
  {Sanchez}, {Santiago}, {Scarpine}, {Smith}, {Soares-Santos}, {Sobreira},
  {Suchyta}, {Swanson}, {Tarle}, {Walker}, \& {DES Collaboration}}]{Hansen+17}
{Hansen}, T.~T., {Simon}, J.~D., {Marshall}, J.~L., {et~al.} 2017, \apj, 838,
  44, \dodoi{10.3847/1538-4357/aa634a}

\bibitem[{{Harris} {et~al.}(2020){Harris}, {Millman}, {van der Walt},
  {Gommers}, {Virtanen}, {Cournapeau}, {Wieser}, {Taylor}, {Berg}, {Smith},
  {Kern}, {Picus}, {Hoyer}, {van Kerkwijk}, {Brett}, {Haldane}, {del R{\'\i}o},
  {Wiebe}, {Peterson}, {G{\'e}rard-Marchant}, {Sheppard}, {Reddy}, {Weckesser},
  {Abbasi}, {Gohlke}, \& {Oliphant}}]{numpy}
{Harris}, C.~R., {Millman}, K.~J., {van der Walt}, S.~J., {et~al.} 2020, \nat,
  585, 357, \dodoi{10.1038/s41586-020-2649-2}

\bibitem[{{Hjorth} {et~al.}(2003){Hjorth}, {Sollerman}, {M{\o}ller}, {Fynbo},
  {Woosley}, {Kouveliotou}, {Tanvir}, {Greiner}, {Andersen}, {Castro-Tirado},
  {Castro Cer{\'o}n}, {Fruchter}, {Gorosabel}, {Jakobsson}, {Kaper}, {Klose},
  {Masetti}, {Pedersen}, {Pedersen}, {Pian}, {Palazzi}, {Rhoads}, {Rol}, {van
  den Heuvel}, {Vreeswijk}, {Watson}, \& {Wijers}}]{Hjorth+03}
{Hjorth}, J., {Sollerman}, J., {M{\o}ller}, P., {et~al.} 2003, \nat, 423, 847,
  \dodoi{10.1038/nature01750}

\bibitem[{{Ho} {et~al.}(2020){Ho}, {Kulkarni}, {Perley}, {Cenko}, {Corsi},
  {Schulze}, {Lunnan}, {Sollerman}, {Gal-Yam}, {Anand}, {Barbarino}, {Bellm},
  {Bruch}, {Burns}, {De}, {Dekany}, {Delacroix}, {Duev}, {Frederiks},
  {Fremling}, {Goldstein}, {Golkhou}, {Graham}, {Hale}, {Kasliwal}, {Kupfer},
  {Laher}, {Martikainen}, {Masci}, {Neill}, {Ridnaia}, {Rusholme}, {Savchenko},
  {Shupe}, {Soumagnac}, {Strotjohann}, {Svinkin}, {Taggart}, {Tartaglia},
  {Yan}, \& {Zolkower}}]{Ho+20}
{Ho}, A. Y.~Q., {Kulkarni}, S.~R., {Perley}, D.~A., {et~al.} 2020, \apj, 902,
  86, \dodoi{10.3847/1538-4357/aba630}

\bibitem[{{Hoffman} {et~al.}(1997){Hoffman}, {Woosley}, \&
  {Qian}}]{HoffmanWoosley97}
{Hoffman}, R.~D., {Woosley}, S.~E., \& {Qian}, Y.~Z. 1997, \apj, 482, 951,
  \dodoi{10.1086/304181}

\bibitem[{{Holmbeck} \& {Andrews}(2023)}]{HolmbeckAndrews23}
{Holmbeck}, E.~M., \& {Andrews}, J.~J. 2023, arXiv e-prints, arXiv:2310.03847,
  \dodoi{10.48550/arXiv.2310.03847}

\bibitem[{{Hosseinzadeh} {et~al.}(2023){Hosseinzadeh}, {Sand}, {Jencson},
  {Andrews}, {Shivaei}, {Bostroem}, {Valenti}, {Szalai}, {Burke}, {Howell},
  {McCully}, {Newsome}, {Gonzalez}, {Pellegrino}, \&
  {Terreran}}]{Hosseinzadeh+23}
{Hosseinzadeh}, G., {Sand}, D.~J., {Jencson}, J.~E., {et~al.} 2023, \apjl, 942,
  L18, \dodoi{10.3847/2041-8213/aca64e}

\bibitem[{{Hotokezaka} {et~al.}(2018){Hotokezaka}, {Beniamini}, \&
  {Piran}}]{Hotokezaka+18}
{Hotokezaka}, K., {Beniamini}, P., \& {Piran}, T. 2018, International Journal
  of Modern Physics D, 27, 1842005, \dodoi{10.1142/S0218271818420051}

\bibitem[{{Hotokezaka} {et~al.}(2023){Hotokezaka}, {Tanaka}, {Kato}, \&
  {Gaigalas}}]{Hotokezaka+23}
{Hotokezaka}, K., {Tanaka}, M., {Kato}, D., \& {Gaigalas}, G. 2023, arXiv
  e-prints, arXiv:2307.00988, \dodoi{10.48550/arXiv.2307.00988}

\bibitem[{{Hu} {et~al.}(2021){Hu}, {Castro-Tirado}, {Kumar}, {Gupta}, {Valeev},
  {Pandey}, {Kann}, {Castell{\'o}n}, {Agudo}, {Aryan}, {Caballero-Garc{\'\i}a},
  {Guziy}, {Martin-Carrillo}, {Oates}, {Pian}, {S{\'a}nchez-Ram{\'\i}rez},
  {Sokolov}, \& {Zhang}}]{Hu+21}
{Hu}, Y.~D., {Castro-Tirado}, A.~J., {Kumar}, A., {et~al.} 2021, \aap, 646,
  A50, \dodoi{10.1051/0004-6361/202039349}

\bibitem[{{Huang} {et~al.}(2023){Huang}, {Huang}, {Cheng}, {Ren}, {Zhang}, \&
  {Liang}}]{huang+23}
{Huang}, J.-K., {Huang}, X.-L., {Cheng}, J.-G., {et~al.} 2023, \apj, 947, 84,
  \dodoi{10.3847/1538-4357/acc85f}

\bibitem[{{Hunter}(2007)}]{Hunter+07}
{Hunter}, J.~D. 2007, Computing in Science and Engineering, 9, 90,
  \dodoi{10.1109/MCSE.2007.55}

\bibitem[{{Izzo} {et~al.}(2020){Izzo}, {Auchettl}, {Hjorth}, {De Colle},
  {Gall}, {Angus}, {Raimundo}, \& {Ramirez-Ruiz}}]{Izzo+20}
{Izzo}, L., {Auchettl}, K., {Hjorth}, J., {et~al.} 2020, \aap, 639, L11,
  \dodoi{10.1051/0004-6361/202038152}

\bibitem[{{Ji} \& {Frebel}(2018)}]{JiFrebel18}
{Ji}, A.~P., \& {Frebel}, A. 2018, \apj, 856, 138,
  \dodoi{10.3847/1538-4357/aab14a}

\bibitem[{{Ji} {et~al.}(2016){Ji}, {Frebel}, {Chiti}, \& {Simon}}]{Ji+16}
{Ji}, A.~P., {Frebel}, A., {Chiti}, A., \& {Simon}, J.~D. 2016, \nat, 531, 610,
  \dodoi{10.1038/nature17425}

\bibitem[{{Ji} {et~al.}(2023){Ji}, {Simon}, {Roederer}, {Magg}, {Frebel},
  {Johnson}, {Klessen}, {Magg}, {Cescutti}, {Mateo}, {Bergemann}, \&
  {Bailey}}]{Ji+23}
{Ji}, A.~P., {Simon}, J.~D., {Roederer}, I.~U., {et~al.} 2023, \aj, 165, 100,
  \dodoi{10.3847/1538-3881/acad84}

\bibitem[{{Jin} {et~al.}(2021){Jin}, {Yoon}, \& {Blinnikov}}]{Jin+21}
{Jin}, H., {Yoon}, S.-C., \& {Blinnikov}, S. 2021, \apj, 910, 68,
  \dodoi{10.3847/1538-4357/abe0b1}

\bibitem[{{Just} {et~al.}(2022){Just}, {Aloy}, {Obergaulinger}, \&
  {Nagataki}}]{Just+22}
{Just}, O., {Aloy}, M.~A., {Obergaulinger}, M., \& {Nagataki}, S. 2022, \apjl,
  934, L30, \dodoi{10.3847/2041-8213/ac83a1}

\bibitem[{{Kamble} {et~al.}(2009){Kamble}, {Horst}, {Bhattacharya}, {Wijers},
  {Chandra}, {Resmi}, {Rol}, {Kouveliotou}, \& {Strom}}]{Kamble+09}
{Kamble}, A., {Horst}, A.~J.~V., {Bhattacharya}, D., {et~al.} 2009, in
  Astronomical Society of the Pacific Conference Series, Vol. 407, The
  Low-Frequency Radio Universe, ed. D.~J. {Saikia}, D.~A. {Green}, Y.~{Gupta},
  \& T.~{Venturi}, 295

\bibitem[{{Kann} {et~al.}(2006){Kann}, {Klose}, \& {Zeh}}]{Kann+06}
{Kann}, D.~A., {Klose}, S., \& {Zeh}, A. 2006, \apj, 641, 993,
  \dodoi{10.1086/500652}

\bibitem[{{Kasen} {et~al.}(2017){Kasen}, {Metzger}, {Barnes}, {Quataert}, \&
  {Ramirez-Ruiz}}]{kasen+17}
{Kasen}, D., {Metzger}, B., {Barnes}, J., {Quataert}, E., \& {Ramirez-Ruiz}, E.
  2017, Nature, 551, 80, \dodoi{10.1038/nature24453}

\bibitem[{{Kashiyama} {et~al.}(2016){Kashiyama}, {Murase}, {Bartos}, {Kiuchi},
  \& {Margutti}}]{Kashiyama+16}
{Kashiyama}, K., {Murase}, K., {Bartos}, I., {Kiuchi}, K., \& {Margutti}, R.
  2016, \apj, 818, 94, \dodoi{10.3847/0004-637X/818/1/94}

\bibitem[{{Kawabata} {et~al.}(2003){Kawabata}, {Deng}, {Wang}, {Mazzali},
  {Nomoto}, {Maeda}, {Tominaga}, {Umeda}, {Iye}, {Kosugi}, {Ohyama}, {Sasaki},
  {H{\"o}flich}, {Wheeler}, {Jeffery}, {Aoki}, {Kashikawa}, {Takata}, {Kawai},
  {Sakamoto}, {Urata}, {Yoshida}, {Tamagawa}, {Torii}, {Aoki}, {Kobayashi},
  {Komiyama}, {Mizumoto}, {Noumaru}, {Ogasawara}, {Sekiguchi}, {Shirasaki},
  {Totani}, {Watanabe}, \& {Yamada}}]{Kawabata+03}
{Kawabata}, K.~S., {Deng}, J., {Wang}, L., {et~al.} 2003, \apjl, 593, L19,
  \dodoi{10.1086/378148}

\bibitem[{{Kawaguchi} {et~al.}(2020){Kawaguchi}, {Shibata}, \&
  {Tanaka}}]{Kawaguchi+20b}
{Kawaguchi}, K., {Shibata}, M., \& {Tanaka}, M. 2020, \apj, 889, 171,
  \dodoi{10.3847/1538-4357/ab61f6}

\bibitem[{{Kilpatrick} {et~al.}(2023){Kilpatrick}, {Izzo}, {Bentley},
  {Chambers}, {Coulter}, {Drout}, {de Boer}, {Foley}, {Gall}, {Halford},
  {Jones}, {Langeroodi}, {Lin}, {Magnier}, {McGill}, {O'Grady}, {Pan},
  {Ramirez-Ruiz}, {Rest}, {Swift}, {Tinyanont}, {Villar}, {Wainscoat},
  {Wasserman}, {Yadavalli}, \& {Yang}}]{Kilpatrick+23_20jfo}
{Kilpatrick}, C.~D., {Izzo}, L., {Bentley}, R.~O., {et~al.} 2023, \mnras, 524,
  2161, \dodoi{10.1093/mnras/stad1954}

\bibitem[{Kirby {et~al.}(2020)Kirby, Duggan, {Ramirez-Ruiz}, \& {et
  al.}}]{Kirby+20}
Kirby, E.~N., Duggan, G., {Ramirez-Ruiz}, E., \& {et al.} 2020, {\apjl}, 891,
  L13, \dodoi{10.3847/2041-8213/ab78a1}

\bibitem[{Kirby {et~al.}(2023)Kirby, Ji, \& Kovalev}]{Kirby+23}
Kirby, E.~N., Ji, A.~P., \& Kovalev, M. 2023, {\apj}, 958, 45,
  \dodoi{10.3847/1538-4357/acf309}

\bibitem[{{Laskar} {et~al.}(2013){Laskar}, {Berger}, {Zauderer}, {Margutti},
  {Soderberg}, {Chakraborti}, {Lunnan}, {Chornock}, {Chandra}, \&
  {Ray}}]{Laskar+13}
{Laskar}, T., {Berger}, E., {Zauderer}, B.~A., {et~al.} 2013, \apj, 776, 119,
  \dodoi{10.1088/0004-637X/776/2/119}

\bibitem[{{Lattimer} \& {Schramm}(1974)}]{LattimerSchramm74}
{Lattimer}, J.~M., \& {Schramm}, D.~N. 1974, \apjl, 192, L145,
  \dodoi{10.1086/181612}

\bibitem[{{Lee} {et~al.}(2022){Lee}, {Bartos}, {Eddins}, {Corsi}, {M{\'a}rka},
  {Privon}, \& {M{\'a}rka}}]{Lee+22}
{Lee}, K.~H., {Bartos}, I., {Eddins}, A., {et~al.} 2022, \apjl, 934, L5,
  \dodoi{10.3847/2041-8213/ac7ff0}

\bibitem[{{Levan} {et~al.}(2023{\natexlab{a}}){Levan}, {Gompertz}, {Salafia},
  {Bulla}, {Burns}, {Hotokezaka}, {Izzo}, {Lamb}, {Malesani}, {Oates},
  {Ravasio}, {Rouco Escorial}, {Schneider}, {Sarin}, {Schulze}, {Tanvir},
  {Ackley}, {Anderson}, {Brammer}, {Christensen}, {Dhillon}, {Evans},
  {Fausnaugh}, {Fong}, {Fruchter}, {Fryer}, {Fynbo}, {Gaspari}, {Heintz},
  {Hjorth}, {Kennea}, {Kennedy}, {Laskar}, {Leloudas}, {Mandel},
  {Martin-Carrillo}, {Metzger}, {Nicholl}, {Nugent}, {Palmerio}, {Pugliese},
  {Rastinejad}, {Rhodes}, {Rossi}, {Smartt}, {Stevance}, {Tohuvavohu}, {van der
  Horst}, {Vergani}, {Watson}, {Barclay}, {Bhirombhakdi}, {Breedt}, {Breeveld},
  {Brown}, {Campana}, {Chrimes}, {D'Avanzo}, {D'Elia}, {De Pasquale}, {Dyer},
  {Galloway}, {Garbutt}, {Green}, {Hartmann}, {Jakobsson}, {Kerry},
  {Langeroodi}, {Leung}, {Littlefair}, {Munday}, {O'Brien}, {Parsons},
  {Pelisoli}, {Saccardi}, {Sahman}, {Salvaterra}, {Sbarufatti}, {Steeghs},
  {Tagliaferri}, {Th{\"o}ne}, {de Ugarte Postigo}, \&
  {Kann}}]{Levan+23_230307a}
{Levan}, A., {Gompertz}, B.~P., {Salafia}, O.~S., {et~al.} 2023{\natexlab{a}},
  arXiv e-prints, arXiv:2307.02098, \dodoi{10.48550/arXiv.2307.02098}

\bibitem[{{Levan} {et~al.}(2013){Levan}, {Cenko}, {Perley}, \&
  {Tanvir}}]{Levan+13_130427a}
{Levan}, A.~J., {Cenko}, S.~B., {Perley}, D.~A., \& {Tanvir}, N.~R. 2013, GRB
  Coordinates Network, 14455, 1

\bibitem[{{Levan} {et~al.}(2014{\natexlab{a}}){Levan}, {Tanvir}, {Fruchter},
  {Hjorth}, {Pian}, {Mazzali}, {Hounsell}, {Perley}, {Cano}, {Graham}, {Cenko},
  {Fynbo}, {Kouveliotou}, {Pe'er}, {Misra}, \& {Wiersema}}]{Levan+14}
{Levan}, A.~J., {Tanvir}, N.~R., {Fruchter}, A.~S., {et~al.}
  2014{\natexlab{a}}, \apj, 792, 115, \dodoi{10.1088/0004-637X/792/2/115}

\bibitem[{{Levan} {et~al.}(2014{\natexlab{b}}){Levan}, {Tanvir}, {Starling},
  {Wiersema}, {Page}, {Perley}, {Schulze}, {Wynn}, {Chornock}, {Hjorth},
  {Cenko}, {Fruchter}, {O'Brien}, {Brown}, {Tunnicliffe}, {Malesani},
  {Jakobsson}, {Watson}, {Berger}, {Bersier}, {Cobb}, {Covino}, {Cucchiara},
  {de Ugarte Postigo}, {Fox}, {Gal-Yam}, {Goldoni}, {Gorosabel}, {Kaper},
  {Kr{\"u}hler}, {Karjalainen}, {Osborne}, {Pian}, {S{\'a}nchez-Ram{\'\i}rez},
  {Schmidt}, {Skillen}, {Tagliaferri}, {Th{\"o}ne}, {Vaduvescu}, {Wijers}, \&
  {Zauderer}}]{Levan+14_ULGRBs}
{Levan}, A.~J., {Tanvir}, N.~R., {Starling}, R.~L.~C., {et~al.}
  2014{\natexlab{b}}, \apj, 781, 13, \dodoi{10.1088/0004-637X/781/1/13}

\bibitem[{{Levan} {et~al.}(2023{\natexlab{b}}){Levan}, {Lamb}, {Schneider},
  {Hjorth}, {Zafar}, {de Ugarte Postigo}, {Sargent}, {Mullally}, {Izzo},
  {D'Avanzo}, {Burns}, {Ag{\"u}{\'\i} Fern{\'a}ndez}, {Barclay}, {Bernardini},
  {Bhirombhakdi}, {Bremer}, {Brivio}, {Campana}, {Chrimes}, {D'Elia}, {Della
  Valle}, {De Pasquale}, {Ferro}, {Fong}, {Fruchter}, {Fynbo}, {Gaspari},
  {Gompertz}, {Hartmann}, {Hedges}, {Heintz}, {Hotokezaka}, {Jakobsson},
  {Kann}, {Kennea}, {Laskar}, {Le Floc'h}, {Malesani}, {Melandri}, {Metzger},
  {Oates}, {Pian}, {Piranomonte}, {Pugliese}, {Racusin}, {Rastinejad},
  {Ravasio}, {Rossi}, {Saccardi}, {Salvaterra}, {Sbarufatti}, {Starling},
  {Tanvir}, {Th{\"o}ne}, {van der Horst}, {Vergani}, {Watson}, {Wiersema},
  {Wijers}, \& {Xu}}]{Levan+23_221009a}
{Levan}, A.~J., {Lamb}, G.~P., {Schneider}, B., {et~al.} 2023{\natexlab{b}},
  \apjl, 946, L28, \dodoi{10.3847/2041-8213/acc2c1}

\bibitem[{{Li} {et~al.}(2023){Li}, {Zhong}, \& {Dai}}]{Li+23}
{Li}, L., {Zhong}, S.-Q., \& {Dai}, Z.-G. 2023, arXiv e-prints,
  arXiv:2307.09917.
\newblock \doarXiv{2307.09917}

\bibitem[{{Lipkin} {et~al.}(2004){Lipkin}, {Ofek}, {Gal-Yam}, {Leibowitz},
  {Poznanski}, {Kaspi}, {Polishook}, {Kulkarni}, {Fox}, {Berger}, {Mirabal},
  {Halpern}, {Bureau}, {Fathi}, {Price}, {Peterson}, {Frebel}, {Schmidt},
  {Orosz}, {Fitzgerald}, {Bloom}, {van Dokkum}, {Bailyn}, {Buxton}, \&
  {Barsony}}]{Lipkin+04}
{Lipkin}, Y.~M., {Ofek}, E.~O., {Gal-Yam}, A., {et~al.} 2004, \apj, 606, 381,
  \dodoi{10.1086/383000}

\bibitem[{{Lipunov} {et~al.}(2017){Lipunov}, {Gorbovskoy}, {Kornilov},
  {.~Tyurina}, {Balanutsa}, {Kuznetsov}, {Vlasenko}, {Kuvshinov}, {Gorbunov},
  {Buckley}, {Krylov}, {Podesta}, {Lopez}, {Podesta}, {Levato}, {Saffe},
  {Mallamachi}, {Potter}, {Budnev}, {Gress}, {Ishmuhametova}, {Vladimirov},
  {Zimnukhov}, {Yurkov}, {Sergienko}, {Gabovich}, {Rebolo}, {Serra-Ricart},
  {Israelyan}, {Chazov}, {Wang}, {Tlatov}, \& {Panchenko}}]{Lipunov+17}
{Lipunov}, V.~M., {Gorbovskoy}, E., {Kornilov}, V.~G., {et~al.} 2017, \apjl,
  850, L1, \dodoi{10.3847/2041-8213/aa92c0}

\bibitem[{{Lyman} {et~al.}(2017){Lyman}, {Levan}, {Tanvir}, {Fynbo}, {McGuire},
  {Perley}, {Angus}, {Bloom}, {Conselice}, {Fruchter}, {Hjorth}, {Jakobsson},
  \& {Starling}}]{Lyman+17}
{Lyman}, J.~D., {Levan}, A.~J., {Tanvir}, N.~R., {et~al.} 2017, \mnras, 467,
  1795, \dodoi{10.1093/mnras/stx220}

\bibitem[{{MacFadyen} \& {Woosley}(1999)}]{MacFadyenWoosley99}
{MacFadyen}, A.~I., \& {Woosley}, S.~E. 1999, \apj, 524, 262,
  \dodoi{10.1086/307790}

\bibitem[{{Macias} \& {Ramirez-Ruiz}(2018)}]{MaciasRamirezRuiz18}
{Macias}, P., \& {Ramirez-Ruiz}, E. 2018, \apj, 860, 89,
  \dodoi{10.3847/1538-4357/aac3e0}

\bibitem[{{Mandel} \& {Broekgaarden}(2022)}]{MandelBroekgaarden22}
{Mandel}, I., \& {Broekgaarden}, F.~S. 2022, Living Reviews in Relativity, 25,
  1, \dodoi{10.1007/s41114-021-00034-3}

\bibitem[{{Mandhai} {et~al.}(2022){Mandhai}, {Lamb}, {Tanvir}, {Bray}, {Nixon},
  {Eyles-Ferris}, {Levan}, \& {Gompertz}}]{Mandhai+22}
{Mandhai}, S., {Lamb}, G.~P., {Tanvir}, N.~R., {et~al.} 2022, \mnras, 514,
  2716, \dodoi{10.1093/mnras/stac1473}

\bibitem[{{Margutti} {et~al.}(2013){Margutti}, {Soderberg}, {Wieringa},
  {Edwards}, {Chevalier}, {Morsony}, {Barniol Duran}, {Sironi}, {Zauderer},
  {Milisavljevic}, {Kamble}, \& {Pian}}]{Margutti+13}
{Margutti}, R., {Soderberg}, A.~M., {Wieringa}, M.~H., {et~al.} 2013, \apj,
  778, 18, \dodoi{10.1088/0004-637X/778/1/18}

\bibitem[{{Margutti} {et~al.}(2023){Margutti}, {Bright}, {Matthews},
  {Coppejans}, {Alexander}, {Berger}, {Bietenholz}, {Chornock}, {DeMarchi},
  {Drout}, {Eftekhari}, {Jacobson-Gal{\'a}n}, {Laskar}, {Milisavljevic},
  {Murase}, {Nicholl}, {Omand}, {Stroh}, {Terreran}, \&
  {VanderLey}}]{Margutti+23}
{Margutti}, R., {Bright}, J.~S., {Matthews}, D.~J., {et~al.} 2023, \apjl, 954,
  L45, \dodoi{10.3847/2041-8213/acf1fd}

\bibitem[{{Mart{\'\i}nez-Pinedo} {et~al.}(2012){Mart{\'\i}nez-Pinedo},
  {Fischer}, {Lohs}, \& {Huther}}]{MartinezPinedo+12}
{Mart{\'\i}nez-Pinedo}, G., {Fischer}, T., {Lohs}, A., \& {Huther}, L. 2012,
  \prl, 109, 251104, \dodoi{10.1103/PhysRevLett.109.251104}

\bibitem[{{Mart{\'\i}nez-Pinedo} {et~al.}(2014){Mart{\'\i}nez-Pinedo}, {Lam},
  {Langanke}, {Zegers}, \& {Sullivan}}]{MartinezPinedo+14}
{Mart{\'\i}nez-Pinedo}, G., {Lam}, Y.~H., {Langanke}, K., {Zegers}, R.~G.~T.,
  \& {Sullivan}, C. 2014, \prc, 89, 045806, \dodoi{10.1103/PhysRevC.89.045806}

\bibitem[{{Maselli} {et~al.}(2014){Maselli}, {Melandri}, {Nava}, {Mundell},
  {Kawai}, {Campana}, {Covino}, {Cummings}, {Cusumano}, {Evans}, {Ghirlanda},
  {Ghisellini}, {Guidorzi}, {Kobayashi}, {Kuin}, {La Parola}, {Mangano},
  {Oates}, {Sakamoto}, {Serino}, {Virgili}, {Zhang}, {Barthelmy}, {Beardmore},
  {Bernardini}, {Bersier}, {Burrows}, {Calderone}, {Capalbi}, {Chiang},
  {D'Avanzo}, {D'Elia}, {De Pasquale}, {Fugazza}, {Gehrels}, {Gomboc},
  {Harrison}, {Hanayama}, {Japelj}, {Kennea}, {Kopac}, {Kouveliotou}, {Kuroda},
  {Levan}, {Malesani}, {Marshall}, {Nousek}, {O'Brien}, {Osborne}, {Pagani},
  {Page}, {Page}, {Perri}, {Pritchard}, {Romano}, {Saito}, {Sbarufatti},
  {Salvaterra}, {Steele}, {Tanvir}, {Vianello}, {Weigand}, {Wiersema}, {Yatsu},
  {Yoshii}, \& {Tagliaferri}}]{Maselli+14}
{Maselli}, A., {Melandri}, A., {Nava}, L., {et~al.} 2014, Science, 343, 48,
  \dodoi{10.1126/science.1242279}

\bibitem[{{Matheson} {et~al.}(2003){Matheson}, {Garnavich}, {Stanek},
  {Bersier}, {Holland}, {Krisciunas}, {Caldwell}, {Berlind}, {Bloom}, {Bolte},
  {Bonanos}, {Brown}, {Brown}, {Calkins}, {Challis}, {Chornock}, {Echevarria},
  {Eisenstein}, {Everett}, {Filippenko}, {Flint}, {Foley}, {Freedman}, {Hamuy},
  {Harding}, {Hathi}, {Hicken}, {Hoopes}, {Impey}, {Jannuzi}, {Jansen}, {Jha},
  {Kaluzny}, {Kannappan}, {Kirshner}, {Latham}, {Lee}, {Leonard}, {Li},
  {Luhman}, {Martini}, {Mathis}, {Maza}, {Megeath}, {Miller}, {Minniti},
  {Olszewski}, {Papenkova}, {Phillips}, {Pindor}, {Sasselov}, {Schild},
  {Schweiker}, {Spahr}, {Thomas-Osip}, {Thompson}, {Weisz}, {Windhorst}, \&
  {Zaritsky}}]{Matheson+03}
{Matheson}, T., {Garnavich}, P.~M., {Stanek}, K.~Z., {et~al.} 2003, \apj, 599,
  394, \dodoi{10.1086/379228}

\bibitem[{{Mazzali} {et~al.}(2021){Mazzali}, {Pian}, {Bufano}, \&
  {Ashall}}]{Mazzali+21}
{Mazzali}, P.~A., {Pian}, E., {Bufano}, F., \& {Ashall}, C. 2021, \mnras, 505,
  4106, \dodoi{10.1093/mnras/stab1594}

\bibitem[{{Mazzali} {et~al.}(2003){Mazzali}, {Deng}, {Tominaga}, {Maeda},
  {Nomoto}, {Matheson}, {Kawabata}, {Stanek}, \& {Garnavich}}]{Mazzali+03}
{Mazzali}, P.~A., {Deng}, J., {Tominaga}, N., {et~al.} 2003, \apjl, 599, L95,
  \dodoi{10.1086/381259}

\bibitem[{{Meegan} {et~al.}(2009){Meegan}, {Lichti}, {Bhat}, {Bissaldi},
  {Briggs}, {Connaughton}, {Diehl}, {Fishman}, {Greiner}, {Hoover}, {van der
  Horst}, {von Kienlin}, {Kippen}, {Kouveliotou}, {McBreen}, {Paciesas},
  {Preece}, {Steinle}, {Wallace}, {Wilson}, \& {Wilson-Hodge}}]{Meegan+09}
{Meegan}, C., {Lichti}, G., {Bhat}, P.~N., {et~al.} 2009, \apj, 702, 791,
  \dodoi{10.1088/0004-637X/702/1/791}

\bibitem[{{Melandri} {et~al.}(2014){Melandri}, {Pian}, {D'Elia}, {D'Avanzo},
  {Della Valle}, {Mazzali}, {Tagliaferri}, {Cano}, {Levan}, {M{\o}oller},
  {Amati}, {Bernardini}, {Bersier}, {Bufano}, {Campana}, {Castro-Tirado},
  {Covino}, {Ghirlanda}, {Hurley}, {Malesani}, {Masetti}, {Palazzi},
  {Piranomonte}, {Rossi}, {Salvaterra}, {Starling}, {Tanaka}, {Tanvir}, \&
  {Vergani}}]{Melandri+14}
{Melandri}, A., {Pian}, E., {D'Elia}, V., {et~al.} 2014, \aap, 567, A29,
  \dodoi{10.1051/0004-6361/201423572}

\bibitem[{{Metzger}(2019)}]{metzger19}
{Metzger}, B.~D. 2019, Living Reviews in Relativity, 23, 1,
  \dodoi{10.1007/s41114-019-0024-0}

\bibitem[{{Metzger} {et~al.}(2009){Metzger}, {Piro}, \&
  {Quataert}}]{Metzger+09}
{Metzger}, B.~D., {Piro}, A.~L., \& {Quataert}, E. 2009, \mnras, 396, 304,
  \dodoi{10.1111/j.1365-2966.2008.14380.x}

\bibitem[{{Metzger} {et~al.}(2007){Metzger}, {Thompson}, \&
  {Quataert}}]{Metzger+07}
{Metzger}, B.~D., {Thompson}, T.~A., \& {Quataert}, E. 2007, \apj, 659, 561,
  \dodoi{10.1086/512059}

\bibitem[{{Miller} {et~al.}(2020){Miller}, {Sprouse}, {Fryer}, {Ryan},
  {Dolence}, {Mumpower}, \& {Surman}}]{Miller+20}
{Miller}, J.~M., {Sprouse}, T.~M., {Fryer}, C.~L., {et~al.} 2020, \apj, 902,
  66, \dodoi{10.3847/1538-4357/abb4e3}

\bibitem[{{Modjaz} {et~al.}(2016){Modjaz}, {Liu}, {Bianco}, \&
  {Graur}}]{Modjaz+16}
{Modjaz}, M., {Liu}, Y.~Q., {Bianco}, F.~B., \& {Graur}, O. 2016, \apj, 832,
  108, \dodoi{10.3847/0004-637X/832/2/108}

\bibitem[{{Moss} {et~al.}(2023){Moss}, {Mochkovitch}, {Daigne}, {Beniamini}, \&
  {Guiriec}}]{Moss+23}
{Moss}, M.~J., {Mochkovitch}, R., {Daigne}, F., {Beniamini}, P., \& {Guiriec},
  S. 2023, arXiv e-prints, arXiv:2306.00815, \dodoi{10.48550/arXiv.2306.00815}

\bibitem[{{M{\"o}sta} {et~al.}(2015){M{\"o}sta}, {Ott}, {Radice}, {Roberts},
  {Schnetter}, \& {Haas}}]{Mosta+15}
{M{\"o}sta}, P., {Ott}, C.~D., {Radice}, D., {et~al.} 2015, \nat, 528, 376,
  \dodoi{10.1038/nature15755}

\bibitem[{{M{\"o}sta} {et~al.}(2018){M{\"o}sta}, {Roberts}, {Halevi}, {Ott},
  {Lippuner}, {Haas}, \& {Schnetter}}]{Mosta+18}
{M{\"o}sta}, P., {Roberts}, L.~F., {Halevi}, G., {et~al.} 2018, \apj, 864, 171,
  \dodoi{10.3847/1538-4357/aad6ec}

\bibitem[{{M{\"o}sta} {et~al.}(2014){M{\"o}sta}, {Richers}, {Ott}, {Haas},
  {Piro}, {Boydstun}, {Abdikamalov}, {Reisswig}, \& {Schnetter}}]{Mosta+14}
{M{\"o}sta}, P., {Richers}, S., {Ott}, C.~D., {et~al.} 2014, \apjl, 785, L29,
  \dodoi{10.1088/2041-8205/785/2/L29}

\bibitem[{{Naidu} {et~al.}(2022){Naidu}, {Ji}, {Conroy}, {Bonaca}, {Ting},
  {Zaritsky}, {van Son}, {Broekgaarden}, {Tacchella}, {Chandra}, {Caldwell},
  {Cargile}, \& {Speagle}}]{Naidu+22}
{Naidu}, R.~P., {Ji}, A.~P., {Conroy}, C., {et~al.} 2022, \apjl, 926, L36,
  \dodoi{10.3847/2041-8213/ac5589}

\bibitem[{{Nugent} {et~al.}(2023){Nugent}, {Fong}, {Castrejon}, {Leja},
  {Zevin}, \& {Ji}}]{Nugent+23}
{Nugent}, A.~E., {Fong}, W.-f., {Castrejon}, C., {et~al.} 2023, arXiv e-prints,
  arXiv:2310.12202.
\newblock \doarXiv{2310.12202}

\bibitem[{{Nugent} {et~al.}(2022){Nugent}, {Fong}, {Dong}, {Leja}, {Berger},
  {Zevin}, {Chornock}, {Cobb}, {Kelley}, {Kilpatrick}, {Levan}, {Margutti},
  {Paterson}, {Perley}, {Escorial}, {Smith}, \& {Tanvir}}]{Nugent+22}
{Nugent}, A.~E., {Fong}, W.-F., {Dong}, Y., {et~al.} 2022, \apj, 940, 57,
  \dodoi{10.3847/1538-4357/ac91d1}

\bibitem[{{O'Connor} {et~al.}(2022){O'Connor}, {Troja}, {Dichiara},
  {Beniamini}, {Cenko}, {Kouveliotou}, {Gonz{\'a}lez}, {Durbak}, {Gatkine},
  {Kutyrev}, {Sakamoto}, {S{\'a}nchez-Ram{\'\i}rez}, \&
  {Veilleux}}]{O'Connor+22}
{O'Connor}, B., {Troja}, E., {Dichiara}, S., {et~al.} 2022, \mnras, 515, 4890,
  \dodoi{10.1093/mnras/stac1982}

\bibitem[{{Olivares E.} {et~al.}(2012){Olivares E.}, {Greiner}, {Schady},
  {Rau}, {Klose}, {Kr{\"u}hler}, {Afonso}, {Updike}, {Nardini}, {Filgas},
  {Nicuesa Guelbenzu}, {Clemens}, {Elliott}, {Kann}, {Rossi}, \&
  {Sudilovsky}}]{Olivares+12}
{Olivares E.}, F., {Greiner}, J., {Schady}, P., {et~al.} 2012, \aap, 539, A76,
  \dodoi{10.1051/0004-6361/201117929}

\bibitem[{{{\"O}stlin} {et~al.}(2008){{\"O}stlin}, {Zackrisson}, {Sollerman},
  {Mattila}, \& {Hayes}}]{Ostlin+08}
{{\"O}stlin}, G., {Zackrisson}, E., {Sollerman}, J., {Mattila}, S., \& {Hayes},
  M. 2008, \mnras, 387, 1227, \dodoi{10.1111/j.1365-2966.2008.13319.x}

\bibitem[{{Perley} {et~al.}(2014){Perley}, {Cenko}, {Corsi}, {Tanvir}, {Levan},
  {Kann}, {Sonbas}, {Wiersema}, {Zheng}, {Zhao}, {Bai}, {Bremer},
  {Castro-Tirado}, {Chang}, {Clubb}, {Frail}, {Fruchter},
  {G{\"o}{\u{g}}{\"u}{\c{s}}}, {Greiner}, {G{\"u}ver}, {Horesh}, {Filippenko},
  {Klose}, {Mao}, {Morgan}, {Pozanenko}, {Schmidl}, {Stecklum}, {Tanga},
  {Volnova}, {Volvach}, {Wang}, {Winters}, \& {Xin}}]{Perley+14}
{Perley}, D.~A., {Cenko}, S.~B., {Corsi}, A., {et~al.} 2014, \apj, 781, 37,
  \dodoi{10.1088/0004-637X/781/1/37}

\bibitem[{{Perley} {et~al.}(2020){Perley}, {Fremling}, {Sollerman}, {Miller},
  {Dahiwale}, {Sharma}, {Bellm}, {Biswas}, {Brink}, {Bruch}, {De}, {Dekany},
  {Drake}, {Duev}, {Filippenko}, {Gal-Yam}, {Goobar}, {Graham}, {Graham}, {Ho},
  {Irani}, {Kasliwal}, {Kim}, {Kulkarni}, {Mahabal}, {Masci}, {Modak}, {Neill},
  {Nordin}, {Riddle}, {Soumagnac}, {Strotjohann}, {Schulze}, {Taggart},
  {Tzanidakis}, {Walters}, \& {Yan}}]{Perley+20}
{Perley}, D.~A., {Fremling}, C., {Sollerman}, J., {et~al.} 2020, \apj, 904, 35,
  \dodoi{10.3847/1538-4357/abbd98}

\bibitem[{{Piro} {et~al.}(2014){Piro}, {Troja}, {Gendre}, {Ghisellini},
  {Ricci}, {Bannister}, {Fiore}, {Kidd}, {Piranomonte}, \&
  {Wieringa}}]{Piro+14}
{Piro}, L., {Troja}, E., {Gendre}, B., {et~al.} 2014, \apjl, 790, L15,
  \dodoi{10.1088/2041-8205/790/2/L15}

\bibitem[{{Prentice} {et~al.}(2019){Prentice}, {Ashall}, {James}, {Short},
  {Mazzali}, {Bersier}, {Crowther}, {Barbarino}, {Chen}, {Copperwheat},
  {Darnley}, {Denneau}, {Elias-Rosa}, {Fraser}, {Galbany}, {Gal-Yam},
  {Harmanen}, {Howell}, {Hosseinzadeh}, {Inserra}, {Kankare}, {Karamehmetoglu},
  {Lamb}, {Limongi}, {Maguire}, {McCully}, {Olivares E}, {Piascik}, {Pignata},
  {Reichart}, {Rest}, {Reynolds}, {Rodr{\'\i}guez}, {Saario}, {Schulze},
  {Smartt}, {Smith}, {Sollerman}, {Stalder}, {Sullivan}, {Taddia}, {Valenti},
  {Vergani}, {Williams}, \& {Young}}]{Prentice2019}
{Prentice}, S.~J., {Ashall}, C., {James}, P.~A., {et~al.} 2019, \mnras, 485,
  1559, \dodoi{10.1093/mnras/sty3399}

\bibitem[{{Rastinejad} {et~al.}(2021){Rastinejad}, {Fong}, {Kilpatrick},
  {Paterson}, {Tanvir}, {Levan}, {Metzger}, {Berger}, {Chornock}, {Cobb},
  {Laskar}, {Milne}, {Nugent}, \& {Smith}}]{Rastinejad+21}
{Rastinejad}, J.~C., {Fong}, W., {Kilpatrick}, C.~D., {et~al.} 2021, \apj, 916,
  89, \dodoi{10.3847/1538-4357/ac04b4}

\bibitem[{{Rastinejad} {et~al.}(2022){Rastinejad}, {Gompertz}, {Levan}, {Fong},
  {Nicholl}, {Lamb}, {Malesani}, {Nugent}, {Oates}, {Tanvir}, {de Ugarte
  Postigo}, {Kilpatrick}, {Moore}, {Metzger}, {Ravasio}, {Rossi}, {Schroeder},
  {Jencson}, {Sand}, {Smith}, {Ag{\"u}{\'\i} Fern{\'a}ndez}, {Berger},
  {Blanchard}, {Chornock}, {Cobb}, {De Pasquale}, {Fynbo}, {Izzo}, {Kann},
  {Laskar}, {Marini}, {Paterson}, {Escorial}, {Sears}, \&
  {Th{\"o}ne}}]{Rastinejad+22}
{Rastinejad}, J.~C., {Gompertz}, B.~P., {Levan}, A.~J., {et~al.} 2022, \nat,
  612, 223, \dodoi{10.1038/s41586-022-05390-w}

\bibitem[{{Reichert} {et~al.}(2023){Reichert}, {Obergaulinger}, {Aloy},
  {Gabler}, {Arcones}, \& {Thielemann}}]{REichert+23}
{Reichert}, M., {Obergaulinger}, M., {Aloy}, M.~{\'A}., {et~al.} 2023, \mnras,
  518, 1557, \dodoi{10.1093/mnras/stac3185}

\bibitem[{{Rhodes} {et~al.}(2020){Rhodes}, {van der Horst}, {Fender},
  {Monageng}, {Anderson}, {Antoniadis}, {Bietenholz}, {B{\"o}ttcher}, {Bright},
  {Green}, {Kouveliotou}, {Kramer}, {Motta}, {Wijers}, {Williams}, \&
  {Woudt}}]{Rhodes+20}
{Rhodes}, L., {van der Horst}, A.~J., {Fender}, R., {et~al.} 2020, \mnras, 496,
  3326, \dodoi{10.1093/mnras/staa1715}

\bibitem[{{Rosswog} \& {Korobkin}(2022)}]{Rosswog+22}
{Rosswog}, S., \& {Korobkin}, O. 2022, arXiv e-prints, arXiv:2208.14026,
  \dodoi{10.48550/arXiv.2208.14026}

\bibitem[{{Rosswog} {et~al.}(1999){Rosswog}, {Liebend{\"o}rfer}, {Thielemann},
  {Davies}, {Benz}, \& {Piran}}]{Rosswog+99}
{Rosswog}, S., {Liebend{\"o}rfer}, M., {Thielemann}, F.~K., {et~al.} 1999,
  \aap, 341, 499, \dodoi{10.48550/arXiv.astro-ph/9811367}

\bibitem[{{Rosswog} {et~al.}(2018){Rosswog}, {Sollerman}, {Feindt}, {Goobar},
  {Korobkin}, {Wollaeger}, {Fremling}, \& {Kasliwal}}]{Rosswog+18}
{Rosswog}, S., {Sollerman}, J., {Feindt}, U., {et~al.} 2018, \aap, 615, A132,
  \dodoi{10.1051/0004-6361/201732117}

\bibitem[{{Rouco Escorial} {et~al.}(2022){Rouco Escorial}, {Fong}, {Berger},
  {Laskar}, {Margutti}, {Schroeder}, {Rastinejad}, {Cornish}, {Popp}, {Lally},
  {Nugent}, {Paterson}, {Metzger}, {Chornock}, {Alexander}, {Cendes}, \&
  {Eftekhari}}]{Rouco+23}
{Rouco Escorial}, A., {Fong}, W.-f., {Berger}, E., {et~al.} 2022, arXiv
  e-prints, arXiv:2210.05695, \dodoi{10.48550/arXiv.2210.05695}

\bibitem[{{Sakamoto} {et~al.}(2010){Sakamoto}, {Barthelmy}, {Baumgartner},
  {Cummings}, {Gehrels}, {Krimm}, {Markwardt}, {Palmer}, {Stamatikos},
  {Tueller}, \& {Ukwatta}}]{Sakamoto+!0}
{Sakamoto}, T., {Barthelmy}, S.~D., {Baumgartner}, W.~H., {et~al.} 2010, GRB
  Coordinates Network, 10511, 1

\bibitem[{{Savchenko} {et~al.}(2017){Savchenko}, {Ferrigno}, {Kuulkers},
  {Bazzano}, {Bozzo}, {Brandt}, {Chenevez}, {Courvoisier}, {Diehl}, {Domingo},
  {Hanlon}, {Jourdain}, {von Kienlin}, {Laurent}, {Lebrun}, {Lutovinov},
  {Martin-Carrillo}, {Mereghetti}, {Natalucci}, {Rodi}, {Roques}, {Sunyaev}, \&
  {Ubertini}}]{Savchenko+17}
{Savchenko}, V., {Ferrigno}, C., {Kuulkers}, E., {et~al.} 2017, \apjl, 848,
  L15, \dodoi{10.3847/2041-8213/aa8f94}

\bibitem[{{Schady} {et~al.}(2012){Schady}, {Dwelly}, {Page}, {Kr{\"u}hler},
  {Greiner}, {Oates}, {de Pasquale}, {Nardini}, {Roming}, {Rossi}, \&
  {Still}}]{Schady+12}
{Schady}, P., {Dwelly}, T., {Page}, M.~J., {et~al.} 2012, \aap, 537, A15,
  \dodoi{10.1051/0004-6361/201117414}

\bibitem[{{Schlafly} \& {Finkbeiner}(2011)}]{SchlaflyFinkbeiner11}
{Schlafly}, E.~F., \& {Finkbeiner}, D.~P. 2011, \apj, 737, 103,
  \dodoi{10.1088/0004-637X/737/2/103}

\bibitem[{{Shahbandeh} {et~al.}(2023){Shahbandeh}, {Sarangi}, {Temim},
  {Szalai}, {Fox}, {Tinyanont}, {Dwek}, {Dessart}, {Filippenko}, {Brink},
  {Foley}, {Jencson}, {Pierel}, {Zs{\'\i}ros}, {Rest}, {Zheng}, {Andrews},
  {Clayton}, {De}, {Engesser}, {Gezari}, {Gomez}, {Gonzaga}, {Johansson},
  {Kasliwal}, {Lau}, {De Looze}, {Marston}, {Milisavljevic}, {O'Steen},
  {Siebert}, {Skrutskie}, {Smith}, {Strolger}, {Van Dyk}, {Wang}, {Williams},
  {Williams}, {Xiao}, \& {Yang}}]{Shahbandeh+23}
{Shahbandeh}, M., {Sarangi}, A., {Temim}, T., {et~al.} 2023, \mnras, 523, 6048,
  \dodoi{10.1093/mnras/stad1681}

\bibitem[{{Shen} {et~al.}(2015){Shen}, {Cooke}, {Ramirez-Ruiz}, {Madau},
  {Mayer}, \& {Guedes}}]{Shen+15}
{Shen}, S., {Cooke}, R.~J., {Ramirez-Ruiz}, E., {et~al.} 2015, \apj, 807, 115,
  \dodoi{10.1088/0004-637X/807/2/115}

\bibitem[{{Siegel} {et~al.}(2019){Siegel}, {Barnes}, \&
  {Metzger}}]{SiegelBarnesMetzger2019}
{Siegel}, D.~M., {Barnes}, J., \& {Metzger}, B.~D. 2019, \nat, 569, 241,
  \dodoi{10.1038/s41586-019-1136-0}

\bibitem[{{Simon} {et~al.}(2023){Simon}, {Brown}, {Mutlu-Pakdil}, {Ji},
  {Drlica-Wagner}, {Avila}, {Mart{\'\i}nez-V{\'a}zquez}, {Li}, {Balbinot},
  {Bechtol}, {Frebel}, {Geha}, {Hansen}, {James}, {Pace}, {Aguena}, {Alves},
  {Andrade-Oliveira}, {Annis}, {Bacon}, {Bertin}, {Brooks}, {Burke}, {Carnero
  Rosell}, {Carrasco Kind}, {Carretero}, {Costanzi}, {da Costa}, {De Vicente},
  {Desai}, {Doel}, {Everett}, {Ferrero}, {Frieman}, {Garc{\'\i}a-Bellido},
  {Gatti}, {Gerdes}, {Gruen}, {Gruendl}, {Gschwend}, {Gutierrez}, {Hinton},
  {Hollowood}, {Honscheid}, {Kuehn}, {Kuropatkin}, {Marshall},
  {Mena-Fern{\'a}ndez}, {Miquel}, {Palmese}, {Paz-Chinch{\'o}n}, {Pereira},
  {Pieres}, {Plazas Malag{\'o}n}, {Raveri}, {Rodriguez-Monroy}, {Sanchez},
  {Santiago}, {Scarpine}, {Sevilla-Noarbe}, {Smith}, {Suchyta}, {Swanson},
  {Tarle}, {To}, {Vincenzi}, {Weaverdyck}, \& {Wilkinson}}]{Simon+23}
{Simon}, J.~D., {Brown}, T.~M., {Mutlu-Pakdil}, B., {et~al.} 2023, \apj, 944,
  43, \dodoi{10.3847/1538-4357/aca9d1}

\bibitem[{{Skrutskie} {et~al.}(2006){Skrutskie}, {Cutri}, {Stiening},
  {Weinberg}, {Schneider}, {Carpenter}, {Beichman}, {Capps}, {Chester},
  {Elias}, {Huchra}, {Liebert}, {Lonsdale}, {Monet}, {Price}, {Seitzer},
  {Jarrett}, {Kirkpatrick}, {Gizis}, {Howard}, {Evans}, {Fowler}, {Fullmer},
  {Hurt}, {Light}, {Kopan}, {Marsh}, {McCallon}, {Tam}, {Van Dyk}, \&
  {Wheelock}}]{2MASS}
{Skrutskie}, M.~F., {Cutri}, R.~M., {Stiening}, R., {et~al.} 2006, \aj, 131,
  1163, \dodoi{10.1086/498708}

\bibitem[{{Soares-Santos} {et~al.}(2017){Soares-Santos}, {Holz}, {Annis},
  {Chornock}, {Herner}, {Berger}, {Brout}, {Chen}, {Kessler}, {Sako}, {Allam},
  {Tucker}, {Butler}, {Palmese}, {Doctor}, {Diehl}, {Frieman}, {Yanny}, {Lin},
  {Scolnic}, {Cowperthwaite}, {Neilsen}, {Marriner}, {Kuropatkin}, {Hartley},
  {Paz-Chinch{\'o}n}, {Alexander}, {Balbinot}, {Blanchard}, {Brown}, {Carlin},
  {Conselice}, {Cook}, {Drlica-Wagner}, {Drout}, {Durret}, {Eftekhari}, {Farr},
  {Finley}, {Foley}, {Fong}, {Fryer}, {Garc{\'\i}a-Bellido}, {Gill}, {Gruendl},
  {Hanna}, {Kasen}, {Li}, {Lopes}, {Louren{\c{c}}o}, {Margutti}, {Marshall},
  {Matheson}, {Medina}, {Metzger}, {Mu{\~n}oz}, {Muir}, {Nicholl}, {Quataert},
  {Rest}, {Sauseda}, {Schlegel}, {Secco}, {Sobreira}, {Stebbins}, {Villar},
  {Vivas}, {Walker}, {Wester}, {Williams}, {Zenteno}, {Zhang}, {Abbott},
  {Abdalla}, {Banerji}, {Bechtol}, {Benoit-L{\'e}vy}, {Bertin}, {Brooks},
  {Buckley-Geer}, {Burke}, {Carnero Rosell}, {Carrasco Kind}, {Carretero},
  {Castander}, {Crocce}, {Cunha}, {D'Andrea}, {da Costa}, {Davis}, {Desai},
  {Dietrich}, {Doel}, {Eifler}, {Fernand ez}, {Flaugher}, {Fosalba},
  {Gaztanaga}, {Gerdes}, {Giannantonio}, {Goldstein}, {Gruen}, {Gschwend},
  {Gutierrez}, {Honscheid}, {Jain}, {James}, {Jeltema}, {Johnson}, {Johnson},
  {Kent}, {Krause}, {Kron}, {Kuehn}, {Kuhlmann}, {Lahav}, {Lima}, {Maia},
  {March}, {McMahon}, {Menanteau}, {Miquel}, {Mohr}, {Nichol}, {Nord}, {Ogand
  o}, {Petravick}, {Plazas}, {Romer}, {Roodman}, {Rykoff}, {Sanchez},
  {Scarpine}, {Schubnell}, {Sevilla-Noarbe}, {Smith}, {Smith}, {Suchyta},
  {Swanson}, {Tarle}, {Thomas}, {Thomas}, {Troxel}, {Vikram}, {Wechsler},
  {Weller}, {Dark Energy Survey}, \& {Dark Energy Camera GW-EM
  Collaboration}}]{Soares-Santos+17}
{Soares-Santos}, M., {Holz}, D.~E., {Annis}, J., {et~al.} 2017, \apjl, 848,
  L16, \dodoi{10.3847/2041-8213/aa9059}

\bibitem[{{Stanek} {et~al.}(2003){Stanek}, {Matheson}, {Garnavich}, {Martini},
  {Berlind}, {Caldwell}, {Challis}, {Brown}, {Schild}, {Krisciunas}, {Calkins},
  {Lee}, {Hathi}, {Jansen}, {Windhorst}, {Echevarria}, {Eisenstein}, {Pindor},
  {Olszewski}, {Harding}, {Holland}, \& {Bersier}}]{Stanek+03}
{Stanek}, K.~Z., {Matheson}, T., {Garnavich}, P.~M., {et~al.} 2003, \apjl, 591,
  L17, \dodoi{10.1086/376976}

\bibitem[{{Starling} {et~al.}(2011){Starling}, {Wiersema}, {Levan}, {Sakamoto},
  {Bersier}, {Goldoni}, {Oates}, {Rowlinson}, {Campana}, {Sollerman}, {Tanvir},
  {Malesani}, {Fynbo}, {Covino}, {D'Avanzo}, {O'Brien}, {Page}, {Osborne},
  {Vergani}, {Barthelmy}, {Burrows}, {Cano}, {Curran}, {de Pasquale}, {D'Elia},
  {Evans}, {Flores}, {Fruchter}, {Garnavich}, {Gehrels}, {Gorosabel}, {Hjorth},
  {Holland}, {van der Horst}, {Hurkett}, {Jakobsson}, {Kamble}, {Kouveliotou},
  {Kuin}, {Kaper}, {Mazzali}, {Nugent}, {Pian}, {Stamatikos}, {Th{\"o}ne}, \&
  {Woosley}}]{Starling+11}
{Starling}, R.~L.~C., {Wiersema}, K., {Levan}, A.~J., {et~al.} 2011, \mnras,
  411, 2792, \dodoi{10.1111/j.1365-2966.2010.17879.x}

\bibitem[{{Sun} {et~al.}(2022){Sun}, {Xiao}, \& {Li}}]{Sun+22}
{Sun}, L., {Xiao}, L., \& {Li}, G. 2022, \mnras, 513, 4057,
  \dodoi{10.1093/mnras/stac1121}

\bibitem[{{Szalai} {et~al.}(2021){Szalai}, {Fox}, {Arendt}, {Dwek}, {Andrews},
  {Clayton}, {Filippenko}, {Johansson}, {Kelly}, {Krafton}, {Marston},
  {Mauerhan}, \& {Van Dyk}}]{Szalai+21}
{Szalai}, T., {Fox}, O.~D., {Arendt}, R.~G., {et~al.} 2021, \apj, 919, 17,
  \dodoi{10.3847/1538-4357/ac0e2b}

\bibitem[{{Tanvir} {et~al.}(2017){Tanvir}, {Levan},
  {Gonz{\'a}lez-Fern{\'a}ndez}, {Korobkin}, {Mandel}, {Rosswog}, {Hjorth},
  {D'Avanzo}, {Fruchter}, {Fryer}, {Kangas}, {Milvang-Jensen}, {Rosetti},
  {Steeghs}, {Wollaeger}, {Cano}, {Copperwheat}, {Covino}, {D'Elia}, {de Ugarte
  Postigo}, {Evans}, {Even}, {Fairhurst}, {Figuera Jaimes}, {Fontes}, {Fujii},
  {Fynbo}, {Gompertz}, {Greiner}, {Hodosan}, {Irwin}, {Jakobsson},
  {J{\o}rgensen}, {Kann}, {Lyman}, {Malesani}, {McMahon}, {Melandri},
  {O'Brien}, {Osborne}, {Palazzi}, {Perley}, {Pian}, {Piranomonte}, {Rabus},
  {Rol}, {Rowlinson}, {Schulze}, {Sutton}, {Th{\"o}ne}, {Ulaczyk}, {Watson},
  {Wiersema}, \& {Wijers}}]{Tanvir+17}
{Tanvir}, N.~R., {Levan}, A.~J., {Gonz{\'a}lez-Fern{\'a}ndez}, C., {et~al.}
  2017, \apjl, 848, L27, \dodoi{10.3847/2041-8213/aa90b6}

\bibitem[{{Tarumi} {et~al.}(2023){Tarumi}, {Hotokezaka}, {Domoto}, \&
  {Tanaka}}]{Tarumi+23}
{Tarumi}, Y., {Hotokezaka}, K., {Domoto}, N., \& {Tanaka}, M. 2023, arXiv
  e-prints, arXiv:2302.13061, \dodoi{10.48550/arXiv.2302.13061}

\bibitem[{{Tauris} {et~al.}(2017){Tauris}, {Kramer}, {Freire}, {Wex}, {Janka},
  {Langer}, {Podsiadlowski}, {Bozzo}, {Chaty}, {Kruckow}, {van den Heuvel},
  {Antoniadis}, {Breton}, \& {Champion}}]{Tauris+17}
{Tauris}, T.~M., {Kramer}, M., {Freire}, P.~C.~C., {et~al.} 2017, \apj, 846,
  170, \dodoi{10.3847/1538-4357/aa7e89}

\bibitem[{{The LIGO Scientific Collaboration} {et~al.}(2021){The LIGO
  Scientific Collaboration}, {the Virgo Collaboration}, {the KAGRA
  Collaboration}, {Abbott}, {Abbott}, {Acernese}, {Ackley}, {Adams},
  {Adhikari}, {Adhikari}, {Adya}, {Affeldt}, {Agarwal}, {Agathos}, {Agatsuma},
  {Aggarwal}, {Aguiar}, {Aiello}, {Ain}, {Ajith}, {Akcay}, {Akutsu},
  {Albanesi}, {Allocca}, {Altin}, {Amato}, {Anand}, {Anand}, {Ananyeva},
  {Anderson}, {Anderson}, {Ando}, {Andrade}, {Andres}, {Andri{\'c}}, \&
  {Angelova}}]{GWTC-3}
{The LIGO Scientific Collaboration}, {the Virgo Collaboration}, {the KAGRA
  Collaboration}, {et~al.} 2021, arXiv e-prints, arXiv:2111.03606.
\newblock \doarXiv{2111.03606}

\bibitem[{{Thielemann} {et~al.}(2011){Thielemann}, {Arcones}, {K{\"a}ppeli},
  {Liebend{\"o}rfer}, {Rauscher}, {Winteler}, {Fr{\"o}hlich}, {Dillmann},
  {Fischer}, {Martinez-Pinedo}, {Langanke}, {Farouqi}, {Kratz}, {Panov}, \&
  {Korneev}}]{Thielemann+11}
{Thielemann}, F.~K., {Arcones}, A., {K{\"a}ppeli}, R., {et~al.} 2011, Progress
  in Particle and Nuclear Physics, 66, 346, \dodoi{10.1016/j.ppnp.2011.01.032}

\bibitem[{{Thompson} {et~al.}(2004){Thompson}, {Chang}, \&
  {Quataert}}]{Thompson+04}
{Thompson}, T.~A., {Chang}, P., \& {Quataert}, E. 2004, \apj, 611, 380,
  \dodoi{10.1086/421969}

\bibitem[{{Thompson} \& {ud-Doula}(2018)}]{Thompson+18}
{Thompson}, T.~A., \& {ud-Doula}, A. 2018, \mnras, 476, 5502,
  \dodoi{10.1093/mnras/sty480}

\bibitem[{{Tiengo} {et~al.}(2004){Tiengo}, {Mereghetti}, {Ghisellini},
  {Tavecchio}, \& {Ghirlanda}}]{Tiengo+04}
{Tiengo}, A., {Mereghetti}, S., {Ghisellini}, G., {Tavecchio}, F., \&
  {Ghirlanda}, G. 2004, \aap, 423, 861, \dodoi{10.1051/0004-6361:20041027}

\bibitem[{{Tinyanont} {et~al.}(2022){Tinyanont}, {Ridden-Harper}, {Foley},
  {Morozova}, {Kilpatrick}, {Dimitriadis}, {DeMarchi}, {Gagliano},
  {Jacobson-Gal{\'a}n}, {Messick}, {Pierel}, {Piro}, {Ramirez-Ruiz}, {Siebert},
  {Chambers}, {Clever}, {Coulter}, {De}, {Hankins}, {Hung}, {Jha}, {Jimenez
  Angel}, {Jones}, {Kasliwal}, {Lin}, {Marques-Chaves}, {Margutti}, {Moore},
  {P{\'e}rez-Fournon}, {Poidevin}, {Rest}, {Shirley}, {Smith}, {Strasburger},
  {Swift}, {Wainscoat}, {Wang}, \& {Zenati}}]{Tinyamont+22}
{Tinyanont}, S., {Ridden-Harper}, R., {Foley}, R.~J., {et~al.} 2022, \mnras,
  512, 2777, \dodoi{10.1093/mnras/stab2887}

\bibitem[{{Tody}(1986)}]{Tody86}
{Tody}, D. 1986, in Society of Photo-Optical Instrumentation Engineers (SPIE)
  Conference Series, Vol. 627, Instrumentation in astronomy VI, ed. D.~L.
  {Crawford}, 733, \dodoi{10.1117/12.968154}

\bibitem[{{Tody}(1993)}]{Tody93}
{Tody}, D. 1993, in Astronomical Society of the Pacific Conference Series,
  Vol.~52, Astronomical Data Analysis Software and Systems II, ed. R.~J.
  {Hanisch}, R.~J.~V. {Brissenden}, \& J.~{Barnes}, 173

\bibitem[{{Troja} {et~al.}(2022){Troja}, {Fryer}, {O'Connor}, {Ryan},
  {Dichiara}, {Kumar}, {Ito}, {Gupta}, {Wollaeger}, {Norris}, {Kawai},
  {Butler}, {Aryan}, {Misra}, {Hosokawa}, {Murata}, {Niwano}, {Pandey},
  {Kutyrev}, {van Eerten}, {Chase}, {Hu}, {Caballero-Garcia}, \&
  {Castro-Tirado}}]{Troja+22}
{Troja}, E., {Fryer}, C.~L., {O'Connor}, B., {et~al.} 2022, \nat, 612, 228,
  \dodoi{10.1038/s41586-022-05327-3}

\bibitem[{{Tsujimoto} \& {Shigeyama}(2001)}]{Tsujimoto+01}
{Tsujimoto}, T., \& {Shigeyama}, T. 2001, \apjl, 561, L97,
  \dodoi{10.1086/324441}

\bibitem[{{Tsvetkova} {et~al.}(2019){Tsvetkova}, {Golenetskii}, {Aptekar},
  {Frederiks}, {Ulanov}, {Svinkin}, {Lysenko}, {Kozlova}, {Cline}, \&
  {Konus-Wind Team}}]{konus_190829a}
{Tsvetkova}, A., {Golenetskii}, S., {Aptekar}, R., {et~al.} 2019, GRB
  Coordinates Network, 25660, 1

\bibitem[{{Valenti} {et~al.}(2017){Valenti}, {Sand}, {Yang}, {Cappellaro},
  {Tartaglia}, {Corsi}, {Jha}, {Reichart}, {Haislip}, \&
  {Kouprianov}}]{Valenti+17}
{Valenti}, S., {Sand}, D.~J., {Yang}, S., {et~al.} 2017, \apjl, 848, L24,
  \dodoi{10.3847/2041-8213/aa8edf}

\bibitem[{{van de Voort} {et~al.}(2022){van de Voort}, {Pakmor}, {Bieri}, \&
  {Grand}}]{vandeVoort+22}
{van de Voort}, F., {Pakmor}, R., {Bieri}, R., \& {Grand}, R. J.~J. 2022,
  \mnras, 512, 5258, \dodoi{10.1093/mnras/stac710}

\bibitem[{{van der Walt} {et~al.}(2011){van der Walt}, {Colbert}, \&
  {Varoquaux}}]{numpy11}
{van der Walt}, S., {Colbert}, S.~C., \& {Varoquaux}, G. 2011, Computing in
  Science and Engineering, 13, 22, \dodoi{10.1109/MCSE.2011.37}

\bibitem[{{Vanderspek} {et~al.}(2003){Vanderspek}, {Crew}, {Doty},
  {Villasenor}, {Monnelly}, {Butler}, {Cline}, {Jernigan}, {Levine}, {Martel},
  {Morgan}, {Prigozhin}, {Azzibrouck}, {Braga}, {Manchanda}, {Pizzichini},
  {Ricker}, {Atteia}, {Kawai}, {Lamb}, {Woosley}, {Donaghy}, {Suzuki},
  {Shirasaki}, {Graziani}, {Matsuoka}, {Tamagawa}, {Torii}, {Sakamoto},
  {Yoshida}, {Fenimore}, {Galassi}, {Tavenner}, {Nakagawa}, {Takahashi},
  {Satoh}, {Urata}, {Boer}, {Olive}, {Dezalay}, {Barraud}, \&
  {Hurley}}]{030329_gcn}
{Vanderspek}, R., {Crew}, G., {Doty}, J., {et~al.} 2003, GRB Coordinates
  Network, 1997, 1

\bibitem[{{Venn} {et~al.}(2004){Venn}, {Irwin}, {Shetrone}, {Tout}, {Hill}, \&
  {Tolstoy}}]{Venn+04}
{Venn}, K.~A., {Irwin}, M., {Shetrone}, M.~D., {et~al.} 2004, \aj, 128, 1177,
  \dodoi{10.1086/422734}

\bibitem[{{von Kienlin}(2013)}]{GBM_130427a}
{von Kienlin}, A. 2013, GRB Coordinates Network, 14473, 1

\bibitem[{{Wallner} {et~al.}(2015){Wallner}, {Faestermann}, {Feige},
  {Feldstein}, {Knie}, {Korschinek}, {Kutschera}, {Ofan}, {Paul}, {Quinto},
  {Rugel}, \& {Steier}}]{Wallner+15}
{Wallner}, A., {Faestermann}, T., {Feige}, J., {et~al.} 2015, Nature
  Communications, 6, 5956, \dodoi{10.1038/ncomms6956}

\bibitem[{{Wallner} {et~al.}(2021){Wallner}, {Froehlich}, {Hotchkis},
  {Kinoshita}, {Paul}, {Martschini}, {Pavetich}, {Tims}, {Kivel}, {Schumann},
  {Honda}, {Matsuzaki}, \& {Yamagata}}]{Wallner+21}
{Wallner}, A., {Froehlich}, M.~B., {Hotchkis}, M.~A.~C., {et~al.} 2021,
  Science, 372, 742, \dodoi{10.1126/science.aax3972}

\bibitem[{{Wang} \& {Burrows}(2023)}]{Wang+23}
{Wang}, T., \& {Burrows}, A. 2023, arXiv e-prints, arXiv:2311.03446.
\newblock \doarXiv{2311.03446}

\bibitem[{{Watson} {et~al.}(2019){Watson}, {Hansen}, {Selsing}, {Koch},
  {Malesani}, {Andersen}, {Fynbo}, {Arcones}, {Bauswein}, {Covino}, {Grado},
  {Heintz}, {Hunt}, {Kouveliotou}, {Leloudas}, {Levan}, {Mazzali}, \&
  {Pian}}]{Watson+19}
{Watson}, D., {Hansen}, C.~J., {Selsing}, J., {et~al.} 2019, \nat, 574, 497,
  \dodoi{10.1038/s41586-019-1676-3}

\bibitem[{{Williams} {et~al.}(2023){Williams}, {Kennea}, {Dichiara},
  {Kobayashi}, {Iwakiri}, {Beardmore}, {Evans}, {Heinz}, {Lien}, {Oates},
  {Negoro}, {Cenko}, {Buisson}, {Hartmann}, {Jaisawal}, {Kuin}, {Lesage},
  {Page}, {Parsotan}, {Pasham}, {Sbarufatti}, {Siegel}, {Sugita}, {Younes},
  {Ambrosi}, {Arzoumanian}, {Bernardini}, {Campana}, {Capalbi}, {Caputo},
  {D'A{\`\i}}, {D'Avanzo}, {D'Elia}, {De Pasquale}, {Eyles-Ferris}, {Ferrara},
  {Gendreau}, {Gropp}, {Kawai}, {Klingler}, {Laha}, {Melandri}, {Mihara},
  {Moss}, {O'Brien}, {Osborne}, {Palmer}, {Perri}, {Serino}, {Sonbas},
  {Stamatikos}, {Starling}, {Tagliaferri}, {Tohuvavohu}, {Zane}, \&
  {Ziaeepour}}]{Williams+23}
{Williams}, M.~A., {Kennea}, J.~A., {Dichiara}, S., {et~al.} 2023, \apjl, 946,
  L24, \dodoi{10.3847/2041-8213/acbcd1}

\bibitem[{{Woosley} {et~al.}(1994){Woosley}, {Wilson}, {Mathews}, {Hoffman}, \&
  {Meyer}}]{Woosley+94}
{Woosley}, S.~E., {Wilson}, J.~R., {Mathews}, G.~J., {Hoffman}, R.~D., \&
  {Meyer}, B.~S. 1994, \apj, 433, 229, \dodoi{10.1086/174638}

\bibitem[{{Xu} {et~al.}(2013){Xu}, {de Ugarte Postigo}, {Leloudas},
  {Kr{\"u}hler}, {Cano}, {Hjorth}, {Malesani}, {Fynbo}, {Th{\"o}ne},
  {S{\'a}nchez-Ram{\'\i}rez}, {Schulze}, {Jakobsson}, {Kaper}, {Sollerman},
  {Watson}, {Cabrera-Lavers}, {Cao}, {Covino}, {Flores}, {Geier}, {Gorosabel},
  {Hu}, {Milvang-Jensen}, {Sparre}, {Xin}, {Zhang}, {Zheng}, \& {Zou}}]{Xu+13}
{Xu}, D., {de Ugarte Postigo}, A., {Leloudas}, G., {et~al.} 2013, \apj, 776,
  98, \dodoi{10.1088/0004-637X/776/2/98}

\bibitem[{{Yang} {et~al.}(2022){Yang}, {Ai}, {Zhang}, {Zhang}, {Liu}, {Wang},
  {Yang}, {Yin}, {Li}, \& {L{\"u}}}]{Yang+22}
{Yang}, J., {Ai}, S., {Zhang}, B.-B., {et~al.} 2022, \nat, 612, 232,
  \dodoi{10.1038/s41586-022-05403-8}

\bibitem[{{Yang} {et~al.}(2023){Yang}, {Troja}, {O'Connor}, {Fryer}, {Im},
  {Durbak}, {Paek}, {Ricci}, {De Bom}, {Gillanders}, {Castro-Tirado}, {Peng},
  {Dichiara}, {Ryan}, {van Eerten}, {Dai}, {Chang}, {Choi}, {De}, {Hu},
  {Kilpatrick}, {Kutyrev}, {Jeong}, {Lee}, {Makler}, {Navarete}, \&
  {P{\'e}rez-Garc{\'\i}a}}]{Yang+23}
{Yang}, Y.-H., {Troja}, E., {O'Connor}, B., {et~al.} 2023, arXiv e-prints,
  arXiv:2308.00638, \dodoi{10.48550/arXiv.2308.00638}

\bibitem[{{Zenati} {et~al.}(2020){Zenati}, {Siegel}, {Metzger}, \&
  {Perets}}]{Zenati+20}
{Zenati}, Y., {Siegel}, D.~M., {Metzger}, B.~D., \& {Perets}, H.~B. 2020,
  \mnras, 499, 4097, \dodoi{10.1093/mnras/staa3002}

\bibitem[{{Zevin} {et~al.}(2019){Zevin}, {Kremer}, {Siegel}, {Coughlin},
  {Tsang}, {Berry}, \& {Kalogera}}]{Zevin+19}
{Zevin}, M., {Kremer}, K., {Siegel}, D.~M., {et~al.} 2019, ApJ, 886, 4,
  \dodoi{10.3847/1538-4357/ab498b}

\bibitem[{{Zevin} {et~al.}(2022){Zevin}, {Nugent}, {Adhikari}, {Fong}, {Holz},
  \& {Kelley}}]{Zevin+22}
{Zevin}, M., {Nugent}, A.~E., {Adhikari}, S., {et~al.} 2022, \apjl, 940, L18,
  \dodoi{10.3847/2041-8213/ac91cd}

\end{thebibliography}
